\documentclass[]{interact}
\usepackage[sort&compress,numbers]{natbib}
\usepackage{hyperref}
\usepackage[version=4]{mhchem}
\usepackage{xcolor}
\usepackage{caption}
\usepackage{microtype}
\usepackage[T1]{fontenc}
\usepackage{bm}

\theoremstyle{plain}

\theoremstyle{definition}

\theoremstyle{remark}

\begin{document}
\articletype{INVITED ARTICLE}
\title{Isotope Effects in Liquid Water via Deep Potential Molecular Dynamics}
\author{\name{
Hsin-Yu~Ko,\textsuperscript{a}
Linfeng~Zhang,\textsuperscript{b}\thanks{HYK and LZ contributed equally to this article.} Biswajit~Santra,\textsuperscript{c} Han~Wang,\textsuperscript{d} Weinan~E,\textsuperscript{b,e} Robert~A.~DiStasio~Jr.,\textsuperscript{f} and Roberto~Car\textsuperscript{a,b,g}\thanks{RC is the corresponding author. Email: rcar@princeton.edu}}
\affil{\textsuperscript{a}~Department of Chemistry, Princeton University, Princeton, NJ 08544; \textsuperscript{b}~Program in Applied and Computational Mathematics, Princeton University, Princeton, NJ 08544; \textsuperscript{c}~Department of Physics, Temple University, Philadelphia, PA 19122; \textsuperscript{d}~Laboratory of Computational Physics, Institute of Applied Physics and Computational Mathematics, Huayuan Road 6, Beijing 100088, P.R.~China; \textsuperscript{e}~Department of Mathematics, Princeton University, Princeton, NJ 08544. \textsuperscript{f}~Department of Chemistry and Chemical Biology, Cornell University, Ithaca, NY 14853; \textsuperscript{g}~Department of Physics and Princeton Institute for the Science and Technology of Materials, Princeton University, Princeton, NJ 08544.}
}
\maketitle

\begin{abstract}
A comprehensive microscopic understanding of ambient liquid water is a major challenge for \textit{ab initio} simulations as it simultaneously requires an accurate quantum mechanical description of the underlying potential energy surface (PES) as well as extensive sampling of configuration space.
Due to the presence of light atoms (e.g., \ce{H} or \ce{D}), nuclear quantum fluctuations lead to observable changes in the structural properties of liquid water (e.g., isotope effects), and therefore provide yet another challenge for \textit{ab initio} approaches.
In this work, we demonstrate that the combination of dispersion-inclusive hybrid density functional theory (DFT), the Feynman discretized path-integral (PI) approach, and machine learning (ML) constitutes a versatile \textit{ab initio} based framework that enables extensive sampling of both thermal and nuclear quantum fluctuations on a quite accurate underlying PES.
In particular, we employ the recently developed deep potential molecular dynamics (DPMD) model---a neural-network representation of the \textit{ab initio} PES---in conjunction with a PI approach based on the generalized Langevin equation (PIGLET) to investigate how isotope effects influence the structural properties of ambient liquid \ce{H2O} and \ce{D2O}.
Through a detailed analysis of the interference differential cross sections as well as several radial and angular distribution functions, we demonstrate that this approach can furnish a semi-quantitative prediction of these subtle isotope effects.
%
\end{abstract}


\section{Introduction}

Despite the relatively simple structure of a single water molecule, water has an unmatched complexity in the condensed (liquid) phase.
This complexity mainly originates from the delicate balance between weak non-covalent intermolecular interactions, e.g., the hydrogen bond (HB) network and van der Waals (vdW) dispersion, and thermal (entropic) effects~\cite{distasio_jr._individual_2014}.
On top of this delicate balance, nuclear quantum effects (NQEs)---such as zero-point motion---also affect the structural and dynamical properties of liquid water~\cite{ceriotti_nuclear_2016}.
This is primarily due to the presence of light atoms (such as \ce{H}) and the central role played by the tetrahedral HB network in determining aqueous properties in the condensed phase.

A salient example of how NQEs affect the structure of liquid water is the experimentally observed isotope effect upon substitution of \ce{H} to \ce{D}.
First demonstrated by Soper and Benmore using joint X-ray/neutron scattering experiments~\cite{soper_quantum_2008}, isotope effects were shown to manifest as significant covalent-bond contraction (by $\approx 3$\%) and HB elongation (by $\approx 4$\%) when comparing \ce{H2O} to \ce{D2O}.
In this study, the positions of the individual atoms were assigned based on the so-called empirical potential structure refinement (EPSR) method~\cite{soper_tests_2001,soper_partial_2005}, in which a physically motivated (but empirical) interatomic potential was utilized in conjunction with reverse Monte Carlo (MC) simulations to generate microscopic snapshots of liquid water (\ce{H2O} or \ce{D2O}) that reproduce the experimentally observed interference differential cross section, $F^{(\rm n)}_{\rm int} (Q)$ (\textit{vide infra}).
These EPSR-based atomic positions were then used to generate structural properties such as radial distribution functions (RDFs) and the \ce{OOO} angular distribution function (ADF), from which one can quantify how the \ce{H} $\rightarrow$ \ce{D} isotopic substitution affects the structure of liquid water.

The microscopic structure of liquid water can also be modeled using \textit{ab initio} molecular dynamics (AIMD) simulations, in which the nuclear potential energy surface (PES) is generated ``on the fly'' from the electronic ground state without the need for empirical input~\cite{car_unified_1985,marx_ab_2009}.
By generating a trajectory of configurations based on interatomic forces derived from a first-principles electronic structure theory, this technique allows for a quantum mechanical treatment of the structural and dynamical properties of complex condensed-phase systems (such as liquid water) at a given set of thermodynamic conditions. 
In addition, AIMD simulations can also describe the electronic and dielectric properties of the system, as well as any chemical reactions that may occur (i.e., bond cleavage/formation).

Due to its quite favorable balance between accuracy and computational cost, density functional theory (DFT)~\cite{hohenberg_inhomogeneous_1964,kohn_self-consistent_1965,parr_density-functional_1989} has emerged as the most commonly used electronic structure theory method during AIMD simulations of condensed-phase systems.
While DFT is (in principle) an exact theory, the functional form of the exchange-correlation energy still remains unknown; as such, DFT (in practice) relies on an established hierarchy of approximations which allows for (semi-)systematic improvements in accuracy with a corresponding increase in the computational cost~\cite{perdew_jacobs_2001}.
Previous studies have demonstrated that generalized-gradient approximation (GGA) based DFT functionals~\cite{becke_density-functional_1988,lee_development_1988,perdew_generalized_1996}---due to their propensity for self-interaction error~\cite{cohen_insights_2008} and lack of non-local correlation effects such as vdW dispersion interactions~\cite{hermann_first-principles_2017}---are inadequate for providing an accurate and reliable description of liquid water.
Instead, it is more appropriate to use constraint-based meta-GGA functionals (such as SCAN~\cite{sun_strongly_2015,chen_ab_2017,zheng_structural_2018,calegari_andrade_structure_2018}) or the class of more accurate vdW-inclusive hybrid functionals (such as PBE0-TS~\cite{perdew_rationale_1996,adamo_toward_1999,tkatchenko_accurate_2009,distasio_jr._individual_2014,ferri_electronic_2015,santra_local_2015,chen_hydroxide_2018} or revPBE0-D3~\cite{zhang_comment_1998,perdew_rationale_1996,adamo_toward_1999,goerigk_thorough_2011,cheng_ab_2019})  when describing condensed-phase aqueous systems.

During AIMD simulations, the nuclei are often treated as classical (point) particles moving on the underlying PES generated from an ``on-the-fly'' and fully quantum mechanical description of the electrons.
To rigorously include NQEs while sampling equilibrium structural properties, the Feynman discretized path-integral (PI)~\cite{fosdick_numerical_1962,chandler_exploiting_1981,marx_ab_1996,tuckerman_efficient_1996} approach is typically used.
In this approach, each nucleus is mapped onto a classical ring polymer comprised of $P$ beads (coupled via harmonic springs), allowing for a so-called PI-AIMD simulation of the system in which the electronic and nuclear degrees of freedom are both treated quantum mechanically.
At finite (but non-zero) temperatures, sampling of the exact quantum distribution of a system (in configuration space) can be accomplished to a desired level of accuracy with PI-AIMD simulations employing a finite number of beads.
However, standard PI-AIMD simulations of liquid water at ambient conditions ($300$~K,~$1$~bar) are still computationally demanding for the following reasons: (\textit{i}) they require many beads ($P \geq 32$) in each classical ring polymer to provide a sufficiently converged description of equilibrium structural properties~\cite{stern_quantum_2001,paesani_accurate_2006,habershon_competing_2009,ceriotti_accelerating_2011}; (\textit{ii}) they also require relatively long trajectories ($t > 20$~ps) to obtain sufficient statistical convergence for many physical quantities, e.g., RDFs and the equilibrium density.
The difficulty originating from the required number of beads has been the subject of intense research and can now be alleviated (to varying extents) using a number of different methods, e.g., generalized Langevin equation (GLE) based colored-noise thermostats~\cite{dammak_quantum_2009,ceriotti_nuclear_2009,ceriotti_efficient_2012}, ring polymer contraction~\cite{markland_refined_2008,markland_efficient_2008,fanourgakis_fast_2009,marsalek_ab_2016,kapil_accurate_2016}, and imaginary-time perturbed path integral approach~\cite{poltavsky_modeling_2016,poltavsky_perturbed_2017}, to name a few.
The sampling time issue is an intrinsic limitation of PI-AIMD (and AIMD) simulations that can only be addressed by more efficient algorithms and/or approximations for evaluating the underlying \textit{ab initio} PES.

Recent progress in the application of machine-learning (ML) techniques to represent PES for complex systems containing many atoms provides a way to overcome this last computational hurdle~\cite{behler2007generalized, bartok2010gaussian,rupp2012fast,montavon2013machine,chmiela2017machine,schutt_schnet:_2017,smith2017ani,han2017deep,zhang2018deepmd,zhang_end--end_2018}.
After training on high-quality \textit{ab initio} PES data, ML-based methods---with a
linear-scaling associated computational cost that is many orders of magnitude cheaper than AIMD---allow one to perform extensive MD simulations without any significant loss of accuracy.
In fact, such methods have already been used to study the complex behavior of liquid water~\cite{morawietz2016van, wang2018force, zhang2018deepmd, cheng_ab_2019}.
For example, Cheng and coworkers~\cite{cheng_ab_2019} recently used a neural network (NN) based potential to study the thermodynamics of liquid water and ice~I (I$h$ and I$c$) under ambient conditions; in doing so, they accurately predicted the RDFs of liquid water, the melting point of ice~I$h$, and the density maximum in liquid water.
Moreover, suitable extensions of these methods also allow one to further coarse-grain the system into beads representing each water molecule; this has the potential to significantly reduce the computational cost without losing the main structural information in the original system~\cite{zhang2018deepcg}.
Among the suite of existing ML methods, the deep potential molecular dynamics (DPMD) approach~\cite{han2017deep,zhang2018deepmd,zhang_end--end_2018} utilizes a deep NN to represent the many-body potential energy as a sum of auxiliary ``atomic'' energies associated with individual atoms in the system.
For a given atom, this ``atomic'' energy depends differentiably on its local environment, i.e., the relative coordinates of its vicinal surrounding atoms within a smooth radial distance cutoff.
A novel symmetry-preserving atomic local coordinate system is then introduced as input to the deep NN and an adaptive training procedure on high-quality \textit{ab initio} PES data is performed to optimize the NN parameters.
By reproducing the total potential energy, atomic forces, and stress tensor to well within the accuracy of the \textit{ab initio} data~\cite{han2017deep,zhang2018deepmd,zhang_end--end_2018}, the resulting DPMD model can recover structural properties that are essentially indistinguishable when compared to the AIMD results for a wide variety of systems (including liquid water).

In this work, we utilize DPMD and PI-DPMD simulations to investigate the experimentally observed structural changes~\cite{soper_quantum_2008} in liquid water upon isotopic substitution of \ce{H2O} with \ce{D2O}.
In particular, we demonstrate that the DPMD model---when trained on a single, relatively short ($\approx 8$~ps) PI-AIMD simulation of the lighter isotope (liquid \ce{H2O})---can be used to provide a (semi-)quantitative description of the isotope effects found in $F^{(\rm n)}_{\rm int} (Q)$, the oxygen--oxygen (\ce{OO}) RDF, and the oxygen--oxygen--oxygen (\ce{OOO}) ADF.
We rationalize this choice of training data by the fact that \ce{H2O} and \ce{D2O} share the same underlying PES, but are subject to different quantum fluctuations; hence, a PI-AIMD simulation of the single lighter isotope (liquid \ce{H2O}) covers the relevant sector of configuration space for both isotopes.
This conclusion is supported by independently testing that the DPMD model trained on quantum \ce{H2O} describes with similar accuracy classical \ce{H2O}.
In doing so, we show that the combination of vdW-inclusive hybrid DFT as the source of the underlying PES, the Feynman discretized PI approach for treating NQEs, and the highly efficient DPMD model constitutes a versatile \textit{ab initio} based framework that enables one to attack problems in chemistry, physics, and materials science that require extensive and simultaneous sampling of both thermal and nuclear quantum fluctuations on a quite accurate PES.

The remainder of the manuscript is organized as follows.
In Sec.~\ref{sec:method}, we describe the computational details required for obtaining the initial PES using vdW-inclusive hybrid DFT and the subsequent generation of the DPMD model parameters needed for performing DPMD and PI-DPMD simulations of liquid \ce{H2O} and \ce{D2O}.
In Sec.~\ref{sec:result-discussion}, we perform a detailed case study of how isotope effects manifest in the equilibrium structural properties of liquid water, including a comparative analysis of our theoretical findings with the experimental work of the Soper and Benmore~\cite{soper_quantum_2008}.
In Sec.~\ref{sec:conclusion}, we provide a brief summary of this work accompanied by some remarks about potential improvements to this approach for the study of liquid water and its many unusual properties.

\section{Computational methods\label{sec:method}}

Our study of the isotope effects in liquid \ce{H2O} and \ce{D2O} via DPMD and PI-DPMD simulations was performed in three stages.
The first stage is the generation of training data for the DPMD model based on a PI-AIMD simulation of liquid \ce{H2O} with vdW-inclusive hybrid DFT (Sec.~\ref{method:aimd}).
The second stage is the training of the DPMD model parameters based on this PI-AIMD trajectory (Sec.~\ref{method:dp_train}).
The third and final stage is the utilization of this model in DPMD and PI-DPMD simulations to equilibrate and sample the statistical ensembles of liquid \ce{H2O} and \ce{D2O} (Sec.~\ref{method:dp_md}).

\subsection{Generation of DPMD training data via PI-AIMD simulation of liquid \ce{H2O} \label{method:aimd}}

To generate the \textit{ab initio} PES data for liquid water that is required for training the DPMD model, we performed a single isobaric-isothermal ($NpT$) Born-Oppenheimer PI-AIMD simulation (with a time step of $\Delta t \approx 0.5$~fs) of a cubic periodic cell containing $64$~\ce{H2O} molecules at ambient ($300$~K and $1.0$~bar) conditions.
We started from a previously equilibrated $NpT$ PI-AIMD trajectory at the PBE0-TS level carried out at $330$~K and $1.0$~bar~\cite{santra_manuscript_water_nqe}, and performed an initial $\approx 1$~ps $NpT$ PI-AIMD simulation in which the temperature was gradually lowered from $330$~K to $300$~K.
A final $NpT$ PI-AIMD production run was then performed for $\approx 8$~ps under ambient conditions of $300$~K and $1$~bar; this \textit{ab initio} PES data will be used to train the DPMD model as described in Sec.~\ref{method:dp_train}.

In the PI-AIMD simulation, the nuclei were modeled with an $8$-bead ring polymer supplemented with a colored-noise generalized Langevin equation thermostat (i.e., PIGLET)~\cite{ceriotti_nuclear_2009,ceriotti_efficient_2012} to accelerate convergence with respect to the Trotter dimension.
The cell was treated classically using the Raiteri-Gale-Bussi approach~\cite{raiteri_reactive_2011} and was thermostatted with an additional generalized Langevin equation thermostat~\cite{ceriotti_colored-noise_2010}.
The cell mass was chosen to be consistent with a $200$~fs characteristic timescale, and the corresponding equations of motion for the ionic and cell degrees of freedom were integrated using the \texttt{i-PI} package~\cite{ceriotti_i-pi:_2014}.
To prevent shearing of the liquid, cubic cell symmetry was enforced during all $NpT$ simulations.

The ionic forces and stress tensor (which are required by \texttt{i-PI}) were computed at the PBE0-TS level~\cite{perdew_rationale_1996,adamo_toward_1999,tkatchenko_accurate_2009} using a linear-scaling exact-exchange algorithm~\cite{wu_order-n_2009,distasio_jr._individual_2014,HYK-paperI,HYK-paperII} and a self-consistent implementation of the Tkatchenko-Scheffler (TS) vdW correction~\cite{tkatchenko_accurate_2009,ferri_electronic_2015}, as provided in the \texttt{CP} package of \texttt{Quantum ESPRESSO}~\cite{giannozzi_quantum_2009,giannozzi_advanced_2017}.
The core electrons were treated with Hamann-Schl\"uter-Chiang-Vanderbilt (HSCV) norm-conserving pseudopotentials~\cite{hamann_norm-conserving_1979,vanderbilt_optimally_1985}, as distributed with the \texttt{Qbox} suite of programs~\cite{gygi_architecture_2008}, while the valence (pseudo-)wavefunctions were represented explicitly with a planewave basis set.
To maintain a constant planewave kinetic energy cutoff of $115$~Ry during the $NpT$ simulation, we followed the procedure of Bernasconi \textit{et al.}~\cite{bernasconi_first-principle-constant_1995} by choosing: (\textit{i}) a cubic reference cell (with $L = 24.7$~Bohr) that is large enough to cover the fluctuations along each lattice vector of the simulation cell throughout the $NpT$ trajectory, and (\textit{ii}) a corresponding planewave basis set with a larger kinetic energy cutoff of $125$~Ry.
During the $NpT$ simulation, planewaves with a kinetic energy beyond the desired cutoff of $115$~Ry were smoothly penalized by changing 
\begin{equation}
  G^{2} \rightarrow G^{2} + A \left[ 1 + \text{erf} \, \left( \frac{\frac{G^{2}}{2} - E_{0}}{\sigma} \right) \right] .
  \label{eq:ecutfix}
\end{equation}
With a judicious choice of parameters ($A=200$~Ry, $\sigma=15$~Ry, $E_{0}=115$~Ry), this modification to $G^2$ causes the higher-energy ($> 115$~Ry) planewaves to become essentially inactive basis functions in the description of the valence (pseudo-)wavefunctions, and thereby leads to $NpT$ dynamics which mimic a constant planewave cutoff of $\approx 115$~Ry.

The electronic ground state was obtained using nested second-order damped Car-Parrinello dynamics~\cite{tassone_acceleration_1994} in which the ionic and cell degrees of freedom were kept fixed.
During the electron minimization, we employed a time step of $\approx 0.125$~fs, a fictitious electron mass of $200$~au, and a mass preconditioning cutoff of $6$~Ry.
Additional computational details regarding this PI-AIMD simulation will be provided in a forthcoming paper~\cite{santra_manuscript_water_nqe}.
The electronic structure calculation is considered converged when the successive second-order-damped CP dynamics has a change in total potential energy below $1.1\times 10^{-6}$~au and a change of mean absolute ionic forces within $3.0\times 10^{-4}$~au.

\subsection{Training of the DPMD model \label{method:dp_train}}

To train the DPMD model, we follow Ref~\cite{zhang2018deepmd} to determine a deep NN potential based on the total potential energy $E$, the ionic forces $\{\bm{F}_i\}$ (for all $N$ ions in the system), and the virial $\bm \Xi$ (or equivalently, the stress tensor) at each step in the $\approx 8$~ps production PI-AIMD simulation described above.
During the training of the DPMD model, we minimize the following loss function $\mathcal{L}$, which depends on the tunable parameters $p_\epsilon$, $p_f$, and $p_\xi$, and is defined as follows:
\begin{align}\label{eqn:loss}
  \mathcal L(p_\epsilon, p_f, p_\xi) = {p_\epsilon} \Delta \epsilon^2 
  + \frac{p_f}{3N} \sum_i \vert \Delta\bm{F}_i\vert^2 
  + \frac{p_\xi}{9} \sum_{ij} \vert \Delta \xi_{ij} \vert^2 .
\end{align}
In this expression, $\Delta \epsilon$, $\Delta \bm{F}_i$, and $\Delta \bm{\xi}$ are the differences between the current DPMD prediction and the training data for the quantities $\epsilon \equiv E/N$, $\bm{F}_i$, and $\bm{\xi} \equiv \bm{\Xi}/N$, respectively.
To optimize the parameters in the DPMD model, we used the Adam method~\cite{Kingma2015adam} with a learning rate that decays exponentially with the number of training steps.
In practice, we observed that more efficient training can be achieved by linearly varying the parameters with respect to the learning rate (i.e., increasing $p_\epsilon$ and $p_\xi$ while decreasing $p_f$); in doing so, one can achieve a well-balanced training procedure in which the energy, ionic forces, and virial are mutually consistent.
All DPMD model training was performed using the \texttt{DeePMD-kit} package~\cite{wang2018kit} interfaced with the \texttt{TensorFlow} library~\cite{tensorflow2015-whitepaper}.

\subsection{DPMD and PI-DPMD simulations of liquid \ce{H2O} and \ce{D2O} \label{method:dp_md}}

To reduce finite-size effects and the statistical errors associated with relatively short PI-AIMD simulations, we extended the system size by one order of magnitude to include $512$ \ce{H2O} molecules and the simulation timescale by two orders of magnitude to span $\approx 1$~ns during the DPMD and PI-DPMD simulations described below.
A single $NpT$ DPMD simulation (in which the nuclei are treated classically) of liquid water at ambient ($300$~K and $1$~bar) conditions was performed using the LAMMPS package~\cite{plimpton_fast_1995}.
The temperature and pressure were controlled by Nos\'e-Hoover thermostat and barostat chains~\cite{tuckerman_liouville-operator_2006} (with characteristic timescales of $0.1$~ps and $0.5$~ps, respectively), and an integration time step of $\approx 0.5$~fs.
To perform PI-DPMD simulations of liquid \ce{H2O} and \ce{D2O}, we integrated the \texttt{DeePMD-kit} and \texttt{i-PI} packages.
All PI-DPMD simulations were performed using exactly the same settings as the original PI-AIMD simulation (see Sec.~\ref{method:aimd}), including identical thermodynamic conditions, sampling techniques, and time step.
We then made the substitution of \ce{H} to \ce{D}, and repeated the PI-DPMD simulation for heavy water.
For both the DPMD and PI-DPMD simulations, we enforced cubic cell symmetry as described in Sec.~\ref{method:aimd}.

\section{Results and discussion\label{sec:result-discussion}}

By performing DPMD and PI-DPMD simulations trained on \textit{ab initio} PES data with extended time ($\approx 1$~ns) and length (\ce{(H2O)512}) scales, this approach mitigates several of the computational challenges associated with modeling complex condensed-phase systems like liquid water. 
As an application, we use this \textit{ab initio} based framework to investigate how isotope effects influence the structural properties of ambient liquid \ce{H2O} and \ce{D2O}.
Based on extensive $NpT$ PI-DPMD simulations of ambient liquid \ce{H2O} and \ce{D2O}, we obtained essentially identical equilibrium atomic number densities of $\rho = 0.1022\pm 0.0001$~atoms/\AA{}$^3$ for these systems.
This value for $\rho$ is slightly higher than the experimental observations of $\rho = 0.10007$~atoms/\AA{}$^3$ (\ce{H2O}) and $0.10000$~atoms/\AA{}$^3$ (\ce{D2O}) at $298$~K.
While the isotope effects on macroscopic properties such as $\rho$ are negligible, the isotope effects on the microscopic structural properties of liquid water are still small (on the order of $3\mathrm{-}4\%$ from experiment~\cite{soper_quantum_2008}), but well above the statistical uncertainty of our simulations.
In this section, we investigate the isotope effects on the microscopic structural properties of liquid water predicted by the PI-DPMD simulations.
In particular, we first compare against the interference differential cross section $F^{(\rm n)}_{\rm int} (Q)$---a reciprocal-space quantity that is directly measured by experiment---to initially assess the accuracy of our simulations (Sec.~\ref{result:fq}).
This is followed by a detailed comparative analysis with the following experimentally (EPSR) determined real-space structural quantities: the oxygen--oxygen (\ce{OO}), oxygen--hydrogen (\ce{OH}/\ce{OD}), and hydrogen--hydrogen (\ce{HH}/\ce{DD}) radial distribution functions (RDFs) in Sec.~\ref{result:rdf}, and the oxygen--oxygen--oxygen (\ce{OOO}) angular distribution function (ADF) in Sec.~\ref{result:adf}.
We complete this section by describing a potential route towards improving the predictions of our approach using other electronic structure theory models (Sec.~\ref{result:improve_es}).

\subsection{Isotope effects on the microscopic structure of liquid water: Interference differential cross section \label{result:fq}}

To provide a comparison with a directly measurable experimental quantity (and sidestep any potential bias from the EPSR post-processing of the raw experimental data), we first computed $F^{(\rm n)}_{\rm int} (Q)$ from our PI-DPMD trajectories as follows:
\begin{equation}
    F^{(\rm n)}_{\rm int} (Q) =
    \begin{cases}
        c^2_{\ce{O}}b^2_{\ce{O}} S_{\ce{O}\ce{O}}(Q)+2c_{\ce{O}}c_{\ce{H}}b_{\ce{O}}b_{\ce{H}} S_{\ce{O}\ce{H}}(Q) + c^2_{\ce{H}}b^2_{\ce{H}} S_{\ce{H}\ce{H}}(Q) \qquad \text{for \ce{H2O}} \\[0.4em]
        c^2_{\ce{O}}b^2_{\ce{O}} S_{\ce{O}\ce{O}}(Q)+2c_{\ce{O}}c_{\ce{D}}b_{\ce{O}}b_{\ce{D}} S_{\ce{O}\ce{D}}(Q) + c^2_{\ce{D}}b^2_{\ce{D}} S_{\ce{D}\ce{D}}(Q) \qquad \text{for \ce{D2O}}  \\
    \end{cases}
    \label{eq:fq}
\end{equation}
in which $c_{\alpha}$ is the atom fraction for the $\alpha$-th species ($c_{\ce{O}}= 1/3$, $c_{\ce{H}}= 2/3$, and $c_{\ce{D}}= 2/3$), $b_{\alpha}$ is the neutron scattering length of the $\alpha$-th species ($b_{\ce{O}} = 5.80$~fm, $b_{\ce{H}} = -3.74$~fm, and $b_{\ce{D}} = 6.67$~fm),\footnote{Data source: https://www.ncnr.nist.gov/resources/n-lengths/elements/} and $S_{\alpha\beta}(Q)$ is the partial structure factor for the $\alpha\beta$ species pair. In this work, the reciprocal-space $S_{\alpha\beta}(Q)$ quantity was computed based on the corresponding RDF for the $\alpha\beta$ species pair ($g_{\alpha\beta}(r)$) as follows:
\begin{equation}
    S_{\alpha\beta}(Q) = 4 \pi \rho \int \text{d}r \, r^2 \left[ g_{\alpha\beta}(r) - 1 \right] \frac{\sin Q r}{Q r} .
    \label{eq:sq}
\end{equation}

%
%
\begin{figure}[t!]
    \centering
    \includegraphics[width=0.7\linewidth]{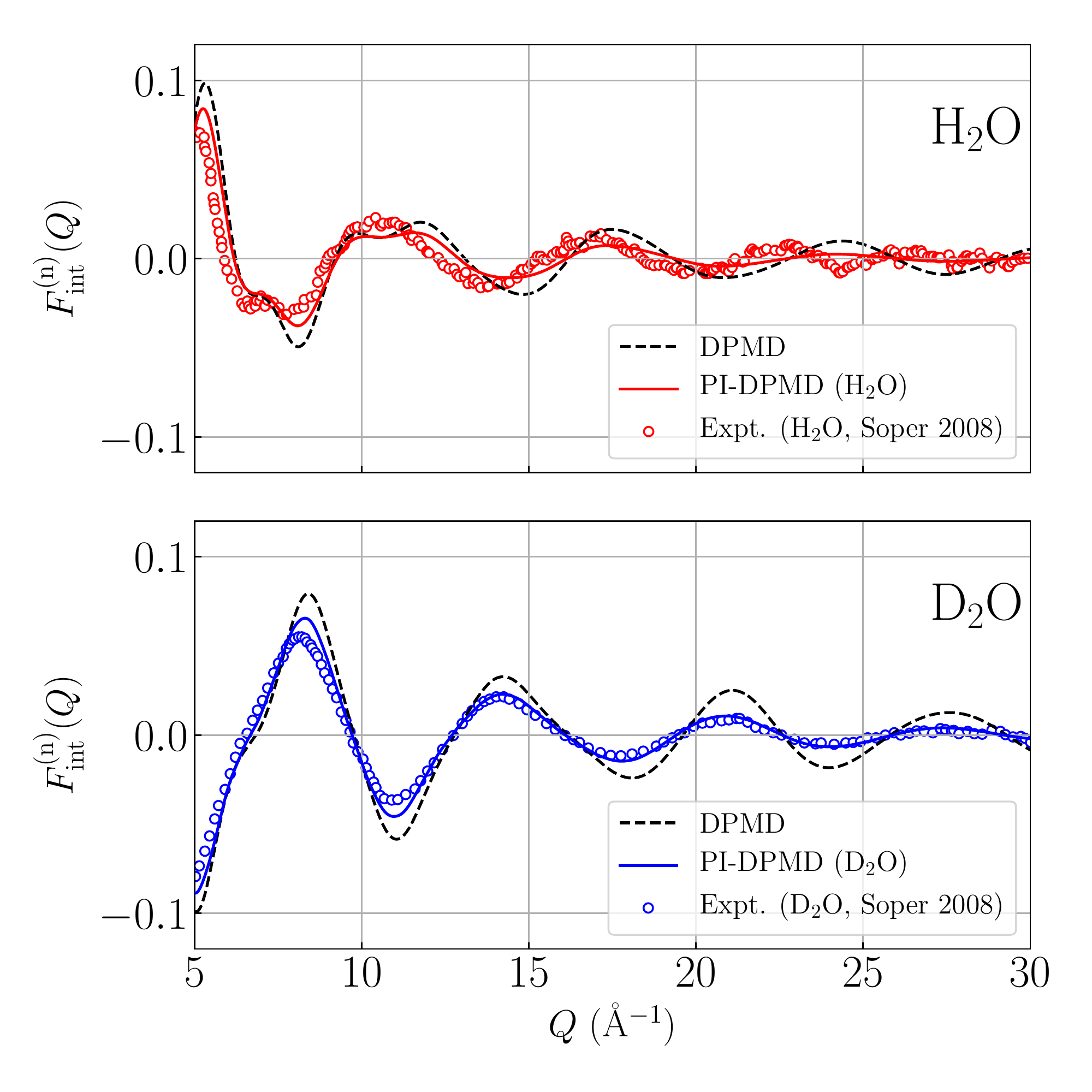}
    \caption{
    Comparison of DPMD and PI-DPMD in predicting the experimentally measured $F^{(\rm n)}_{\rm int} (Q)$ of \ce{H2O} (\textit{top panel}) and \ce{D2O} (\textit{bottom panel}) under ambient thermodynamic conditions~\cite{soper_quantum_2008}.
    All PI-DPMD predictions (solid lines) and experimental results (open circles) are colored red for \ce{H2O} and blue for \ce{D2O}.
    The DPMD simulation results (in which the nuclei are treated classically) are plotted as dashed black lines in both panels.
    }
    \label{fig:fq}
\end{figure}
%
%
In Fig.~\ref{fig:fq}, we compare the $F^{(\rm n)}_{\rm int} (Q)$ computed from DPMD simulations (in which the nuclei are treated classically) and 
PI-DPMD simulations (in which the nuclei are treated quantum mechanically) against the experimental results of Soper and Benmore~\cite{soper_quantum_2008}.
For $Q<15$~\AA{}$^{-1}$, the line shape of $F^{(\rm n)}_{\rm int} (Q)$ originates from the combined shapes of $S_{\ce{OO}}$, $S_{\ce{OH}}$ ($S_{\ce{OD}}$), and $S_{\ce{HH}}$ ($S_{\ce{DD}}$).
For higher $Q$, the \ce{OH} (\ce{OD}) contribution dominates $F^{(\rm n)}_{\rm int} (Q)$.
As a consequence, the relative amplitude and (approximately) opposite phase in $F^{(\rm n)}_{\rm int} (Q)$ for $Q>15$~\AA{}$^{-1}$ between \ce{H2O} and \ce{D2O} mainly arise from the ratio in the scattering length $b_{\ce{H}}/b_{\ce{D}} \approx -0.56$.
From the apparent differences in $F^{(\rm n)}_{\rm int} (Q)$ between the DPMD and PI-DPMD simulations (for both \ce{H2O} and \ce{D2O}), it is clear that NQEs play an important role in determining the structural properties of ambient liquid water.
By accounting for NQEs in the PI-DPMD simulations, the theoretical predictions for $F^{(\rm n)}_{\rm int} (Q)$ are in significantly improved agreement with experiment.
Even with extensive sampling of thermal and nuclear quantum fluctuations, however, small differences still remain between the theoretical and experimental results. 
In this regard, more pronounced discrepancies were observed in the $F^{(\rm n)}_{\rm int} (Q)$ for \ce{H2O}; since this quantity is a linear combination of $S_{\ce{OO}}$, $S_{\ce{OH}}$, and $S_{\ce{HH}}$ via Eqs.~\eqref{eq:fq} and \eqref{eq:sq}, one can argue that such differences primarily arise from deficiencies in the theoretical determinations of these structure factors.
Assuming that our sampling of the thermal and nuclear quantum fluctuations in this system is indeed exhaustive and the DPMD model reproduces the underlying \textit{ab initio} PES~\cite{zhang2018deepmd,zhang_end--end_2018}, this finding suggests that further improvement may be needed in the model used to describe the electronic structure.
Although the nuclei in \ce{H2O} and \ce{D2O} move on the same PES in the Born-Oppenheimer approximation, the lighter \ce{H} atoms are more quantum mechanical in nature and will explore more of the PES than the heavier (and more classical) \ce{D} atoms at a given temperature.
As such, PI-DPMD simulations of liquid \ce{H2O} are more sensitive to the finer details of the underlying PES (e.g., surface curvature, barrier heights, and number/location/topology of local minima).
Since small deviations from the true PES (as derived from the exact solution to the time-independent Schr\"odinger equation) will lead to incorrect sampling of this surface during PI simulations (e.g., overestimation/underestimation of anharmonicity, incorrect relative populations of different minima/basins), further improvements in the electronic structure model are expected to improve our theoretical prediction of $F^{(\rm n)}_{\rm int} (Q)$ for liquid water.
Additional support for this conjecture is provided in Sec.~\ref{result:rdf}.

\subsection{Isotope effects on the microscopic structure of liquid water: Radial distribution functions \label{result:rdf}}

\begin{figure}[t!]
    \centering
    \includegraphics[width=0.75\linewidth]{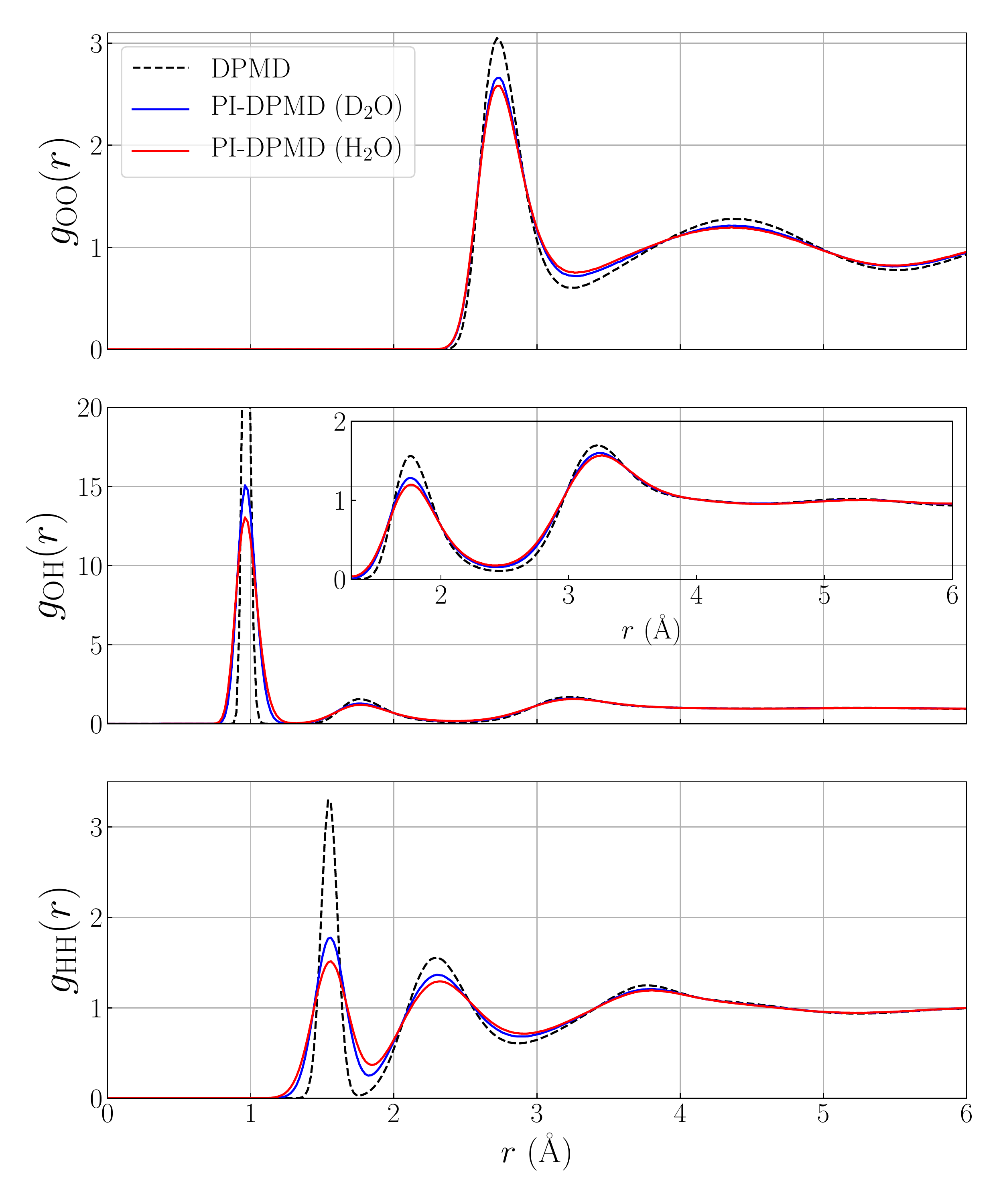}
    \caption{
    Comparison of the DPMD and PI-DPMD predictions of the following RDFs in liquid \ce{H2O} and \ce{D2O} at ambient conditions ($300$~K,~$1$~bar): $g_{\ce{OO}}(\bm r)$ (\textit{top panel}), $g_{\ce{OH}}(\bm r)$ and $g_{\ce{OD}}(\bm r)$ (\textit{middle panel}), and $g_{\ce{HH}}(\bm r)$ and $g_{\ce{DD}}(\bm r)$ (\textit{bottom panel}).
    All PI-DPMD predictions (solid lines) are colored red for \ce{H2O} and blue for \ce{D2O}, while the DPMD results are plotted as dashed black lines in all three panels.
    The inset in the middle panel provides a magnified view of the isotope effects beyond the first peak in the $g_{\ce{OH}}(\bm r)$ and $g_{\ce{OD}}(\bm r)$ RDFs.
    }
    \label{fig:rdf}
\end{figure}

To perform a more intuitive analysis of the isotope effects on the microscopic structure of liquid water, we now consider how the isotopic substitution of \ce{H} to \ce{D} influences the oxygen--oxygen (\ce{OO}), oxygen--hydrogen (\ce{OH}/\ce{OD}), and hydrogen--hydrogen (\ce{HH}/\ce{DD}) radial distribution functions (RDFs) of \ce{H2O} and \ce{D2O}.
RDFs are real-space quantities that measure the probability of finding a pair of atoms (of a given type) as a function of their radial distance; these quantities are related to the structure factors used in determining $F^{(\rm n)}_{\rm int} (Q)$ above (cf. Eqs.~\eqref{eq:fq} and \eqref{eq:sq}), and are commonly used in studying the structure of liquids.
Based on the DPMD and PI-DPMD simulations described above, we computed the $g_{\ce{OO}}(\bm r)$, $g_{\ce{OH}}(\bm r)$, and $g_{\ce{HH}}(\bm r)$ RDFs for liquid \ce{H2O} ($g_{\ce{OO}}(\bm r)$, $g_{\ce{OD}}(\bm r)$, and $g_{\ce{DD}}(\bm r)$ for liquid \ce{D2O}) and plotted them in Fig.~\ref{fig:rdf}.
%
%
%
%
From this figure, one can see that the isotopic substitution \ce{H} $\rightarrow$ \ce{D} slightly increases the structure of each RDF.
This difference primarily originates from the reduced zero-point motion in \ce{D2O}, and is consistent with the changes observed in the RDFs of liquid \ce{H2O} and \ce{D2O} obtained from simulations which account for NQEs~\cite{ceriotti_nuclear_2016}.
Before entering into a detailed discussion on the $g_{\ce{OO}}(\bm r)$---a quantity that can also be determined via X-ray scattering experiments~\cite{skinner_benchmark_2013}---we first analyze the \ce{OH}/\ce{OD} and \ce{HH}/\ce{DD} RDFs.

\begin{figure}[ht!]
    \centering
    \includegraphics[width=0.8\linewidth]{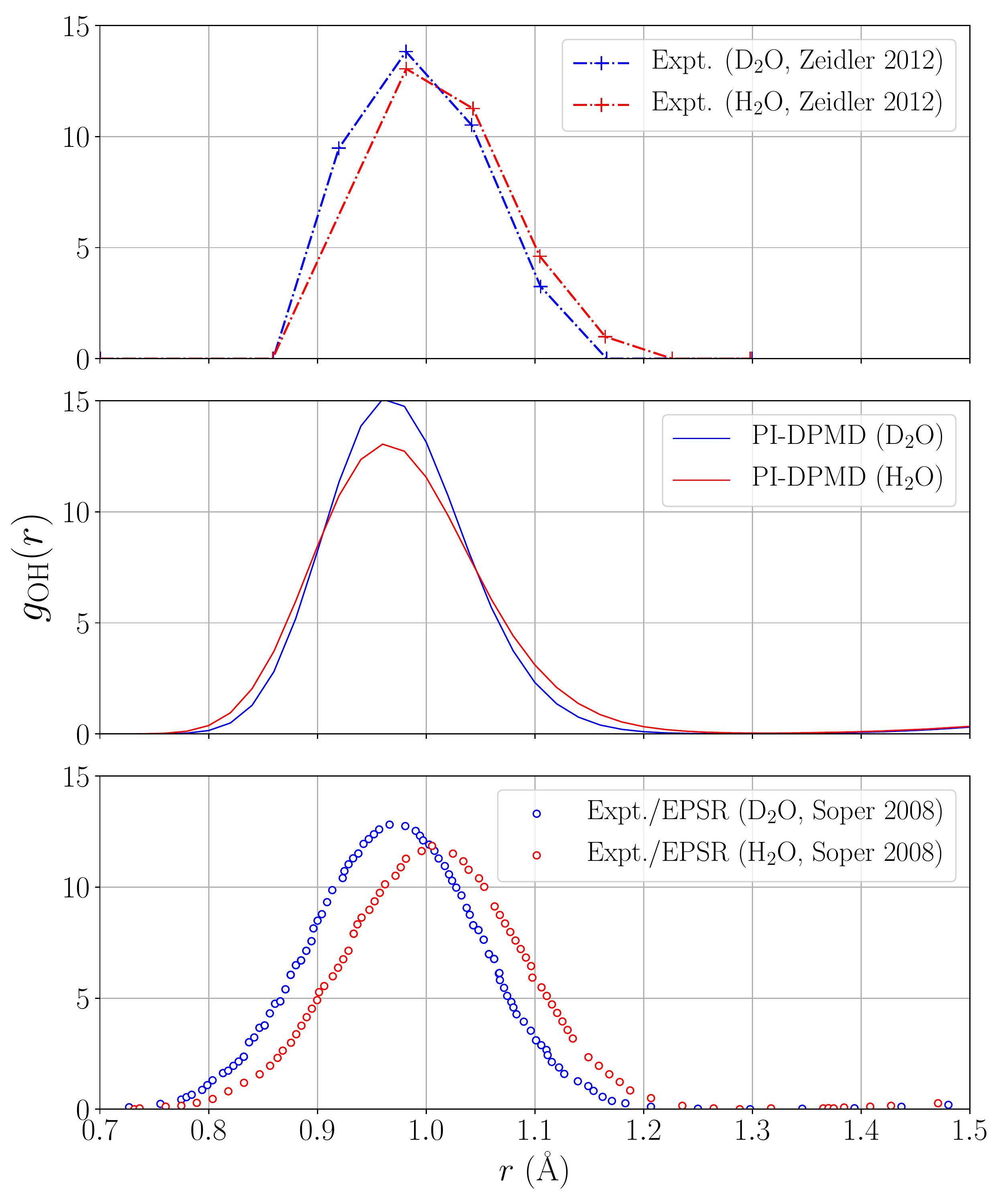}
    \caption{Intramolecular RDF for \ce{OH} (red) and \ce{OD} (blue) from: oxygen substitution experiment~\cite{zeidler_oxygen_2011,zeidler_isotope_2012} (\textit{upper panel}), PI-DPMD (\textit{middle panel}), and the experimental/EPSR~\cite{soper_quantum_2008} data reported in Ref.~\cite{zeidler_isotope_2012} (\textit{lower panel}).}
    \label{fig:compare_goh_expt}
\end{figure}

\begin{table}[ht!]
  \caption{
  Comparison of DPMD, PI-DPMD, and experimental/EPSR assigned~\cite{soper_quantum_2008} average structural properties (in \AA) in liquid \ce{H2O} and \ce{D2O} at ambient conditions ($300$~K,~$1$~bar).
  $d_{\ce{OH}}$ ($d_{\ce{OD}}$): \ce{OH} (\ce{OD}) covalent bond lengths; $d_{\ce{O}\cdots\ce{H}}$ ($d_{\ce{O}\cdots\ce{D}}$): \ce{O}$\cdots$\ce{H} (\ce{O}$\cdots$\ce{D}) HB lengths; $d_{\ce{HH}}$ ($d_{\ce{DD}}$): (shortest) intermolecular \ce{HH} (\ce{DD}) distances.
  All standard errors are smaller than $0.01$~\AA{} and are therefore omitted for clarity.
  }
  \centering
  \begin{tabular}{c|cc|cc|cc}
  \hline\hline
                          & $d_{\ce{OH}}$  & $d_{\ce{OD}}$ & $d_{\ce{O}\cdots\ce{H}}$ & $d_{\ce{O}\cdots\ce{D}}$ & $d_{\ce{HH}}$ & $d_{\ce{DD}}$ \\
  \hline
  DPMD                    & $0.98$ & $0.98$ & 1.76 & 1.76 & 2.30 & 2.30 \\
  PI-DPMD                 & $1.00$ & $0.99$ & $1.76$  & $1.76$ & $2.32$& $2.30$ \\
  Expt./EPSR (Soper 2008) & $1.01$ & $0.98$ & $1.74$  & $1.81$ & $2.42$& $2.37$ \\
  \hline\hline
  \end{tabular}
  \label{tab:covalent_bond}
\end{table}
Following the procedure outlined by Soper and Benmore~\cite{soper_quantum_2008}, we computed the following structural properties based on our PI-DPMD simulations of liquid \ce{H2O} and \ce{D2O}: (\textit{i}) the \ce{OH} and \ce{OD} covalent bond lengths ($d_{\ce{OH}}$ and $d_{\ce{OD}}$), (\textit{ii}) the \ce{O}$\cdots$\ce{H} and \ce{O}$\cdots$\ce{D} HB lengths ($d_{\ce{O}\cdots\ce{H}}$ and $d_{\ce{O}\cdots\ce{D}}$), and (\textit{iii}) the (shortest) intermolecular \ce{HH} and \ce{DD} distances ($d_{\ce{HH}}$ and $d_{\ce{DD}}$).
This data is summarized in Table~\ref{tab:covalent_bond}, where we find that PI-DPMD simulations predict that isotopic substitution from \ce{H} to \ce{D} leads to an $\approx 1\%$ decrease in the covalent bond, a negligible ($< 1\%$) change in the HB length, and an $\approx 1\%$ decrease in the (shortest) intermolecular \ce{HH}/\ce{DD} distance.
These findings are consistent with but smaller than the experimental/EPSR assignment of an $\approx 3\%$ contraction in $d_{\ce{OD}}$, an $\approx 4\%$ elongation in $d_{\ce{O}\cdots\ce{D}}$, and an $\approx 2\%$ contraction in $d_{\ce{DD}}$.
The experimental/EPSR assignment of these structural quantities requires post-processing of the directly observable experimental data (i.e., $F^{(\rm n)}_{\rm int} (Q)$).
%
A direct experimental probe of the intramolecular $g_{\ce{OH}}(r)$ ($g_{\ce{OD}}(r)$) would be required to better assess the intramolecular isotope effect.
This was provided by an elegant oxygen istope substitution experiment by Zeidler \textit{et al.}~\cite{zeidler_oxygen_2011,zeidler_isotope_2012}, in which they measured the difference between the neutron scattering cross section of \ce{H2}$^{18}$\ce{O} (i.e., $F_{\rm int}^{\rm (n, \ce{H2}^{18}\ce{O})} (Q)$) and of naturally occurring \ce{H2O} (i.e., $F_{\rm int}^{\rm (n, \ce{H2O})} (Q)$).
If the oxygen isotope effect on $g^{\ce{(H2O)}}_{\ce{OO}}(r)$ and $g^{\ce{(H2O)}}_{\ce{OH}}(r)$ (as well as on $g^{\ce{(D2O)}}_{\ce{OO}}(r)$ and $g^{\ce{(D2O)}}_{\ce{OD}}(r)$) is negligible, as one would expect, the inverse Fourier transform (\texttt{FT}$^{-1}$) of the measured cross sections would give for light and heavy water, respectively:
\begin{equation}
  \begin{split}
    \Delta G_{\ce{H}} (r) =\,\,& \texttt{FT}^{-1}[F_{\rm int}^{\rm (n, \ce{H2}^{18}\ce{O})} (Q)-F_{\rm int}^{\rm (n, \ce{H2O})} (Q)] \\
    =\,\,& c_{\ce{O}}^2 \left(b^2_{^{18}\ce{O}}-b^2_{^{\rm nat}\ce{O}} \right) \left[ g^{\ce{(H2O)}}_{\ce{OO}}(r) - 1 \right] \\
    &+ 2 c_{\ce{O}} c_{\ce{H}} b_{\ce{H}}\left(b_{^{18}\ce{O}}-b_{^{\rm nat}\ce{O}} \right) \left[ g^{\ce{(H2O)}}_{\ce{OH}}(r) - 1 \right] ,
  \end{split}
  \label{eq:deltaG_H}
\end{equation}
\begin{equation}
  \begin{split}
    \Delta G_{\ce{D}} (r) =\,\,& c_{\ce{O}}^2 \left(b^2_{^{18}\ce{O}}-b^2_{^{\rm nat}\ce{O}} \right) \left[ g^{\ce{(D2O)}}_{\ce{OO}}(r) - 1 \right] \\
    &+ 2 c_{\ce{O}} c_{\ce{D}} b_{\ce{D}}\left(b_{^{18}\ce{O}}-b_{^{\rm nat}\ce{O}} \right) \left[ g^{\ce{(D2O)}}_{\ce{OD}}(r) - 1 \right] ,
  \end{split}
  \label{eq:deltaG_D}
\end{equation}
in which $b_{^{18}\ce{O}} = 6.01$~fm and $b_{^{\rm nat}\ce{O}} = 5.80$~fm are the relevant neutron scattering lengths.
In the $r$ range corresponding intramolecular \ce{OH} (\ce{OD}) bonds, $g^{\ce{(H2O)}}_{\ce{OO}}(r)$ and $g^{\ce{(D2O)}}_{\ce{OO}}(r)$ are both equal to zero.
Thus, the intramolecular $g^{\ce{(H2O)}}_{\ce{OH}}(r)$ and $g^{\ce{(D2O)}}_{\ce{OD}}(r)$ can be extracted from Eqs.~\eqref{eq:deltaG_H} and \eqref{eq:deltaG_D}, respectively.
The results are displayed in the upper panel of Fig.~\ref{fig:compare_goh_expt}, which also shows in the middle panel the results of our simulation, and in the lower panel the corresponding RDFs from EPSR~\cite{soper_quantum_2008} as reported in Ref.~\cite{zeidler_isotope_2012}.       
%
%
%
%
%
The data points from the oxygen substitution experiment are few and affected by uncertainty, reflecting the difficulty of the experiment.
However, the \ce{OH} to \ce{OD} bond contraction is clear and in qualitative agreement with both EPSR and PI-DPMD data.
To quantify the effect, Zeidler \textit{et al.} estimated the peak position of the RDFs by spline fitting obtaining a contraction of $\approx 0.5$\%, smaller than the $\approx 3$\% effect estimated from EPSR but closer to our PI-DPMD estimate of $\approx 1$\%.
Interestingly, the PI-DPMD RDFs are asymmetric around the peak position, reflecting the anharmonicity of the potential experienced by \ce{H} (\ce{D}), which is steeper when \ce{H} (\ce{D}) approaches oxygen.
In spite of the limited resolution, such anharmonicity seems to be present in the oxygen substitution experiment.
It is absent, however, in the EPSR data in which a harmonic approximation was used for the \ce{OH} (\ce{OD}) potential.
Further experiments would seem necessary to quantify accurately the \ce{OH} to \ce{OD} bond contraction.
However, a $3$\% contraction seems excessive in view of the fact that a contraction of only $\approx 0.4$\% is measured in the vapor phase for the molecule in the roto-vibrational ground state~\cite{cook_molecular_1974}, and a similar contraction seems to occur also in ice I$h$, albeit with a larger uncertainty~\cite{franks_structure_1986}.

In spite of the uncertainties that affect both theory and experiment significant physical insight can be gleaned from the comparison of PI-DPMD simulations and EPSR structures.  
Both approaches characterize the isotope effects as small and subtle changes to the microscopic structure of liquid water.
This finding is also consistent with the differences observed between the predicted structural properties from DPMD simulations (in which the nuclei are treated classically) and PI-DPMD simulations of heavy (and hence more classical) liquid water (see Fig.~\ref{fig:rdf}).

%
%
\begin{figure}[ht!]
    \centering
    \includegraphics[width=0.8\linewidth]{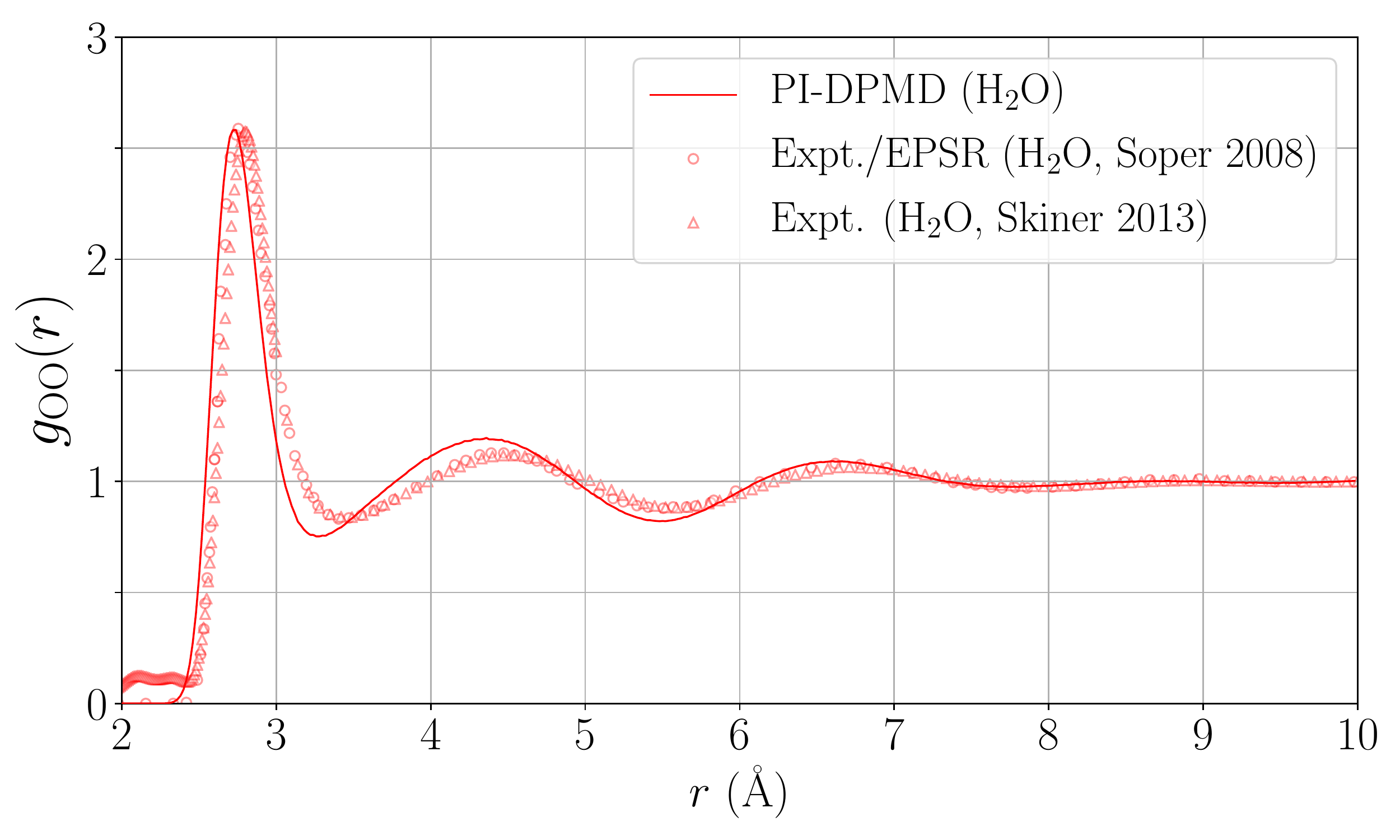}\\
    \includegraphics[width=0.8\linewidth]{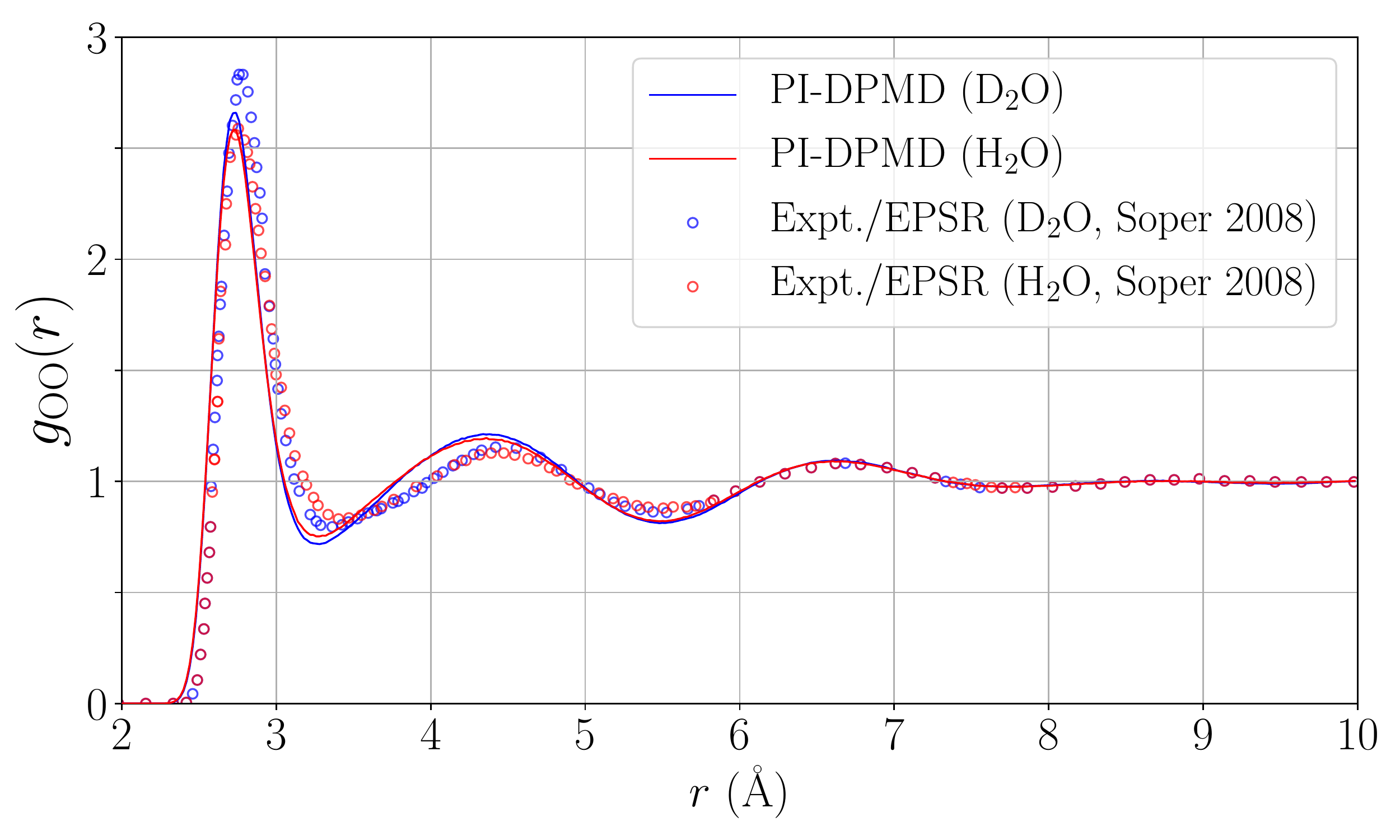}
    \caption{
    (\textit{Top panel})
    Comparison of $g_{\ce{OO}}(\bm r)$ in \ce{H2O} from PI-DPMD simulations (solid lines) and experiments (open circles for the experimental/EPSR assignment~\cite{soper_quantum_2008} and open triangles for the benchmark X-ray diffraction study~\cite{skinner_benchmark_2013}).
    The saturation of the red color in each curve has been adjusted to enhance the visibility.
    (\textit{Bottom panel})
    Comparison of the PI-DPMD predictions of $g_{\ce{OO}}(\bm r)$ with assignments from the experimental/EPSR procedure~\cite{soper_quantum_2008} for both liquid \ce{H2O} and \ce{D2O}.
    All PI-DPMD predictions (solid lines) and experimental results (open circles) are colored red for \ce{H2O} and blue for \ce{D2O}.
    }
    \label{fig:compare_goor_expt}
\end{figure}
%
%
Now we turn our discussion to $g_{\ce{OO}}(\bm r)$ in liquid \ce{H2O} and \ce{D2O}.
In Fig.~\ref{fig:compare_goor_expt}, we plot this quantity as computed from PI-DPMD simulations, the experimental/EPSR procedure~\cite{soper_quantum_2008}, and a benchmark X-ray diffraction experiment~\cite{skinner_benchmark_2013}.
Focusing on liquid \ce{H2O}, PI-DPMD underestimates the \ce{OO} distance, as its $g_{\ce{OO}}(r)$ is characterized by a first peak that is shorter by $\approx 0.07$~\AA{} (around $2.5$\%) than the benchmark X-ray experiment value of $2.80$~\AA{}.
Interestingly, the benchmark X-ray assignment of the first-peak position also agrees very well with a more sophisticated quantum Monte Carlo prediction at $2.80$~\AA{}~\cite{zen_ab_2015}, reflecting the limitation of the PBE0-TS model.
Such underestimation in the shortest \ce{OO} distance is consistent with previous simulations~\cite{distasio_jr._individual_2014,zen_ab_2015}.
This underestimation in the shortest \ce{OO} distance is likely a consequence of the residual self-interaction error that is still present in the PBE0 functional and has also been reported for water clusters~\cite{santra_accuracy_2007,santra_accuracy_2008}.
When compared with the benchmark X-ray result, the experimental/EPSR assignment also slightly underestimates the position of the first peak in the $g_{\ce{OO}}(\bm r)$ at $2.77$~\AA{}.
While $g_{\ce{OO}}(\bm r)$ in both assignments are based on X-ray scattering data, the benchmark result seems more accurate due to its higher resolution---a resolution up to $Q_{\rm max} \approx 24$~\AA{}$^{-1}$~\cite{skinner_benchmark_2013}, which is higher than the measurement with $Q_{\rm max} \approx 18$~\AA{}$^{-1}$~\cite{hart_temperature_2005} used in the experimental/EPSR procedure~\cite{soper_quantum_2008}.
Besides the $Q_{\rm max}$ probed in a given X-ray experiment, the assignment of $g_{\ce{OO}}(\bm r)$ should also be sensitive to the form factors used in deriving $S_{\ce{OO}}(\bm Q)$ from the scattering intensity.
In the benchmark experiment~\cite{skinner_benchmark_2013}, the form factors are derived from the electron density of a gas phase \ce{H2O} molecule~\cite{wang_chemical_1994} instead of one in the liquid phase.
However, such condensed-phase effect on the form factors has been demonstrated negligible ($\approx 1$--$2$\%) and concentrated in the low-$Q$ region ($Q<2$~\AA{}$^{-1}$)~\cite{distasio_jr._individual_2014}, thereby reinforcing the central role played by probing sufficiently high $Q_{\rm max}$ in the accurate experimental determination of the first peak in $g_{\ce{OO}}(\bm r)$.
%
Apart from the slightly overstructured $g_{\ce{OO}}(\bm r)$ within the second (coordination) shell (i.e., $r<5.5$~\AA{}), the PI-DPMD prediction agrees quite well with both experimental assignments.
%
%
Interestingly, the PI-DPMD prediction yields a long-range structural correlation beyond the third shell ($r > 8$~\AA{}) as assigned in the benchmark experiment to be observable up to the sixth shell ($r \approx 15$~\AA{})~\cite{skinner_benchmark_2013}.

When compared with the experimental/EPSR assignment, we observed similar isotope effect in $g_{\ce{OO}}(\bm r)$ due to \ce{H}$\rightarrow$\ce{D} beyond the first peak ($r>3$~\AA{}) as shown in the lower panel of Fig.~\ref{fig:compare_goor_expt}.
However, the experimental/EPSR approach assigns an isotope increase in the height of the first peak from $2.59$ (\ce{H2O}) to $2.84$ (\ce{D2O}) that is significantly larger than PI-DPMD prediction ($2.58\rightarrow 2.66$).
This difference could arise from the limitation of the PBE0-TS model in the PI-DPMD simulation or the challenge to probe the high-$Q$ region in the X-ray scattering experiment.
To resolve this issue, it seems necessary to resort to either a high-resolution X-ray experiment on liquid \ce{D2O} as performed in Ref.~\cite{skinner_benchmark_2013} or a PI-DPMD simulation trained on realistic electronic structure theory beyond the PBE0-TS level (see Sec.~\ref{result:improve_es}).

\subsection{Isotope effects on the microscopic structure of liquid water: Angular distribution function and tetrahedrality \label{result:adf}}

\begin{figure}[ht!]
    \centering
    \includegraphics[width=0.8\linewidth]{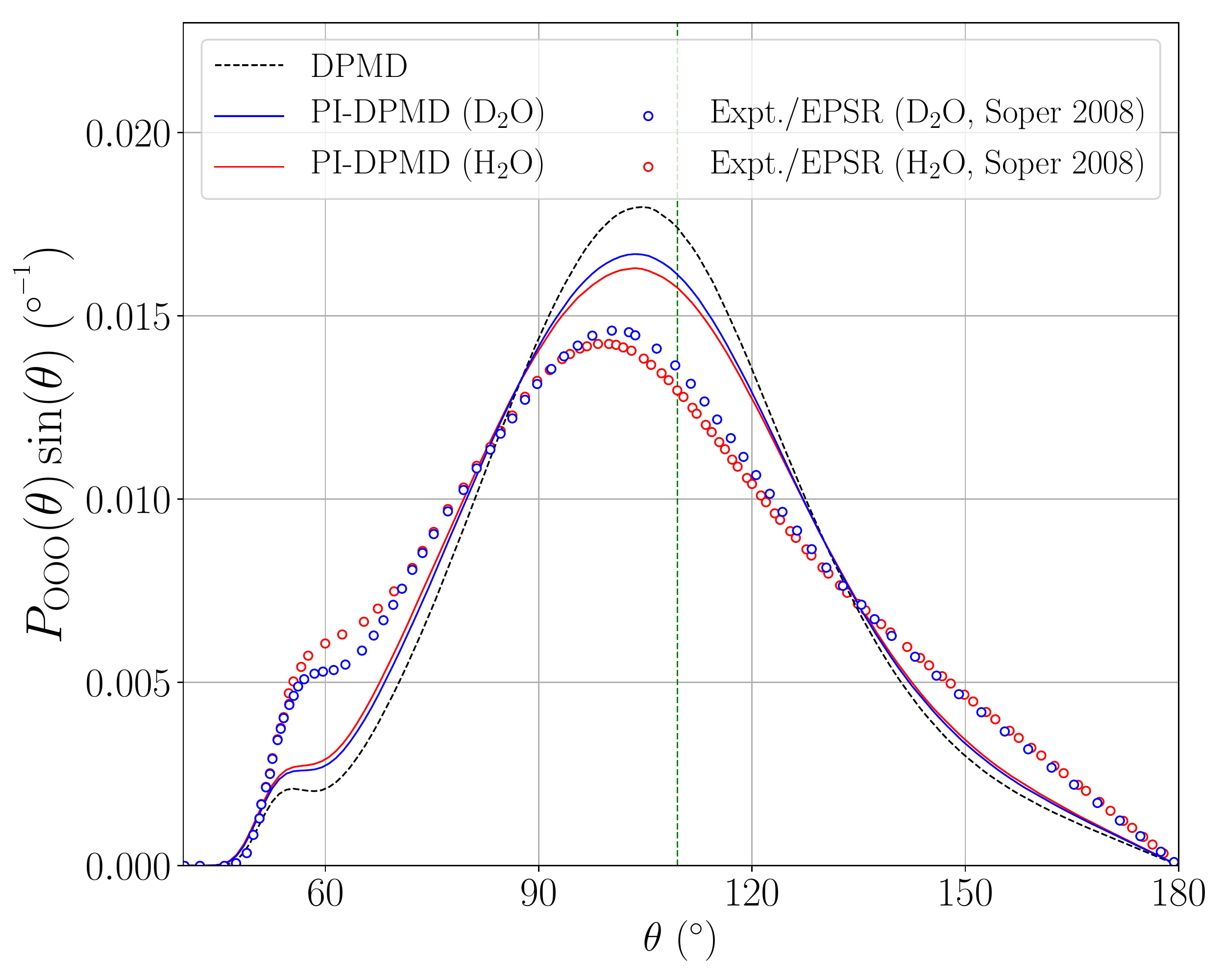}
    \caption{
    Comparison of DPMD and PI-DPMD in predicting the experimentally assigned~\cite{soper_quantum_2008} \ce{OOO} ADF of liquid \ce{H2O} and \ce{D2O} at ambient conditions ($300$~K,~$1$~bar).
    All PI-DPMD predictions (solid lines) and experimental results (open circles) are colored red for \ce{H2O} and blue for \ce{D2O}, while the DPMD result is plotted as a dashed black line.
    All ADFs were normalized such that $\int_{0}^{\pi} \,\text{d}\theta\, P_{OOO}(\theta) \sin(\theta) = 1$. The dashed green vertical line reflects the interior angle of a perfect tetrahedron ($\theta = 109.5^\circ$).
    }
    \label{fig:adf}
\end{figure}

In this section, we consider how the isotopic substitution of \ce{H} to \ce{D} influences the oxygen--oxygen--oxygen (\ce{OOO}) angular distribution function (ADF) of \ce{H2O} and \ce{D2O}.
In this analysis, an \ce{OOO} triplet is defined as any three \ce{O} atoms ($\ce{O}_A$, $\ce{O}_B$, and $\ce{O}_C$) in which $\ce{O}_A$ and $\ce{O}_B$ are located within a prescribed cutoff distance $d_{3}$ from $\ce{O}_C$.
Following the procedure outlined by Soper and Benmore~\cite{soper_quantum_2008}, $d_3$ was chosen to yield an average \ce{O}--\ce{O} coordination number of $4.0$; in this work, we used $d_3 = 3.24$~\AA{} which is consistent with previous AIMD simulations at the PBE0-TS level~\cite{distasio_jr._individual_2014} but larger than the experimental/EPSR value of $3.18$~\AA{}~\cite{soper_quantum_2008}.
In Fig.~\ref{fig:adf}, we plot the $P_{\ce{OOO}}(\theta)$ ADF---the probability of finding the triplet angle ($\theta = \angle\ce{O}_A\ce{O}_C\ce{O}_B$)---from the PI-DPMD predictions and experimental/EPSR procedure~\cite{soper_quantum_2008} for liquid \ce{H2O} and \ce{D2O}.

In the ADF, the PI-DPMD prediction shows consistent isotope effects in which \ce{H2O} is less structured than \ce{D2O}; this manifests as a reduced peak at $\approx 100^\circ$ (due to the tetrahedral HB network) and an enhanced peak at $\approx 50^\circ$  (due to highly distorted broken HB configurations).
Compared with the experiment/EPSR assignment, the isotope effect predicted by the PI-DPMD simulations is once again smaller in extent (particularly for the peak at $\approx 50^\circ$).
While this discrepancy may due in part to the limitation of the PBE0-TS model, it could also reflect the remaining uncertainty from the experimental/EPSR assignment, which seems to overestimate the isotope effect (e.g., in the \ce{OH} to \ce{OD} bond contraction as discussed in Sec.~\ref{result:rdf}).
On an absolute scale, our predicted ADFs are slightly more structured with a higher main peak at $\approx 100^\circ$ and a lower secondary peak at $\approx 50^\circ$.
This phenomenon is consistent with our observations of the slightly overstructured $g_{\ce{OO}}(\bm r)$ within the second shell (Sec.~\ref{result:rdf}).
However, unlike the $g_{\ce{OO}}(\bm r)$, the ADF is not directly probed from the experiment and relies on the EPSR procedure.
As such, more direct experiment or accurate simulation seem necessary to quantify the ADFs and the associated isotope effects in ambient liquid water.
%
%
%
%

\subsection{Potential improvement and future outlook \label{result:improve_es}}

It is well known that AIMD simulations which employ common GGA functionals (e.g., PBE~\cite{perdew_generalized_1996} and BLYP~\cite{becke_density-functional_1988,lee_development_1988}) for the underlying PES yield a liquid water that is substantially overstructured~\cite{schmidt_isobaricisothermal_2009,wang_density_2011,lin_structure_2012,distasio_jr._individual_2014,gaiduk_density_2015,miceli_isobaric_2015,pestana_ab_2017,chen_ab_2017,zheng_structural_2018}.
This effect is reduced but not eliminated with the inclusion of NQEs via PI-AIMD simulations~\cite{morrone_nuclear_2008}.
In the present work, further improvement is achieved by modeling the electronic structure at the PBE0-TS level, which includes a fraction of exact exchange to mitigate the self-interaction error and an effective-pairwise treatment of the non-local vdW interactions.
As described throughout this work, this improved electronic structure is not sufficient to provide a quantitative description of the structural properties of liquid water, and there is still room for further improvement.
At the level of vdW-inclusive hybrid DFT, the revPBE0-D3 functional~\cite{zhang_comment_1998,perdew_rationale_1996,adamo_toward_1999,goerigk_thorough_2011} appears to predict a $g_{\ce{OO}}(r)$ in better overall agreement with experiment~\cite{cheng_ab_2019}.
However, that functional predicts smaller NQEs than PBE0-TS; for example, the differences between the classical and quantum RDFs for \ce{H2O} reported in Ref.~\cite{cheng_ab_2019} are smaller than that reported in this paper.
This observation suggests that revPBE0-D3 should also underestimate the isotope effects in the microscopic structural properties of liquid water.
These general difficulties of vdW-inclusive hybrid DFT most likely originate from the imperfect cancellation of the self-interaction error inherent from their approximated functional forms.
To overcome these difficulties, it seems necessary to apply a more sophisticated electronic structure theory such as self-interaction-free DFT methods or highly accurate quantum chemical (wavefunction theory) approaches.
These approaches are substantially more expensive than hybrid DFT, and can only be used to obtain accurate interatomic interactions (including the total potential energy, atomic forces, and stress tensor) for a small subset of liquid water configurations.
As such, definite progress beyond vdW-inclusive hybrid DFT is possible if a ML-based model (like DPMD) could be refined with a select set of accurate quantum chemical data or direct observations from experiments (with approaches such as EPSR).
Provided with such an accurate and efficient interatomic potential, we expect that DPMD and PI-DPMD simulations can also be used to study structural and thermal properties of molecular crystals bound by HB interactions and dispersion forces~\cite{reilly_report_2016,ko_thermal_2018,hoja_reliable_2019}, which require the use of hybrid DFT corrected with a beyond effective-pairwise many-body dispersion (MBD) model~\cite{tkatchenko_accurate_2012,ambrosetti_long-range_2014,blood-forsythe_analytical_2016}.

\section{Conclusions\label{sec:conclusion}}

In this work, we demonstrate that the deep neural network based DPMD model---when trained on a single, relatively short ($\approx 8$~ps) PI-AIMD simulation of liquid \ce{H2O}---can provide a (semi-)quantitative description of the isotope effects found in $F^{(\rm n)}_{\rm int} (Q)$, the $g_{\ce{OO}}(\bm r)$, $g_{\ce{OH}}(\bm r)$/$g_{\ce{OD}}(\bm r)$, and $g_{\ce{HH}}(\bm r)$/$g_{\ce{DD}}(\bm r)$ RDFs, and the $P_{\ce{OOO}} (\theta)$ ADF.
Like the experimental/EPSR procedure of Soper and Benmore~\cite{soper_quantum_2008} and the oxygen isotope substitution experiment of Zeidler \textit{et al.}~\cite{zeidler_oxygen_2011,zeidler_isotope_2012}, our PI-DPMD simulations characterize these isotope effects as small and subtle changes to the microscopic structure of liquid water.
In doing so, we find that our approach predicts systematically smaller isotope effects in liquid water compared with the experimental/EPSR procedure.
This discrepancy can be attributed to the imperfect description of the underlying PES at the vdW-inclusive hybrid DFT (i.e., PBE0-TS) level of theory, or to the remaining uncertainty due to the experimental challenge.
Looking ahead, this work suggests that the combination of ML-based techniques (such as the DPMD model) and a sufficient amount of highly accurate \textit{ab initio} data (beyond the vdW-inclusive hybrid DFT level) could pave the way towards studying subtle physical effects that were computationally prohibited by the length and time scales accessible to standard AIMD and PI-AIMD simulations.

\section*{Acknowledgements}
HYK, LZ, WE, and RC gratefully acknowledge support from the U.S. Department of Energy (DOE) under Grant No. DE-SC0019394.
HW gratefully acknowledge support from the National Natural Science Foundation of China (NSFC) under Grant Nos. 11501039, 11871110, and 91530322, as well as the National Key Research and Development Program of China under Grant Nos. 2016YFB0201200 and 2016YFB0201203. 
LZ and WE also acknowledge support from the Major Program of the National Natural Science Foundation of China (NNSFC) under Grant Nos. 91130005 and U1430237, as well as the Office of Naval Research (ONR) Grant No. N00014-13-1-0338.
RD acknowledges the support of the Center for Alkaline Based Energy Solutions (CABES), an Energy Frontier Research Center funded by the U.S. Department of Energy, Office of Science, Office of Basic Energy Sciences, under Award No. DE-SC0019445.
This work used resources of the National Energy Research Scientific Computing Center, which is supported by the Office of Science of the U.S. Department of Energy under Contract No. DE-AC02-05CH11231.


\begin{thebibliography}{102}%
\makeatletter
\providecommand \@ifxundefined [1]{%
 \@ifx{#1\undefined}
}%
\providecommand \@ifnum [1]{%
 \ifnum #1\expandafter \@firstoftwo
 \else \expandafter \@secondoftwo
 \fi
}%
\providecommand \@ifx [1]{%
 \ifx #1\expandafter \@firstoftwo
 \else \expandafter \@secondoftwo
 \fi
}%
\providecommand \natexlab [1]{#1}%
\providecommand \enquote  [1]{``#1''}%
\providecommand \bibnamefont  [1]{#1}%
\providecommand \bibfnamefont [1]{#1}%
\providecommand \citenamefont [1]{#1}%
\providecommand \href@noop [0]{\@secondoftwo}%
\providecommand \href [0]{\begingroup \@sanitize@url \@href}%
\providecommand \@href[1]{\@@startlink{#1}\@@href}%
\providecommand \@@href[1]{\endgroup#1\@@endlink}%
\providecommand \@sanitize@url [0]{\catcode `\\12\catcode `\$12\catcode
  `\&12\catcode `\#12\catcode `\^12\catcode `\_12\catcode `\%12\relax}%
\providecommand \@@startlink[1]{}%
\providecommand \@@endlink[0]{}%
\providecommand \url  [0]{\begingroup\@sanitize@url \@url }%
\providecommand \@url [1]{\endgroup\@href {#1}{\urlprefix }}%
\providecommand \urlprefix  [0]{URL }%
\providecommand \Eprint [0]{\href }%
\providecommand \doibase [0]{http://dx.doi.org/}%
\providecommand \selectlanguage [0]{\@gobble}%
\providecommand \bibinfo  [0]{\@secondoftwo}%
\providecommand \bibfield  [0]{\@secondoftwo}%
\providecommand \translation [1]{[#1]}%
\providecommand \BibitemOpen [0]{}%
\providecommand \bibitemStop [0]{}%
\providecommand \bibitemNoStop [0]{.\EOS\space}%
\providecommand \EOS [0]{\spacefactor3000\relax}%
\providecommand \BibitemShut  [1]{\csname bibitem#1\endcsname}%
\let\auto@bib@innerbib\@empty
\bibitem [{\citenamefont {DiStasio~Jr.}\ \emph {et~al.}(2014)\citenamefont
  {DiStasio~Jr.}, \citenamefont {Santra}, \citenamefont {Li}, \citenamefont
  {Wu},\ and\ \citenamefont {Car}}]{distasio_jr._individual_2014}%
  \BibitemOpen
  \bibfield  {author} {\bibinfo {author} {\bibfnamefont {R.~A.}\ \bibnamefont
  {DiStasio~Jr.}}, \bibinfo {author} {\bibfnamefont {B.}~\bibnamefont
  {Santra}}, \bibinfo {author} {\bibfnamefont {Z.}~\bibnamefont {Li}}, \bibinfo
  {author} {\bibfnamefont {X.}~\bibnamefont {Wu}}, \ and\ \bibinfo {author}
  {\bibfnamefont {R.}~\bibnamefont {Car}},\ }\href@noop {} {\bibfield
  {journal} {\bibinfo  {journal} {J. Chem. Phys.}\ }\textbf {\bibinfo {volume}
  {141}},\ \bibinfo {pages} {084502} (\bibinfo {year} {2014})}\BibitemShut
  {NoStop}%
\bibitem [{\citenamefont {Ceriotti}\ \emph {et~al.}(2016)\citenamefont
  {Ceriotti}, \citenamefont {Fang}, \citenamefont {Kusalik}, \citenamefont
  {McKenzie}, \citenamefont {Michaelides}, \citenamefont {Morales},\ and\
  \citenamefont {Markland}}]{ceriotti_nuclear_2016}%
  \BibitemOpen
  \bibfield  {author} {\bibinfo {author} {\bibfnamefont {M.}~\bibnamefont
  {Ceriotti}}, \bibinfo {author} {\bibfnamefont {W.}~\bibnamefont {Fang}},
  \bibinfo {author} {\bibfnamefont {P.~G.}\ \bibnamefont {Kusalik}}, \bibinfo
  {author} {\bibfnamefont {R.~H.}\ \bibnamefont {McKenzie}}, \bibinfo {author}
  {\bibfnamefont {A.}~\bibnamefont {Michaelides}}, \bibinfo {author}
  {\bibfnamefont {M.~A.}\ \bibnamefont {Morales}}, \ and\ \bibinfo {author}
  {\bibfnamefont {T.~E.}\ \bibnamefont {Markland}},\ }\href@noop {} {\bibfield
  {journal} {\bibinfo  {journal} {Chem. Rev.}\ }\textbf {\bibinfo {volume}
  {116}},\ \bibinfo {pages} {7529} (\bibinfo {year} {2016})}\BibitemShut
  {NoStop}%
\bibitem [{\citenamefont {Soper}\ and\ \citenamefont
  {Benmore}(2008)}]{soper_quantum_2008}%
  \BibitemOpen
  \bibfield  {author} {\bibinfo {author} {\bibfnamefont {A.~K.}\ \bibnamefont
  {Soper}}\ and\ \bibinfo {author} {\bibfnamefont {C.~J.}\ \bibnamefont
  {Benmore}},\ }\href@noop {} {\bibfield  {journal} {\bibinfo  {journal} {Phys.
  Rev. Lett.}\ }\textbf {\bibinfo {volume} {101}},\ \bibinfo {pages} {065502}
  (\bibinfo {year} {2008})}\BibitemShut {NoStop}%
\bibitem [{\citenamefont {Soper}(2001)}]{soper_tests_2001}%
  \BibitemOpen
  \bibfield  {author} {\bibinfo {author} {\bibfnamefont {A.~K.}\ \bibnamefont
  {Soper}},\ }\href@noop {} {\bibfield  {journal} {\bibinfo  {journal} {Mol.
  Phys.}\ }\textbf {\bibinfo {volume} {99}},\ \bibinfo {pages} {1503} (\bibinfo
  {year} {2001})}\BibitemShut {NoStop}%
\bibitem [{\citenamefont {Soper}(2005)}]{soper_partial_2005}%
  \BibitemOpen
  \bibfield  {author} {\bibinfo {author} {\bibfnamefont {A.~K.}\ \bibnamefont
  {Soper}},\ }\href@noop {} {\bibfield  {journal} {\bibinfo  {journal} {Phys.
  Rev. B}\ }\textbf {\bibinfo {volume} {72}},\ \bibinfo {pages} {104204}
  (\bibinfo {year} {2005})}\BibitemShut {NoStop}%
\bibitem [{\citenamefont {Car}\ and\ \citenamefont
  {Parrinello}(1985)}]{car_unified_1985}%
  \BibitemOpen
  \bibfield  {author} {\bibinfo {author} {\bibfnamefont {R.}~\bibnamefont
  {Car}}\ and\ \bibinfo {author} {\bibfnamefont {M.}~\bibnamefont
  {Parrinello}},\ }\href@noop {} {\bibfield  {journal} {\bibinfo  {journal}
  {Phys. Rev. Lett.}\ }\textbf {\bibinfo {volume} {55}},\ \bibinfo {pages}
  {2471} (\bibinfo {year} {1985})}\BibitemShut {NoStop}%
\bibitem [{\citenamefont {Marx}\ and\ \citenamefont
  {Hutter}(2009)}]{marx_ab_2009}%
  \BibitemOpen
  \bibfield  {author} {\bibinfo {author} {\bibfnamefont {D.}~\bibnamefont
  {Marx}}\ and\ \bibinfo {author} {\bibfnamefont {J.}~\bibnamefont {Hutter}},\
  }\href@noop {} {\emph {\bibinfo {title} {Ab {Initio} {Molecular} {Dynamics}:
  {Basic} {Theory} and {Advanced} {Methods}}}}\ (\bibinfo  {publisher}
  {Cambridge University Press},\ \bibinfo {address} {Cambridge},\ \bibinfo
  {year} {2009})\BibitemShut {NoStop}%
\bibitem [{\citenamefont {Hohenberg}\ and\ \citenamefont
  {Kohn}(1964)}]{hohenberg_inhomogeneous_1964}%
  \BibitemOpen
  \bibfield  {author} {\bibinfo {author} {\bibfnamefont {P.}~\bibnamefont
  {Hohenberg}}\ and\ \bibinfo {author} {\bibfnamefont {W.}~\bibnamefont
  {Kohn}},\ }\href@noop {} {\bibfield  {journal} {\bibinfo  {journal} {Phys.
  Rev.}\ }\textbf {\bibinfo {volume} {136}},\ \bibinfo {pages} {B864} (\bibinfo
  {year} {1964})}\BibitemShut {NoStop}%
\bibitem [{\citenamefont {Kohn}\ and\ \citenamefont
  {Sham}(1965)}]{kohn_self-consistent_1965}%
  \BibitemOpen
  \bibfield  {author} {\bibinfo {author} {\bibfnamefont {W.}~\bibnamefont
  {Kohn}}\ and\ \bibinfo {author} {\bibfnamefont {L.~J.}\ \bibnamefont
  {Sham}},\ }\href@noop {} {\bibfield  {journal} {\bibinfo  {journal} {Phys.
  Rev.}\ }\textbf {\bibinfo {volume} {140}},\ \bibinfo {pages} {A1133}
  (\bibinfo {year} {1965})}\BibitemShut {NoStop}%
\bibitem [{\citenamefont {Parr}\ and\ \citenamefont
  {Yang}(1989)}]{parr_density-functional_1989}%
  \BibitemOpen
  \bibfield  {author} {\bibinfo {author} {\bibfnamefont {R.~G.}\ \bibnamefont
  {Parr}}\ and\ \bibinfo {author} {\bibfnamefont {W.}~\bibnamefont {Yang}},\
  }\href@noop {} {\emph {\bibinfo {title} {Density-{Functional} {Theory} of
  {Atoms} and {Molecules}}}}\ (\bibinfo  {publisher} {Oxford University
  Press},\ \bibinfo {address} {New York},\ \bibinfo {year} {1989})\BibitemShut
  {NoStop}%
\bibitem [{\citenamefont {Perdew}\ and\ \citenamefont
  {Schmidt}(2001)}]{perdew_jacobs_2001}%
  \BibitemOpen
  \bibfield  {author} {\bibinfo {author} {\bibfnamefont {J.~P.}\ \bibnamefont
  {Perdew}}\ and\ \bibinfo {author} {\bibfnamefont {K.}~\bibnamefont
  {Schmidt}},\ }in\ \href@noop {} {\emph {\bibinfo {booktitle} {Density
  {Functional} {Theory} and {Its} {Application} to {Materials}: {Antwerp},
  {Belgium}, {Jun}. 8-10, 2000}}},\ Vol.\ \bibinfo {volume} {577},\ \bibinfo
  {editor} {edited by\ \bibinfo {editor} {\bibfnamefont {V.~E.}\ \bibnamefont
  {Van~Doren}}, \bibinfo {editor} {\bibfnamefont {C.}~\bibnamefont
  {Van~Alsenoy}}, \ and\ \bibinfo {editor} {\bibfnamefont {P.}~\bibnamefont
  {Geerlings}}}\ (\bibinfo  {publisher} {AIP Publishing},\ \bibinfo {address}
  {Melville},\ \bibinfo {year} {2001})\ pp.\ \bibinfo {pages}
  {1--20}\BibitemShut {NoStop}%
\bibitem [{\citenamefont {Becke}(1988)}]{becke_density-functional_1988}%
  \BibitemOpen
  \bibfield  {author} {\bibinfo {author} {\bibfnamefont {A.~D.}\ \bibnamefont
  {Becke}},\ }\href@noop {} {\bibfield  {journal} {\bibinfo  {journal} {Phys.
  Rev. A}\ }\textbf {\bibinfo {volume} {38}},\ \bibinfo {pages} {3098}
  (\bibinfo {year} {1988})}\BibitemShut {NoStop}%
\bibitem [{\citenamefont {Lee}\ \emph {et~al.}(1988)\citenamefont {Lee},
  \citenamefont {Yang},\ and\ \citenamefont {Parr}}]{lee_development_1988}%
  \BibitemOpen
  \bibfield  {author} {\bibinfo {author} {\bibfnamefont {C.}~\bibnamefont
  {Lee}}, \bibinfo {author} {\bibfnamefont {W.}~\bibnamefont {Yang}}, \ and\
  \bibinfo {author} {\bibfnamefont {R.~G.}\ \bibnamefont {Parr}},\ }\href@noop
  {} {\bibfield  {journal} {\bibinfo  {journal} {Phys. Rev. B}\ }\textbf
  {\bibinfo {volume} {37}},\ \bibinfo {pages} {785} (\bibinfo {year}
  {1988})}\BibitemShut {NoStop}%
\bibitem [{\citenamefont {Perdew}\ \emph
  {et~al.}(1996{\natexlab{a}})\citenamefont {Perdew}, \citenamefont {Burke},\
  and\ \citenamefont {Ernzerhof}}]{perdew_generalized_1996}%
  \BibitemOpen
  \bibfield  {author} {\bibinfo {author} {\bibfnamefont {J.~P.}\ \bibnamefont
  {Perdew}}, \bibinfo {author} {\bibfnamefont {K.}~\bibnamefont {Burke}}, \
  and\ \bibinfo {author} {\bibfnamefont {M.}~\bibnamefont {Ernzerhof}},\
  }\href@noop {} {\bibfield  {journal} {\bibinfo  {journal} {Phys. Rev. Lett.}\
  }\textbf {\bibinfo {volume} {77}},\ \bibinfo {pages} {3865} (\bibinfo {year}
  {1996}{\natexlab{a}})}\BibitemShut {NoStop}%
\bibitem [{\citenamefont {Cohen}\ \emph {et~al.}(2008)\citenamefont {Cohen},
  \citenamefont {Mori-S{\'a}nchez},\ and\ \citenamefont
  {Yang}}]{cohen_insights_2008}%
  \BibitemOpen
  \bibfield  {author} {\bibinfo {author} {\bibfnamefont {A.~J.}\ \bibnamefont
  {Cohen}}, \bibinfo {author} {\bibfnamefont {P.}~\bibnamefont
  {Mori-S{\'a}nchez}}, \ and\ \bibinfo {author} {\bibfnamefont
  {W.}~\bibnamefont {Yang}},\ }\href@noop {} {\bibfield  {journal} {\bibinfo
  {journal} {Science}\ }\textbf {\bibinfo {volume} {321}},\ \bibinfo {pages}
  {792} (\bibinfo {year} {2008})}\BibitemShut {NoStop}%
\bibitem [{\citenamefont {Hermann}\ \emph {et~al.}(2017)\citenamefont
  {Hermann}, \citenamefont {DiStasio~Jr.},\ and\ \citenamefont
  {Tkatchenko}}]{hermann_first-principles_2017}%
  \BibitemOpen
  \bibfield  {author} {\bibinfo {author} {\bibfnamefont {J.}~\bibnamefont
  {Hermann}}, \bibinfo {author} {\bibfnamefont {R.~A.}\ \bibnamefont
  {DiStasio~Jr.}}, \ and\ \bibinfo {author} {\bibfnamefont {A.}~\bibnamefont
  {Tkatchenko}},\ }\href@noop {} {\bibfield  {journal} {\bibinfo  {journal}
  {Chem. Rev.}\ }\textbf {\bibinfo {volume} {117}},\ \bibinfo {pages} {4714}
  (\bibinfo {year} {2017})}\BibitemShut {NoStop}%
\bibitem [{\citenamefont {Sun}\ \emph {et~al.}(2015)\citenamefont {Sun},
  \citenamefont {Ruzsinszky},\ and\ \citenamefont
  {Perdew}}]{sun_strongly_2015}%
  \BibitemOpen
  \bibfield  {author} {\bibinfo {author} {\bibfnamefont {J.}~\bibnamefont
  {Sun}}, \bibinfo {author} {\bibfnamefont {A.}~\bibnamefont {Ruzsinszky}}, \
  and\ \bibinfo {author} {\bibfnamefont {J.~P.}\ \bibnamefont {Perdew}},\
  }\href@noop {} {\bibfield  {journal} {\bibinfo  {journal} {Phys. Rev. Lett.}\
  }\textbf {\bibinfo {volume} {115}},\ \bibinfo {pages} {036402} (\bibinfo
  {year} {2015})}\BibitemShut {NoStop}%
\bibitem [{\citenamefont {Chen}\ \emph {et~al.}(2017)\citenamefont {Chen},
  \citenamefont {Ko}, \citenamefont {Remsing}, \citenamefont {Andrade},
  \citenamefont {Santra}, \citenamefont {Sun}, \citenamefont {Selloni},
  \citenamefont {Car}, \citenamefont {Klein}, \citenamefont {Perdew},\ and\
  \citenamefont {Wu}}]{chen_ab_2017}%
  \BibitemOpen
  \bibfield  {author} {\bibinfo {author} {\bibfnamefont {M.}~\bibnamefont
  {Chen}}, \bibinfo {author} {\bibfnamefont {H.-Y.}\ \bibnamefont {Ko}},
  \bibinfo {author} {\bibfnamefont {R.~C.}\ \bibnamefont {Remsing}}, \bibinfo
  {author} {\bibfnamefont {M.~F.~C.}\ \bibnamefont {Andrade}}, \bibinfo
  {author} {\bibfnamefont {B.}~\bibnamefont {Santra}}, \bibinfo {author}
  {\bibfnamefont {Z.}~\bibnamefont {Sun}}, \bibinfo {author} {\bibfnamefont
  {A.}~\bibnamefont {Selloni}}, \bibinfo {author} {\bibfnamefont
  {R.}~\bibnamefont {Car}}, \bibinfo {author} {\bibfnamefont {M.~L.}\
  \bibnamefont {Klein}}, \bibinfo {author} {\bibfnamefont {J.~P.}\ \bibnamefont
  {Perdew}}, \ and\ \bibinfo {author} {\bibfnamefont {X.}~\bibnamefont {Wu}},\
  }\href@noop {} {\bibfield  {journal} {\bibinfo  {journal} {Proc. Natl. Acad.
  Sci. U.S.A.}\ }\textbf {\bibinfo {volume} {114}},\ \bibinfo {pages} {10846}
  (\bibinfo {year} {2017})}\BibitemShut {NoStop}%
\bibitem [{\citenamefont {Zheng}\ \emph {et~al.}(2018)\citenamefont {Zheng},
  \citenamefont {Chen}, \citenamefont {Sun}, \citenamefont {Ko}, \citenamefont
  {Santra}, \citenamefont {Dhuvad},\ and\ \citenamefont
  {Wu}}]{zheng_structural_2018}%
  \BibitemOpen
  \bibfield  {author} {\bibinfo {author} {\bibfnamefont {L.}~\bibnamefont
  {Zheng}}, \bibinfo {author} {\bibfnamefont {M.}~\bibnamefont {Chen}},
  \bibinfo {author} {\bibfnamefont {Z.}~\bibnamefont {Sun}}, \bibinfo {author}
  {\bibfnamefont {H.-Y.}\ \bibnamefont {Ko}}, \bibinfo {author} {\bibfnamefont
  {B.}~\bibnamefont {Santra}}, \bibinfo {author} {\bibfnamefont
  {P.}~\bibnamefont {Dhuvad}}, \ and\ \bibinfo {author} {\bibfnamefont
  {X.}~\bibnamefont {Wu}},\ }\href@noop {} {\bibfield  {journal} {\bibinfo
  {journal} {J. Chem. Phys.}\ }\textbf {\bibinfo {volume} {148}},\ \bibinfo
  {pages} {164505} (\bibinfo {year} {2018})}\BibitemShut {NoStop}%
\bibitem [{\citenamefont {Calegari~Andrade}\ \emph {et~al.}(2018)\citenamefont
  {Calegari~Andrade}, \citenamefont {Ko}, \citenamefont {Car},\ and\
  \citenamefont {Selloni}}]{calegari_andrade_structure_2018}%
  \BibitemOpen
  \bibfield  {author} {\bibinfo {author} {\bibfnamefont {M.~F.}\ \bibnamefont
  {Calegari~Andrade}}, \bibinfo {author} {\bibfnamefont {H.-Y.}\ \bibnamefont
  {Ko}}, \bibinfo {author} {\bibfnamefont {R.}~\bibnamefont {Car}}, \ and\
  \bibinfo {author} {\bibfnamefont {A.}~\bibnamefont {Selloni}},\ }\href@noop
  {} {\bibfield  {journal} {\bibinfo  {journal} {J. Phys. Chem. Lett.}\
  }\textbf {\bibinfo {volume} {9}},\ \bibinfo {pages} {6716} (\bibinfo {year}
  {2018})}\BibitemShut {NoStop}%
\bibitem [{\citenamefont {Perdew}\ \emph
  {et~al.}(1996{\natexlab{b}})\citenamefont {Perdew}, \citenamefont
  {Ernzerhof},\ and\ \citenamefont {Burke}}]{perdew_rationale_1996}%
  \BibitemOpen
  \bibfield  {author} {\bibinfo {author} {\bibfnamefont {J.~P.}\ \bibnamefont
  {Perdew}}, \bibinfo {author} {\bibfnamefont {M.}~\bibnamefont {Ernzerhof}}, \
  and\ \bibinfo {author} {\bibfnamefont {K.}~\bibnamefont {Burke}},\
  }\href@noop {} {\bibfield  {journal} {\bibinfo  {journal} {J. Chem. Phys.}\
  }\textbf {\bibinfo {volume} {105}},\ \bibinfo {pages} {9982} (\bibinfo {year}
  {1996}{\natexlab{b}})}\BibitemShut {NoStop}%
\bibitem [{\citenamefont {Adamo}\ and\ \citenamefont
  {Barone}(1999)}]{adamo_toward_1999}%
  \BibitemOpen
  \bibfield  {author} {\bibinfo {author} {\bibfnamefont {C.}~\bibnamefont
  {Adamo}}\ and\ \bibinfo {author} {\bibfnamefont {V.}~\bibnamefont {Barone}},\
  }\href@noop {} {\bibfield  {journal} {\bibinfo  {journal} {J. Chem. Phys.}\
  }\textbf {\bibinfo {volume} {110}},\ \bibinfo {pages} {6158} (\bibinfo {year}
  {1999})}\BibitemShut {NoStop}%
\bibitem [{\citenamefont {Tkatchenko}\ and\ \citenamefont
  {Scheffler}(2009)}]{tkatchenko_accurate_2009}%
  \BibitemOpen
  \bibfield  {author} {\bibinfo {author} {\bibfnamefont {A.}~\bibnamefont
  {Tkatchenko}}\ and\ \bibinfo {author} {\bibfnamefont {M.}~\bibnamefont
  {Scheffler}},\ }\href@noop {} {\bibfield  {journal} {\bibinfo  {journal}
  {Phys. Rev. Lett.}\ }\textbf {\bibinfo {volume} {102}},\ \bibinfo {pages}
  {073005} (\bibinfo {year} {2009})}\BibitemShut {NoStop}%
\bibitem [{\citenamefont {Ferri}\ \emph {et~al.}(2015)\citenamefont {Ferri},
  \citenamefont {DiStasio~Jr.}, \citenamefont {Ambrosetti}, \citenamefont
  {Car},\ and\ \citenamefont {Tkatchenko}}]{ferri_electronic_2015}%
  \BibitemOpen
  \bibfield  {author} {\bibinfo {author} {\bibfnamefont {N.}~\bibnamefont
  {Ferri}}, \bibinfo {author} {\bibfnamefont {R.~A.}\ \bibnamefont
  {DiStasio~Jr.}}, \bibinfo {author} {\bibfnamefont {A.}~\bibnamefont
  {Ambrosetti}}, \bibinfo {author} {\bibfnamefont {R.}~\bibnamefont {Car}}, \
  and\ \bibinfo {author} {\bibfnamefont {A.}~\bibnamefont {Tkatchenko}},\
  }\href@noop {} {\bibfield  {journal} {\bibinfo  {journal} {Phys. Rev. Lett.}\
  }\textbf {\bibinfo {volume} {114}},\ \bibinfo {pages} {176802} (\bibinfo
  {year} {2015})}\BibitemShut {NoStop}%
\bibitem [{\citenamefont {Santra}\ \emph {et~al.}(2015)\citenamefont {Santra},
  \citenamefont {DiStasio~Jr.}, \citenamefont {Martelli},\ and\ \citenamefont
  {Car}}]{santra_local_2015}%
  \BibitemOpen
  \bibfield  {author} {\bibinfo {author} {\bibfnamefont {B.}~\bibnamefont
  {Santra}}, \bibinfo {author} {\bibfnamefont {R.~A.}\ \bibnamefont
  {DiStasio~Jr.}}, \bibinfo {author} {\bibfnamefont {F.}~\bibnamefont
  {Martelli}}, \ and\ \bibinfo {author} {\bibfnamefont {R.}~\bibnamefont
  {Car}},\ }\href@noop {} {\bibfield  {journal} {\bibinfo  {journal} {Mol.
  Phys.}\ }\textbf {\bibinfo {volume} {113}},\ \bibinfo {pages} {2829}
  (\bibinfo {year} {2015})}\BibitemShut {NoStop}%
\bibitem [{\citenamefont {Chen}\ \emph {et~al.}(2018)\citenamefont {Chen},
  \citenamefont {Zheng}, \citenamefont {Santra}, \citenamefont {Ko},
  \citenamefont {DiStasio~Jr}, \citenamefont {Klein}, \citenamefont {Car},\
  and\ \citenamefont {Wu}}]{chen_hydroxide_2018}%
  \BibitemOpen
  \bibfield  {author} {\bibinfo {author} {\bibfnamefont {M.}~\bibnamefont
  {Chen}}, \bibinfo {author} {\bibfnamefont {L.}~\bibnamefont {Zheng}},
  \bibinfo {author} {\bibfnamefont {B.}~\bibnamefont {Santra}}, \bibinfo
  {author} {\bibfnamefont {H.-Y.}\ \bibnamefont {Ko}}, \bibinfo {author}
  {\bibfnamefont {R.~A.}\ \bibnamefont {DiStasio~Jr}}, \bibinfo {author}
  {\bibfnamefont {M.~L.}\ \bibnamefont {Klein}}, \bibinfo {author}
  {\bibfnamefont {R.}~\bibnamefont {Car}}, \ and\ \bibinfo {author}
  {\bibfnamefont {X.}~\bibnamefont {Wu}},\ }\href@noop {} {\bibfield  {journal}
  {\bibinfo  {journal} {Nat. Chem.}\ }\textbf {\bibinfo {volume} {10}},\
  \bibinfo {pages} {413} (\bibinfo {year} {2018})}\BibitemShut {NoStop}%
\bibitem [{\citenamefont {Zhang}\ and\ \citenamefont
  {Yang}(1998)}]{zhang_comment_1998}%
  \BibitemOpen
  \bibfield  {author} {\bibinfo {author} {\bibfnamefont {Y.}~\bibnamefont
  {Zhang}}\ and\ \bibinfo {author} {\bibfnamefont {W.}~\bibnamefont {Yang}},\
  }\href@noop {} {\bibfield  {journal} {\bibinfo  {journal} {Phys. Rev. Lett.}\
  }\textbf {\bibinfo {volume} {80}},\ \bibinfo {pages} {890} (\bibinfo {year}
  {1998})}\BibitemShut {NoStop}%
\bibitem [{\citenamefont {Goerigk}\ and\ \citenamefont
  {Grimme}(2011)}]{goerigk_thorough_2011}%
  \BibitemOpen
  \bibfield  {author} {\bibinfo {author} {\bibfnamefont {L.}~\bibnamefont
  {Goerigk}}\ and\ \bibinfo {author} {\bibfnamefont {S.}~\bibnamefont
  {Grimme}},\ }\href@noop {} {\bibfield  {journal} {\bibinfo  {journal} {Phys.
  Chem. Chem. Phys.}\ }\textbf {\bibinfo {volume} {13}},\ \bibinfo {pages}
  {6670} (\bibinfo {year} {2011})}\BibitemShut {NoStop}%
\bibitem [{\citenamefont {Cheng}\ \emph {et~al.}(2019)\citenamefont {Cheng},
  \citenamefont {Engel}, \citenamefont {Behler}, \citenamefont {Dellago},\ and\
  \citenamefont {Ceriotti}}]{cheng_ab_2019}%
  \BibitemOpen
  \bibfield  {author} {\bibinfo {author} {\bibfnamefont {B.}~\bibnamefont
  {Cheng}}, \bibinfo {author} {\bibfnamefont {E.~A.}\ \bibnamefont {Engel}},
  \bibinfo {author} {\bibfnamefont {J.}~\bibnamefont {Behler}}, \bibinfo
  {author} {\bibfnamefont {C.}~\bibnamefont {Dellago}}, \ and\ \bibinfo
  {author} {\bibfnamefont {M.}~\bibnamefont {Ceriotti}},\ }\href@noop {}
  {\bibfield  {journal} {\bibinfo  {journal} {Proc. Natl. Acad. Sci. U.S.A.}\
  }\textbf {\bibinfo {volume} {116}},\ \bibinfo {pages} {1110} (\bibinfo {year}
  {2019})}\BibitemShut {NoStop}%
\bibitem [{\citenamefont {Fosdick}(1962)}]{fosdick_numerical_1962}%
  \BibitemOpen
  \bibfield  {author} {\bibinfo {author} {\bibfnamefont {L.~D.}\ \bibnamefont
  {Fosdick}},\ }\href@noop {} {\bibfield  {journal} {\bibinfo  {journal} {J.
  Math Phys.}\ }\textbf {\bibinfo {volume} {3}},\ \bibinfo {pages} {1251}
  (\bibinfo {year} {1962})}\BibitemShut {NoStop}%
\bibitem [{\citenamefont {Chandler}\ and\ \citenamefont
  {Wolynes}(1981)}]{chandler_exploiting_1981}%
  \BibitemOpen
  \bibfield  {author} {\bibinfo {author} {\bibfnamefont {D.}~\bibnamefont
  {Chandler}}\ and\ \bibinfo {author} {\bibfnamefont {P.~G.}\ \bibnamefont
  {Wolynes}},\ }\href@noop {} {\bibfield  {journal} {\bibinfo  {journal} {J.
  Chem. Phys.}\ }\textbf {\bibinfo {volume} {74}},\ \bibinfo {pages} {4078}
  (\bibinfo {year} {1981})}\BibitemShut {NoStop}%
\bibitem [{\citenamefont {Marx}\ and\ \citenamefont
  {Parrinello}(1996)}]{marx_ab_1996}%
  \BibitemOpen
  \bibfield  {author} {\bibinfo {author} {\bibfnamefont {D.}~\bibnamefont
  {Marx}}\ and\ \bibinfo {author} {\bibfnamefont {M.}~\bibnamefont
  {Parrinello}},\ }\href@noop {} {\bibfield  {journal} {\bibinfo  {journal} {J.
  Chem. Phys.}\ }\textbf {\bibinfo {volume} {104}},\ \bibinfo {pages} {4077}
  (\bibinfo {year} {1996})}\BibitemShut {NoStop}%
\bibitem [{\citenamefont {Tuckerman}\ \emph {et~al.}(1996)\citenamefont
  {Tuckerman}, \citenamefont {Marx}, \citenamefont {Klein},\ and\ \citenamefont
  {Parrinello}}]{tuckerman_efficient_1996}%
  \BibitemOpen
  \bibfield  {author} {\bibinfo {author} {\bibfnamefont {M.~E.}\ \bibnamefont
  {Tuckerman}}, \bibinfo {author} {\bibfnamefont {D.}~\bibnamefont {Marx}},
  \bibinfo {author} {\bibfnamefont {M.~L.}\ \bibnamefont {Klein}}, \ and\
  \bibinfo {author} {\bibfnamefont {M.}~\bibnamefont {Parrinello}},\
  }\href@noop {} {\bibfield  {journal} {\bibinfo  {journal} {J. Chem. Phys.}\
  }\textbf {\bibinfo {volume} {104}},\ \bibinfo {pages} {5579} (\bibinfo {year}
  {1996})}\BibitemShut {NoStop}%
\bibitem [{\citenamefont {Stern}\ and\ \citenamefont
  {Berne}(2001)}]{stern_quantum_2001}%
  \BibitemOpen
  \bibfield  {author} {\bibinfo {author} {\bibfnamefont {H.~A.}\ \bibnamefont
  {Stern}}\ and\ \bibinfo {author} {\bibfnamefont {B.~J.}\ \bibnamefont
  {Berne}},\ }\href@noop {} {\bibfield  {journal} {\bibinfo  {journal} {J.
  Chem. Phys.}\ }\textbf {\bibinfo {volume} {115}},\ \bibinfo {pages} {7622}
  (\bibinfo {year} {2001})}\BibitemShut {NoStop}%
\bibitem [{\citenamefont {Paesani}\ \emph {et~al.}(2006)\citenamefont
  {Paesani}, \citenamefont {Zhang}, \citenamefont {Case}, \citenamefont
  {Cheatham},\ and\ \citenamefont {Voth}}]{paesani_accurate_2006}%
  \BibitemOpen
  \bibfield  {author} {\bibinfo {author} {\bibfnamefont {F.}~\bibnamefont
  {Paesani}}, \bibinfo {author} {\bibfnamefont {W.}~\bibnamefont {Zhang}},
  \bibinfo {author} {\bibfnamefont {D.~A.}\ \bibnamefont {Case}}, \bibinfo
  {author} {\bibfnamefont {T.~E.}\ \bibnamefont {Cheatham}}, \ and\ \bibinfo
  {author} {\bibfnamefont {G.~A.}\ \bibnamefont {Voth}},\ }\href@noop {}
  {\bibfield  {journal} {\bibinfo  {journal} {J. Chem. Phys.}\ }\textbf
  {\bibinfo {volume} {125}},\ \bibinfo {pages} {184507} (\bibinfo {year}
  {2006})}\BibitemShut {NoStop}%
\bibitem [{\citenamefont {Habershon}\ \emph {et~al.}(2009)\citenamefont
  {Habershon}, \citenamefont {Markland},\ and\ \citenamefont
  {Manolopoulos}}]{habershon_competing_2009}%
  \BibitemOpen
  \bibfield  {author} {\bibinfo {author} {\bibfnamefont {S.}~\bibnamefont
  {Habershon}}, \bibinfo {author} {\bibfnamefont {T.~E.}\ \bibnamefont
  {Markland}}, \ and\ \bibinfo {author} {\bibfnamefont {D.~E.}\ \bibnamefont
  {Manolopoulos}},\ }\href@noop {} {\bibfield  {journal} {\bibinfo  {journal}
  {J. Chem. Phys.}\ }\textbf {\bibinfo {volume} {131}},\ \bibinfo {pages}
  {024501} (\bibinfo {year} {2009})}\BibitemShut {NoStop}%
\bibitem [{\citenamefont {Ceriotti}\ \emph {et~al.}(2011)\citenamefont
  {Ceriotti}, \citenamefont {Manolopoulos},\ and\ \citenamefont
  {Parrinello}}]{ceriotti_accelerating_2011}%
  \BibitemOpen
  \bibfield  {author} {\bibinfo {author} {\bibfnamefont {M.}~\bibnamefont
  {Ceriotti}}, \bibinfo {author} {\bibfnamefont {D.~E.}\ \bibnamefont
  {Manolopoulos}}, \ and\ \bibinfo {author} {\bibfnamefont {M.}~\bibnamefont
  {Parrinello}},\ }\href@noop {} {\bibfield  {journal} {\bibinfo  {journal} {J.
  Chem. Phys.}\ }\textbf {\bibinfo {volume} {134}},\ \bibinfo {pages} {084104}
  (\bibinfo {year} {2011})}\BibitemShut {NoStop}%
\bibitem [{\citenamefont {Dammak}\ \emph {et~al.}(2009)\citenamefont {Dammak},
  \citenamefont {Chalopin}, \citenamefont {Laroche}, \citenamefont {Hayoun},\
  and\ \citenamefont {Greffet}}]{dammak_quantum_2009}%
  \BibitemOpen
  \bibfield  {author} {\bibinfo {author} {\bibfnamefont {H.}~\bibnamefont
  {Dammak}}, \bibinfo {author} {\bibfnamefont {Y.}~\bibnamefont {Chalopin}},
  \bibinfo {author} {\bibfnamefont {M.}~\bibnamefont {Laroche}}, \bibinfo
  {author} {\bibfnamefont {M.}~\bibnamefont {Hayoun}}, \ and\ \bibinfo {author}
  {\bibfnamefont {J.-J.}\ \bibnamefont {Greffet}},\ }\href@noop {} {\bibfield
  {journal} {\bibinfo  {journal} {Phys. Rev. Lett.}\ }\textbf {\bibinfo
  {volume} {103}},\ \bibinfo {pages} {190601} (\bibinfo {year}
  {2009})}\BibitemShut {NoStop}%
\bibitem [{\citenamefont {Ceriotti}\ \emph {et~al.}(2009)\citenamefont
  {Ceriotti}, \citenamefont {Bussi},\ and\ \citenamefont
  {Parrinello}}]{ceriotti_nuclear_2009}%
  \BibitemOpen
  \bibfield  {author} {\bibinfo {author} {\bibfnamefont {M.}~\bibnamefont
  {Ceriotti}}, \bibinfo {author} {\bibfnamefont {G.}~\bibnamefont {Bussi}}, \
  and\ \bibinfo {author} {\bibfnamefont {M.}~\bibnamefont {Parrinello}},\
  }\href@noop {} {\bibfield  {journal} {\bibinfo  {journal} {Phys. Rev. Lett.}\
  }\textbf {\bibinfo {volume} {103}},\ \bibinfo {pages} {030603} (\bibinfo
  {year} {2009})}\BibitemShut {NoStop}%
\bibitem [{\citenamefont {Ceriotti}\ and\ \citenamefont
  {Manolopoulos}(2012)}]{ceriotti_efficient_2012}%
  \BibitemOpen
  \bibfield  {author} {\bibinfo {author} {\bibfnamefont {M.}~\bibnamefont
  {Ceriotti}}\ and\ \bibinfo {author} {\bibfnamefont {D.~E.}\ \bibnamefont
  {Manolopoulos}},\ }\href@noop {} {\bibfield  {journal} {\bibinfo  {journal}
  {Phys. Rev. Lett.}\ }\textbf {\bibinfo {volume} {109}},\ \bibinfo {pages}
  {100604} (\bibinfo {year} {2012})}\BibitemShut {NoStop}%
\bibitem [{\citenamefont {Markland}\ and\ \citenamefont
  {Manolopoulos}(2008{\natexlab{a}})}]{markland_refined_2008}%
  \BibitemOpen
  \bibfield  {author} {\bibinfo {author} {\bibfnamefont {T.~E.}\ \bibnamefont
  {Markland}}\ and\ \bibinfo {author} {\bibfnamefont {D.~E.}\ \bibnamefont
  {Manolopoulos}},\ }\href@noop {} {\bibfield  {journal} {\bibinfo  {journal}
  {Chem. Phys. Lett.}\ }\textbf {\bibinfo {volume} {464}},\ \bibinfo {pages}
  {256} (\bibinfo {year} {2008}{\natexlab{a}})}\BibitemShut {NoStop}%
\bibitem [{\citenamefont {Markland}\ and\ \citenamefont
  {Manolopoulos}(2008{\natexlab{b}})}]{markland_efficient_2008}%
  \BibitemOpen
  \bibfield  {author} {\bibinfo {author} {\bibfnamefont {T.~E.}\ \bibnamefont
  {Markland}}\ and\ \bibinfo {author} {\bibfnamefont {D.~E.}\ \bibnamefont
  {Manolopoulos}},\ }\href@noop {} {\bibfield  {journal} {\bibinfo  {journal}
  {J. Chem. Phys.}\ }\textbf {\bibinfo {volume} {129}},\ \bibinfo {pages}
  {024105} (\bibinfo {year} {2008}{\natexlab{b}})}\BibitemShut {NoStop}%
\bibitem [{\citenamefont {Fanourgakis}\ \emph {et~al.}(2009)\citenamefont
  {Fanourgakis}, \citenamefont {Markland},\ and\ \citenamefont
  {Manolopoulos}}]{fanourgakis_fast_2009}%
  \BibitemOpen
  \bibfield  {author} {\bibinfo {author} {\bibfnamefont {G.~S.}\ \bibnamefont
  {Fanourgakis}}, \bibinfo {author} {\bibfnamefont {T.~E.}\ \bibnamefont
  {Markland}}, \ and\ \bibinfo {author} {\bibfnamefont {D.~E.}\ \bibnamefont
  {Manolopoulos}},\ }\href@noop {} {\bibfield  {journal} {\bibinfo  {journal}
  {J. Chem. Phys.}\ }\textbf {\bibinfo {volume} {131}},\ \bibinfo {pages}
  {094102} (\bibinfo {year} {2009})}\BibitemShut {NoStop}%
\bibitem [{\citenamefont {Marsalek}\ and\ \citenamefont
  {Markland}(2016)}]{marsalek_ab_2016}%
  \BibitemOpen
  \bibfield  {author} {\bibinfo {author} {\bibfnamefont {O.}~\bibnamefont
  {Marsalek}}\ and\ \bibinfo {author} {\bibfnamefont {T.~E.}\ \bibnamefont
  {Markland}},\ }\href@noop {} {\bibfield  {journal} {\bibinfo  {journal} {J.
  Chem. Phys.}\ }\textbf {\bibinfo {volume} {144}},\ \bibinfo {pages} {054112}
  (\bibinfo {year} {2016})}\BibitemShut {NoStop}%
\bibitem [{\citenamefont {Kapil}\ \emph {et~al.}(2016)\citenamefont {Kapil},
  \citenamefont {VandeVondele},\ and\ \citenamefont
  {Ceriotti}}]{kapil_accurate_2016}%
  \BibitemOpen
  \bibfield  {author} {\bibinfo {author} {\bibfnamefont {V.}~\bibnamefont
  {Kapil}}, \bibinfo {author} {\bibfnamefont {J.}~\bibnamefont {VandeVondele}},
  \ and\ \bibinfo {author} {\bibfnamefont {M.}~\bibnamefont {Ceriotti}},\
  }\href@noop {} {\bibfield  {journal} {\bibinfo  {journal} {J. Chem. Phys.}\
  }\textbf {\bibinfo {volume} {144}},\ \bibinfo {pages} {054111} (\bibinfo
  {year} {2016})}\BibitemShut {NoStop}%
\bibitem [{\citenamefont {Poltavsky}\ and\ \citenamefont
  {Tkatchenko}(2016)}]{poltavsky_modeling_2016}%
  \BibitemOpen
  \bibfield  {author} {\bibinfo {author} {\bibfnamefont {I.}~\bibnamefont
  {Poltavsky}}\ and\ \bibinfo {author} {\bibfnamefont {A.}~\bibnamefont
  {Tkatchenko}},\ }\href@noop {} {\bibfield  {journal} {\bibinfo  {journal}
  {Chem. Sci.}\ }\textbf {\bibinfo {volume} {7}},\ \bibinfo {pages} {1368}
  (\bibinfo {year} {2016})}\BibitemShut {NoStop}%
\bibitem [{\citenamefont {Poltavsky}\ \emph {et~al.}(2017)\citenamefont
  {Poltavsky}, \citenamefont {DiStasio~Jr.},\ and\ \citenamefont
  {Tkatchenko}}]{poltavsky_perturbed_2017}%
  \BibitemOpen
  \bibfield  {author} {\bibinfo {author} {\bibfnamefont {I.}~\bibnamefont
  {Poltavsky}}, \bibinfo {author} {\bibfnamefont {R.~A.}\ \bibnamefont
  {DiStasio~Jr.}}, \ and\ \bibinfo {author} {\bibfnamefont {A.}~\bibnamefont
  {Tkatchenko}},\ }\href@noop {} {\bibfield  {journal} {\bibinfo  {journal} {J.
  Chem. Phys.}\ }\textbf {\bibinfo {volume} {148}},\ \bibinfo {pages} {102325}
  (\bibinfo {year} {2017})}\BibitemShut {NoStop}%
\bibitem [{\citenamefont {Behler}\ and\ \citenamefont
  {Parrinello}(2007)}]{behler2007generalized}%
  \BibitemOpen
  \bibfield  {author} {\bibinfo {author} {\bibfnamefont {J.}~\bibnamefont
  {Behler}}\ and\ \bibinfo {author} {\bibfnamefont {M.}~\bibnamefont
  {Parrinello}},\ }\href@noop {} {\bibfield  {journal} {\bibinfo  {journal}
  {Phys. Rev. Lett.}\ }\textbf {\bibinfo {volume} {98}},\ \bibinfo {pages}
  {146401} (\bibinfo {year} {2007})}\BibitemShut {NoStop}%
\bibitem [{\citenamefont {Bart{\'o}k}\ \emph {et~al.}(2010)\citenamefont
  {Bart{\'o}k}, \citenamefont {Payne}, \citenamefont {Kondor},\ and\
  \citenamefont {Cs{\'a}nyi}}]{bartok2010gaussian}%
  \BibitemOpen
  \bibfield  {author} {\bibinfo {author} {\bibfnamefont {A.~P.}\ \bibnamefont
  {Bart{\'o}k}}, \bibinfo {author} {\bibfnamefont {M.~C.}\ \bibnamefont
  {Payne}}, \bibinfo {author} {\bibfnamefont {R.}~\bibnamefont {Kondor}}, \
  and\ \bibinfo {author} {\bibfnamefont {G.}~\bibnamefont {Cs{\'a}nyi}},\
  }\href@noop {} {\bibfield  {journal} {\bibinfo  {journal} {Phys. Rev. Lett.}\
  }\textbf {\bibinfo {volume} {104}},\ \bibinfo {pages} {136403} (\bibinfo
  {year} {2010})}\BibitemShut {NoStop}%
\bibitem [{\citenamefont {Rupp}\ \emph {et~al.}(2012)\citenamefont {Rupp},
  \citenamefont {Tkatchenko}, \citenamefont {M{\"u}ller},\ and\ \citenamefont
  {von Lilienfeld}}]{rupp2012fast}%
  \BibitemOpen
  \bibfield  {author} {\bibinfo {author} {\bibfnamefont {M.}~\bibnamefont
  {Rupp}}, \bibinfo {author} {\bibfnamefont {A.}~\bibnamefont {Tkatchenko}},
  \bibinfo {author} {\bibfnamefont {K.-R.}\ \bibnamefont {M{\"u}ller}}, \ and\
  \bibinfo {author} {\bibfnamefont {O.~A.}\ \bibnamefont {von Lilienfeld}},\
  }\href@noop {} {\bibfield  {journal} {\bibinfo  {journal} {Phys. Rev. Lett.}\
  }\textbf {\bibinfo {volume} {108}},\ \bibinfo {pages} {058301} (\bibinfo
  {year} {2012})}\BibitemShut {NoStop}%
\bibitem [{\citenamefont {Montavon}\ \emph {et~al.}(2013)\citenamefont
  {Montavon}, \citenamefont {Rupp}, \citenamefont {Gobre}, \citenamefont
  {Vazquez-Mayagoitia}, \citenamefont {Hansen}, \citenamefont {Tkatchenko},
  \citenamefont {M{\"u}ller},\ and\ \citenamefont {von
  Lilienfeld}}]{montavon2013machine}%
  \BibitemOpen
  \bibfield  {author} {\bibinfo {author} {\bibfnamefont {G.}~\bibnamefont
  {Montavon}}, \bibinfo {author} {\bibfnamefont {M.}~\bibnamefont {Rupp}},
  \bibinfo {author} {\bibfnamefont {V.}~\bibnamefont {Gobre}}, \bibinfo
  {author} {\bibfnamefont {A.}~\bibnamefont {Vazquez-Mayagoitia}}, \bibinfo
  {author} {\bibfnamefont {K.}~\bibnamefont {Hansen}}, \bibinfo {author}
  {\bibfnamefont {A.}~\bibnamefont {Tkatchenko}}, \bibinfo {author}
  {\bibfnamefont {K.-R.}\ \bibnamefont {M{\"u}ller}}, \ and\ \bibinfo {author}
  {\bibfnamefont {O.~A.}\ \bibnamefont {von Lilienfeld}},\ }\href@noop {}
  {\bibfield  {journal} {\bibinfo  {journal} {New J. Phys.}\ }\textbf {\bibinfo
  {volume} {15}},\ \bibinfo {pages} {095003} (\bibinfo {year}
  {2013})}\BibitemShut {NoStop}%
\bibitem [{\citenamefont {Chmiela}\ \emph {et~al.}(2017)\citenamefont
  {Chmiela}, \citenamefont {Tkatchenko}, \citenamefont {Sauceda}, \citenamefont
  {Poltavsky}, \citenamefont {Sch{\"u}tt},\ and\ \citenamefont
  {M{\"u}ller}}]{chmiela2017machine}%
  \BibitemOpen
  \bibfield  {author} {\bibinfo {author} {\bibfnamefont {S.}~\bibnamefont
  {Chmiela}}, \bibinfo {author} {\bibfnamefont {A.}~\bibnamefont {Tkatchenko}},
  \bibinfo {author} {\bibfnamefont {H.~E.}\ \bibnamefont {Sauceda}}, \bibinfo
  {author} {\bibfnamefont {I.}~\bibnamefont {Poltavsky}}, \bibinfo {author}
  {\bibfnamefont {K.~T.}\ \bibnamefont {Sch{\"u}tt}}, \ and\ \bibinfo {author}
  {\bibfnamefont {K.-R.}\ \bibnamefont {M{\"u}ller}},\ }\href@noop {}
  {\bibfield  {journal} {\bibinfo  {journal} {Sci. Adv.}\ }\textbf {\bibinfo
  {volume} {3}},\ \bibinfo {pages} {e1603015} (\bibinfo {year}
  {2017})}\BibitemShut {NoStop}%
\bibitem [{\citenamefont {Sch{\"u}tt}\ \emph {et~al.}(2017)\citenamefont
  {Sch{\"u}tt}, \citenamefont {Kindermans}, \citenamefont {Sauceda~Felix},
  \citenamefont {Chmiela}, \citenamefont {Tkatchenko},\ and\ \citenamefont
  {M{\"u}ller}}]{schutt_schnet:_2017}%
  \BibitemOpen
  \bibfield  {author} {\bibinfo {author} {\bibfnamefont {K.}~\bibnamefont
  {Sch{\"u}tt}}, \bibinfo {author} {\bibfnamefont {P.-J.}\ \bibnamefont
  {Kindermans}}, \bibinfo {author} {\bibfnamefont {H.~E.}\ \bibnamefont
  {Sauceda~Felix}}, \bibinfo {author} {\bibfnamefont {S.}~\bibnamefont
  {Chmiela}}, \bibinfo {author} {\bibfnamefont {A.}~\bibnamefont {Tkatchenko}},
  \ and\ \bibinfo {author} {\bibfnamefont {K.-R.}\ \bibnamefont {M{\"u}ller}},\
  }in\ \href@noop {} {\emph {\bibinfo {booktitle} {Advances in {Neural}
  {Information} {Processing} {Systems} 30}}},\ \bibinfo {editor} {edited by\
  \bibinfo {editor} {\bibfnamefont {I.}~\bibnamefont {Guyon}}, \bibinfo
  {editor} {\bibfnamefont {U.~V.}\ \bibnamefont {Luxburg}}, \bibinfo {editor}
  {\bibfnamefont {S.}~\bibnamefont {Bengio}}, \bibinfo {editor} {\bibfnamefont
  {H.}~\bibnamefont {Wallach}}, \bibinfo {editor} {\bibfnamefont
  {R.}~\bibnamefont {Fergus}}, \bibinfo {editor} {\bibfnamefont
  {S.}~\bibnamefont {Vishwanathan}}, \ and\ \bibinfo {editor} {\bibfnamefont
  {R.}~\bibnamefont {Garnett}}}\ (\bibinfo  {publisher} {Curran Associates},\
  \bibinfo {address} {Red Hook},\ \bibinfo {year} {2017})\ pp.\ \bibinfo
  {pages} {991--1001}\BibitemShut {NoStop}%
\bibitem [{\citenamefont {Smith}\ \emph {et~al.}(2017)\citenamefont {Smith},
  \citenamefont {Isayev},\ and\ \citenamefont {Roitberg}}]{smith2017ani}%
  \BibitemOpen
  \bibfield  {author} {\bibinfo {author} {\bibfnamefont {J.~S.}\ \bibnamefont
  {Smith}}, \bibinfo {author} {\bibfnamefont {O.}~\bibnamefont {Isayev}}, \
  and\ \bibinfo {author} {\bibfnamefont {A.~E.}\ \bibnamefont {Roitberg}},\
  }\href@noop {} {\bibfield  {journal} {\bibinfo  {journal} {Chem. Sci.}\
  }\textbf {\bibinfo {volume} {8}},\ \bibinfo {pages} {3192} (\bibinfo {year}
  {2017})}\BibitemShut {NoStop}%
\bibitem [{\citenamefont {Han}\ \emph {et~al.}(2018)\citenamefont {Han},
  \citenamefont {Zhang}, \citenamefont {Car},\ and\ \citenamefont
  {E}}]{han2017deep}%
  \BibitemOpen
  \bibfield  {author} {\bibinfo {author} {\bibfnamefont {J.}~\bibnamefont
  {Han}}, \bibinfo {author} {\bibfnamefont {L.}~\bibnamefont {Zhang}}, \bibinfo
  {author} {\bibfnamefont {R.}~\bibnamefont {Car}}, \ and\ \bibinfo {author}
  {\bibfnamefont {W.}~\bibnamefont {E}},\ }\href@noop {} {\bibfield  {journal}
  {\bibinfo  {journal} {Commun. Comput. Phys.}\ }\textbf {\bibinfo {volume}
  {23}},\ \bibinfo {pages} {629} (\bibinfo {year} {2018})}\BibitemShut
  {NoStop}%
\bibitem [{\citenamefont {Zhang}\ \emph
  {et~al.}(2018{\natexlab{a}})\citenamefont {Zhang}, \citenamefont {Han},
  \citenamefont {Wang}, \citenamefont {Car},\ and\ \citenamefont
  {E}}]{zhang2018deepmd}%
  \BibitemOpen
  \bibfield  {author} {\bibinfo {author} {\bibfnamefont {L.}~\bibnamefont
  {Zhang}}, \bibinfo {author} {\bibfnamefont {J.}~\bibnamefont {Han}}, \bibinfo
  {author} {\bibfnamefont {H.}~\bibnamefont {Wang}}, \bibinfo {author}
  {\bibfnamefont {R.}~\bibnamefont {Car}}, \ and\ \bibinfo {author}
  {\bibfnamefont {W.}~\bibnamefont {E}},\ }\href@noop {} {\bibfield  {journal}
  {\bibinfo  {journal} {Phys. Rev. Lett.}\ }\textbf {\bibinfo {volume} {120}},\
  \bibinfo {pages} {143001} (\bibinfo {year} {2018}{\natexlab{a}})}\BibitemShut
  {NoStop}%
\bibitem [{\citenamefont {Zhang}\ \emph
  {et~al.}(2018{\natexlab{b}})\citenamefont {Zhang}, \citenamefont {Han},
  \citenamefont {Wang}, \citenamefont {Saidi}, \citenamefont {Car},\ and\
  \citenamefont {E}}]{zhang_end--end_2018}%
  \BibitemOpen
  \bibfield  {author} {\bibinfo {author} {\bibfnamefont {L.}~\bibnamefont
  {Zhang}}, \bibinfo {author} {\bibfnamefont {J.}~\bibnamefont {Han}}, \bibinfo
  {author} {\bibfnamefont {H.}~\bibnamefont {Wang}}, \bibinfo {author}
  {\bibfnamefont {W.}~\bibnamefont {Saidi}}, \bibinfo {author} {\bibfnamefont
  {R.}~\bibnamefont {Car}}, \ and\ \bibinfo {author} {\bibfnamefont
  {W.}~\bibnamefont {E}},\ }in\ \href@noop {} {\emph {\bibinfo {booktitle}
  {Advances in {Neural} {Information} {Processing} {Systems} 31}}},\ \bibinfo
  {editor} {edited by\ \bibinfo {editor} {\bibfnamefont {S.}~\bibnamefont
  {Bengio}}, \bibinfo {editor} {\bibfnamefont {H.}~\bibnamefont {Wallach}},
  \bibinfo {editor} {\bibfnamefont {H.}~\bibnamefont {Larochelle}}, \bibinfo
  {editor} {\bibfnamefont {K.}~\bibnamefont {Grauman}}, \bibinfo {editor}
  {\bibfnamefont {N.}~\bibnamefont {Cesa-Bianchi}}, \ and\ \bibinfo {editor}
  {\bibfnamefont {R.}~\bibnamefont {Garnett}}}\ (\bibinfo  {publisher} {Curran
  Associates},\ \bibinfo {address} {Red Hook},\ \bibinfo {year} {2018})\ pp.\
  \bibinfo {pages} {4436--4446}\BibitemShut {NoStop}%
\bibitem [{\citenamefont {Morawietz}\ \emph {et~al.}(2016)\citenamefont
  {Morawietz}, \citenamefont {Singraber}, \citenamefont {Dellago},\ and\
  \citenamefont {Behler}}]{morawietz2016van}%
  \BibitemOpen
  \bibfield  {author} {\bibinfo {author} {\bibfnamefont {T.}~\bibnamefont
  {Morawietz}}, \bibinfo {author} {\bibfnamefont {A.}~\bibnamefont
  {Singraber}}, \bibinfo {author} {\bibfnamefont {C.}~\bibnamefont {Dellago}},
  \ and\ \bibinfo {author} {\bibfnamefont {J.}~\bibnamefont {Behler}},\
  }\href@noop {} {\bibfield  {journal} {\bibinfo  {journal} {Proc. Natl. Acad.
  Sci. U.S.A.}\ }\textbf {\bibinfo {volume} {113}},\ \bibinfo {pages} {8368}
  (\bibinfo {year} {2016})}\BibitemShut {NoStop}%
\bibitem [{\citenamefont {Wang}\ and\ \citenamefont
  {Yang}(2018)}]{wang2018force}%
  \BibitemOpen
  \bibfield  {author} {\bibinfo {author} {\bibfnamefont {H.}~\bibnamefont
  {Wang}}\ and\ \bibinfo {author} {\bibfnamefont {W.}~\bibnamefont {Yang}},\
  }\href@noop {} {\bibfield  {journal} {\bibinfo  {journal} {J. Phys. Chem.
  Lett.}\ }\textbf {\bibinfo {volume} {9}},\ \bibinfo {pages} {3232} (\bibinfo
  {year} {2018})}\BibitemShut {NoStop}%
\bibitem [{\citenamefont {Zhang}\ \emph
  {et~al.}(2018{\natexlab{c}})\citenamefont {Zhang}, \citenamefont {Han},
  \citenamefont {Wang}, \citenamefont {Car},\ and\ \citenamefont
  {E}}]{zhang2018deepcg}%
  \BibitemOpen
  \bibfield  {author} {\bibinfo {author} {\bibfnamefont {L.}~\bibnamefont
  {Zhang}}, \bibinfo {author} {\bibfnamefont {J.}~\bibnamefont {Han}}, \bibinfo
  {author} {\bibfnamefont {H.}~\bibnamefont {Wang}}, \bibinfo {author}
  {\bibfnamefont {R.}~\bibnamefont {Car}}, \ and\ \bibinfo {author}
  {\bibfnamefont {W.}~\bibnamefont {E}},\ }\href@noop {} {\bibfield  {journal}
  {\bibinfo  {journal} {J. Chem. Phys.}\ }\textbf {\bibinfo {volume} {149}},\
  \bibinfo {pages} {034101} (\bibinfo {year} {2018}{\natexlab{c}})}\BibitemShut
  {NoStop}%
\bibitem [{\citenamefont {Santra}\ \emph {et~al.}(prep)\citenamefont {Santra},
  \citenamefont {Ko}, \citenamefont {Zhang}, \citenamefont {DiStasio~Jr.},\
  and\ \citenamefont {Car}}]{santra_manuscript_water_nqe}%
  \BibitemOpen
  \bibfield  {author} {\bibinfo {author} {\bibfnamefont {B.}~\bibnamefont
  {Santra}}, \bibinfo {author} {\bibfnamefont {H.-Y.}\ \bibnamefont {Ko}},
  \bibinfo {author} {\bibfnamefont {L.}~\bibnamefont {Zhang}}, \bibinfo
  {author} {\bibfnamefont {R.~A.}\ \bibnamefont {DiStasio~Jr.}}, \ and\
  \bibinfo {author} {\bibfnamefont {R.}~\bibnamefont {Car}},\ }\href@noop {}
  {\enquote {\bibinfo {title} {{Nuclear} {Quantum} {Effects} in {Liquid}
  {Water} and {Ice}},}\ } (\bibinfo {year} {in prep.})\BibitemShut {NoStop}%
\bibitem [{\citenamefont {Raiteri}\ \emph {et~al.}(2011)\citenamefont
  {Raiteri}, \citenamefont {Gale},\ and\ \citenamefont
  {Bussi}}]{raiteri_reactive_2011}%
  \BibitemOpen
  \bibfield  {author} {\bibinfo {author} {\bibfnamefont {P.}~\bibnamefont
  {Raiteri}}, \bibinfo {author} {\bibfnamefont {J.~D.}\ \bibnamefont {Gale}}, \
  and\ \bibinfo {author} {\bibfnamefont {G.}~\bibnamefont {Bussi}},\
  }\href@noop {} {\bibfield  {journal} {\bibinfo  {journal} {J. Phys.: Condens.
  Matter}\ }\textbf {\bibinfo {volume} {23}},\ \bibinfo {pages} {334213}
  (\bibinfo {year} {2011})}\BibitemShut {NoStop}%
\bibitem [{\citenamefont {Ceriotti}\ \emph {et~al.}(2010)\citenamefont
  {Ceriotti}, \citenamefont {Bussi},\ and\ \citenamefont
  {Parrinello}}]{ceriotti_colored-noise_2010}%
  \BibitemOpen
  \bibfield  {author} {\bibinfo {author} {\bibfnamefont {M.}~\bibnamefont
  {Ceriotti}}, \bibinfo {author} {\bibfnamefont {G.}~\bibnamefont {Bussi}}, \
  and\ \bibinfo {author} {\bibfnamefont {M.}~\bibnamefont {Parrinello}},\
  }\href@noop {} {\bibfield  {journal} {\bibinfo  {journal} {J. Chem. Theory
  Comput.}\ }\textbf {\bibinfo {volume} {6}},\ \bibinfo {pages} {1170}
  (\bibinfo {year} {2010})}\BibitemShut {NoStop}%
\bibitem [{\citenamefont {Ceriotti}\ \emph {et~al.}(2014)\citenamefont
  {Ceriotti}, \citenamefont {More},\ and\ \citenamefont
  {Manolopoulos}}]{ceriotti_i-pi:_2014}%
  \BibitemOpen
  \bibfield  {author} {\bibinfo {author} {\bibfnamefont {M.}~\bibnamefont
  {Ceriotti}}, \bibinfo {author} {\bibfnamefont {J.}~\bibnamefont {More}}, \
  and\ \bibinfo {author} {\bibfnamefont {D.~E.}\ \bibnamefont {Manolopoulos}},\
  }\href@noop {} {\bibfield  {journal} {\bibinfo  {journal} {Comput. Phys.
  Commun.}\ }\textbf {\bibinfo {volume} {185}},\ \bibinfo {pages} {1019}
  (\bibinfo {year} {2014})}\BibitemShut {NoStop}%
\bibitem [{\citenamefont {Wu}\ \emph {et~al.}(2009)\citenamefont {Wu},
  \citenamefont {Selloni},\ and\ \citenamefont {Car}}]{wu_order-n_2009}%
  \BibitemOpen
  \bibfield  {author} {\bibinfo {author} {\bibfnamefont {X.}~\bibnamefont
  {Wu}}, \bibinfo {author} {\bibfnamefont {A.}~\bibnamefont {Selloni}}, \ and\
  \bibinfo {author} {\bibfnamefont {R.}~\bibnamefont {Car}},\ }\href@noop {}
  {\bibfield  {journal} {\bibinfo  {journal} {Phys. Rev. B}\ }\textbf {\bibinfo
  {volume} {79}},\ \bibinfo {pages} {085102} (\bibinfo {year}
  {2009})}\BibitemShut {NoStop}%
\bibitem [{\citenamefont {Ko}\ \emph {et~al.}(prep{\natexlab{a}})\citenamefont
  {Ko}, \citenamefont {Jia}, \citenamefont {Santra}, \citenamefont {Wu},
  \citenamefont {Car},\ and\ \citenamefont {DiStasio~Jr.}}]{HYK-paperI}%
  \BibitemOpen
  \bibfield  {author} {\bibinfo {author} {\bibfnamefont {H.-Y.}\ \bibnamefont
  {Ko}}, \bibinfo {author} {\bibfnamefont {J.}~\bibnamefont {Jia}}, \bibinfo
  {author} {\bibfnamefont {B.}~\bibnamefont {Santra}}, \bibinfo {author}
  {\bibfnamefont {X.}~\bibnamefont {Wu}}, \bibinfo {author} {\bibfnamefont
  {R.}~\bibnamefont {Car}}, \ and\ \bibinfo {author} {\bibfnamefont {R.~A.}\
  \bibnamefont {DiStasio~Jr.}},\ }\href@noop {} {\enquote {\bibinfo {title}
  {{Enabling} {Large}-{Scale} {Condensed}-{Phase} {Hybrid} {Density}
  {Functional} {Theory} {Based} \textit{Ab Initio} {Molecular} {Dynamics} {I}:
  {Theory}, {Algorithm}, and {Performance}},}\ } (\bibinfo {year} {in
  prep.}{\natexlab{a}})\BibitemShut {NoStop}%
\bibitem [{\citenamefont {Ko}\ \emph {et~al.}(prep{\natexlab{b}})\citenamefont
  {Ko}, \citenamefont {Santra}, \citenamefont {Car},\ and\ \citenamefont
  {DiStasio~Jr.}}]{HYK-paperII}%
  \BibitemOpen
  \bibfield  {author} {\bibinfo {author} {\bibfnamefont {H.-Y.}\ \bibnamefont
  {Ko}}, \bibinfo {author} {\bibfnamefont {B.}~\bibnamefont {Santra}}, \bibinfo
  {author} {\bibfnamefont {R.}~\bibnamefont {Car}}, \ and\ \bibinfo {author}
  {\bibfnamefont {R.~A.}\ \bibnamefont {DiStasio~Jr.}},\ }\href@noop {}
  {\enquote {\bibinfo {title} {{Enabling} {Large}-{Scale} {Condensed}-{Phase}
  {Hybrid} {Density} {Functional} {Theory} {Based} \textit{Ab Initio}
  {Molecular} {Dynamics} {II}: {Extensions} to the {Isobaric}-{Isothermal}
  {Ensemble}},}\ } (\bibinfo {year} {in prep.}{\natexlab{b}})\BibitemShut
  {NoStop}%
\bibitem [{\citenamefont {Giannozzi}\ \emph {et~al.}(2009)\citenamefont
  {Giannozzi}, \citenamefont {Baroni}, \citenamefont {Bonini}, \citenamefont
  {Calandra}, \citenamefont {Car}, \citenamefont {Cavazzoni}, \citenamefont
  {Ceresoli}, \citenamefont {Chiarotti}, \citenamefont {Cococcioni},
  \citenamefont {Dabo}, \citenamefont {Dal~Corso}, \citenamefont
  {de~Gironcoli}, \citenamefont {Fabris}, \citenamefont {Fratesi},
  \citenamefont {Gebauer}, \citenamefont {Gerstmann}, \citenamefont
  {Gougoussis}, \citenamefont {Kokalj}, \citenamefont {Lazzeri}, \citenamefont
  {Martin-Samos}, \citenamefont {Marzari}, \citenamefont {Mauri}, \citenamefont
  {Mazzarello}, \citenamefont {Paolini}, \citenamefont {Pasquarello},
  \citenamefont {Paulatto}, \citenamefont {Sbraccia}, \citenamefont {Scandolo},
  \citenamefont {Sclauzero}, \citenamefont {Seitsonen}, \citenamefont
  {Smogunov}, \citenamefont {Umari},\ and\ \citenamefont
  {Wentzcovitch}}]{giannozzi_quantum_2009}%
  \BibitemOpen
  \bibfield  {author} {\bibinfo {author} {\bibfnamefont {P.}~\bibnamefont
  {Giannozzi}}, \bibinfo {author} {\bibfnamefont {S.}~\bibnamefont {Baroni}},
  \bibinfo {author} {\bibfnamefont {N.}~\bibnamefont {Bonini}}, \bibinfo
  {author} {\bibfnamefont {M.}~\bibnamefont {Calandra}}, \bibinfo {author}
  {\bibfnamefont {R.}~\bibnamefont {Car}}, \bibinfo {author} {\bibfnamefont
  {C.}~\bibnamefont {Cavazzoni}}, \bibinfo {author} {\bibfnamefont
  {D.}~\bibnamefont {Ceresoli}}, \bibinfo {author} {\bibfnamefont {G.~L.}\
  \bibnamefont {Chiarotti}}, \bibinfo {author} {\bibfnamefont {M.}~\bibnamefont
  {Cococcioni}}, \bibinfo {author} {\bibfnamefont {I.}~\bibnamefont {Dabo}},
  \bibinfo {author} {\bibfnamefont {A.}~\bibnamefont {Dal~Corso}}, \bibinfo
  {author} {\bibfnamefont {S.}~\bibnamefont {de~Gironcoli}}, \bibinfo {author}
  {\bibfnamefont {S.}~\bibnamefont {Fabris}}, \bibinfo {author} {\bibfnamefont
  {G.}~\bibnamefont {Fratesi}}, \bibinfo {author} {\bibfnamefont
  {R.}~\bibnamefont {Gebauer}}, \bibinfo {author} {\bibfnamefont
  {U.}~\bibnamefont {Gerstmann}}, \bibinfo {author} {\bibfnamefont
  {C.}~\bibnamefont {Gougoussis}}, \bibinfo {author} {\bibfnamefont
  {A.}~\bibnamefont {Kokalj}}, \bibinfo {author} {\bibfnamefont
  {M.}~\bibnamefont {Lazzeri}}, \bibinfo {author} {\bibfnamefont
  {L.}~\bibnamefont {Martin-Samos}}, \bibinfo {author} {\bibfnamefont
  {N.}~\bibnamefont {Marzari}}, \bibinfo {author} {\bibfnamefont
  {F.}~\bibnamefont {Mauri}}, \bibinfo {author} {\bibfnamefont
  {R.}~\bibnamefont {Mazzarello}}, \bibinfo {author} {\bibfnamefont
  {S.}~\bibnamefont {Paolini}}, \bibinfo {author} {\bibfnamefont
  {A.}~\bibnamefont {Pasquarello}}, \bibinfo {author} {\bibfnamefont
  {L.}~\bibnamefont {Paulatto}}, \bibinfo {author} {\bibfnamefont
  {C.}~\bibnamefont {Sbraccia}}, \bibinfo {author} {\bibfnamefont
  {S.}~\bibnamefont {Scandolo}}, \bibinfo {author} {\bibfnamefont
  {G.}~\bibnamefont {Sclauzero}}, \bibinfo {author} {\bibfnamefont {A.~P.}\
  \bibnamefont {Seitsonen}}, \bibinfo {author} {\bibfnamefont {A.}~\bibnamefont
  {Smogunov}}, \bibinfo {author} {\bibfnamefont {P.}~\bibnamefont {Umari}}, \
  and\ \bibinfo {author} {\bibfnamefont {R.~M.}\ \bibnamefont {Wentzcovitch}},\
  }\href@noop {} {\bibfield  {journal} {\bibinfo  {journal} {J. Phys.: Condens.
  Matter}\ }\textbf {\bibinfo {volume} {21}},\ \bibinfo {pages} {395502}
  (\bibinfo {year} {2009})}\BibitemShut {NoStop}%
\bibitem [{\citenamefont {Giannozzi}\ \emph {et~al.}(2017)\citenamefont
  {Giannozzi}, \citenamefont {Andreussi}, \citenamefont {Brumme}, \citenamefont
  {Bunau}, \citenamefont {Nardelli}, \citenamefont {Calandra}, \citenamefont
  {Car}, \citenamefont {Cavazzoni}, \citenamefont {Ceresoli}, \citenamefont
  {Cococcioni}, \citenamefont {Colonna}, \citenamefont {Carnimeo},
  \citenamefont {Corso}, \citenamefont {de~Gironcoli}, \citenamefont {Delugas},
  \citenamefont {DiStasio~Jr.}, \citenamefont {Ferretti}, \citenamefont
  {Floris}, \citenamefont {Fratesi}, \citenamefont {Fugallo}, \citenamefont
  {Gebauer}, \citenamefont {Gerstmann}, \citenamefont {Giustino}, \citenamefont
  {Gorni}, \citenamefont {Jia}, \citenamefont {Kawamura}, \citenamefont {Ko},
  \citenamefont {Kokalj}, \citenamefont {K{\"u}{\c c}{\"u}kbenli},
  \citenamefont {Lazzeri}, \citenamefont {Marsili}, \citenamefont {Marzari},
  \citenamefont {Mauri}, \citenamefont {Nguyen}, \citenamefont {Nguyen},
  \citenamefont {Otero-de-la Roza}, \citenamefont {Paulatto}, \citenamefont
  {Ponc{\'e}}, \citenamefont {Rocca}, \citenamefont {Sabatini}, \citenamefont
  {Santra}, \citenamefont {Schlipf}, \citenamefont {Seitsonen}, \citenamefont
  {Smogunov}, \citenamefont {Timrov}, \citenamefont {Thonhauser}, \citenamefont
  {Umari}, \citenamefont {Vast}, \citenamefont {Wu},\ and\ \citenamefont
  {Baroni}}]{giannozzi_advanced_2017}%
  \BibitemOpen
  \bibfield  {author} {\bibinfo {author} {\bibfnamefont {P.}~\bibnamefont
  {Giannozzi}}, \bibinfo {author} {\bibfnamefont {O.}~\bibnamefont
  {Andreussi}}, \bibinfo {author} {\bibfnamefont {T.}~\bibnamefont {Brumme}},
  \bibinfo {author} {\bibfnamefont {O.}~\bibnamefont {Bunau}}, \bibinfo
  {author} {\bibfnamefont {M.~B.}\ \bibnamefont {Nardelli}}, \bibinfo {author}
  {\bibfnamefont {M.}~\bibnamefont {Calandra}}, \bibinfo {author}
  {\bibfnamefont {R.}~\bibnamefont {Car}}, \bibinfo {author} {\bibfnamefont
  {C.}~\bibnamefont {Cavazzoni}}, \bibinfo {author} {\bibfnamefont
  {D.}~\bibnamefont {Ceresoli}}, \bibinfo {author} {\bibfnamefont
  {M.}~\bibnamefont {Cococcioni}}, \bibinfo {author} {\bibfnamefont
  {N.}~\bibnamefont {Colonna}}, \bibinfo {author} {\bibfnamefont
  {I.}~\bibnamefont {Carnimeo}}, \bibinfo {author} {\bibfnamefont {A.~D.}\
  \bibnamefont {Corso}}, \bibinfo {author} {\bibfnamefont {S.}~\bibnamefont
  {de~Gironcoli}}, \bibinfo {author} {\bibfnamefont {P.}~\bibnamefont
  {Delugas}}, \bibinfo {author} {\bibfnamefont {R.~A.}\ \bibnamefont
  {DiStasio~Jr.}}, \bibinfo {author} {\bibfnamefont {A.}~\bibnamefont
  {Ferretti}}, \bibinfo {author} {\bibfnamefont {A.}~\bibnamefont {Floris}},
  \bibinfo {author} {\bibfnamefont {G.}~\bibnamefont {Fratesi}}, \bibinfo
  {author} {\bibfnamefont {G.}~\bibnamefont {Fugallo}}, \bibinfo {author}
  {\bibfnamefont {R.}~\bibnamefont {Gebauer}}, \bibinfo {author} {\bibfnamefont
  {U.}~\bibnamefont {Gerstmann}}, \bibinfo {author} {\bibfnamefont
  {F.}~\bibnamefont {Giustino}}, \bibinfo {author} {\bibfnamefont
  {T.}~\bibnamefont {Gorni}}, \bibinfo {author} {\bibfnamefont
  {J.}~\bibnamefont {Jia}}, \bibinfo {author} {\bibfnamefont {M.}~\bibnamefont
  {Kawamura}}, \bibinfo {author} {\bibfnamefont {H.-Y.}\ \bibnamefont {Ko}},
  \bibinfo {author} {\bibfnamefont {A.}~\bibnamefont {Kokalj}}, \bibinfo
  {author} {\bibfnamefont {E.}~\bibnamefont {K{\"u}{\c c}{\"u}kbenli}},
  \bibinfo {author} {\bibfnamefont {M.}~\bibnamefont {Lazzeri}}, \bibinfo
  {author} {\bibfnamefont {M.}~\bibnamefont {Marsili}}, \bibinfo {author}
  {\bibfnamefont {N.}~\bibnamefont {Marzari}}, \bibinfo {author} {\bibfnamefont
  {F.}~\bibnamefont {Mauri}}, \bibinfo {author} {\bibfnamefont {N.~L.}\
  \bibnamefont {Nguyen}}, \bibinfo {author} {\bibfnamefont {H.-V.}\
  \bibnamefont {Nguyen}}, \bibinfo {author} {\bibfnamefont {A.}~\bibnamefont
  {Otero-de-la Roza}}, \bibinfo {author} {\bibfnamefont {L.}~\bibnamefont
  {Paulatto}}, \bibinfo {author} {\bibfnamefont {S.}~\bibnamefont {Ponc{\'e}}},
  \bibinfo {author} {\bibfnamefont {D.}~\bibnamefont {Rocca}}, \bibinfo
  {author} {\bibfnamefont {R.}~\bibnamefont {Sabatini}}, \bibinfo {author}
  {\bibfnamefont {B.}~\bibnamefont {Santra}}, \bibinfo {author} {\bibfnamefont
  {M.}~\bibnamefont {Schlipf}}, \bibinfo {author} {\bibfnamefont {A.~P.}\
  \bibnamefont {Seitsonen}}, \bibinfo {author} {\bibfnamefont {A.}~\bibnamefont
  {Smogunov}}, \bibinfo {author} {\bibfnamefont {I.}~\bibnamefont {Timrov}},
  \bibinfo {author} {\bibfnamefont {T.}~\bibnamefont {Thonhauser}}, \bibinfo
  {author} {\bibfnamefont {P.}~\bibnamefont {Umari}}, \bibinfo {author}
  {\bibfnamefont {N.}~\bibnamefont {Vast}}, \bibinfo {author} {\bibfnamefont
  {X.}~\bibnamefont {Wu}}, \ and\ \bibinfo {author} {\bibfnamefont
  {S.}~\bibnamefont {Baroni}},\ }\href@noop {} {\bibfield  {journal} {\bibinfo
  {journal} {J. Phys.: Condens. Matter}\ }\textbf {\bibinfo {volume} {29}},\
  \bibinfo {pages} {465901} (\bibinfo {year} {2017})}\BibitemShut {NoStop}%
\bibitem [{\citenamefont {Hamann}\ \emph {et~al.}(1979)\citenamefont {Hamann},
  \citenamefont {Schl{\"u}ter},\ and\ \citenamefont
  {Chiang}}]{hamann_norm-conserving_1979}%
  \BibitemOpen
  \bibfield  {author} {\bibinfo {author} {\bibfnamefont {D.~R.}\ \bibnamefont
  {Hamann}}, \bibinfo {author} {\bibfnamefont {M.}~\bibnamefont
  {Schl{\"u}ter}}, \ and\ \bibinfo {author} {\bibfnamefont {C.}~\bibnamefont
  {Chiang}},\ }\href@noop {} {\bibfield  {journal} {\bibinfo  {journal} {Phys.
  Rev. Lett.}\ }\textbf {\bibinfo {volume} {43}},\ \bibinfo {pages} {1494}
  (\bibinfo {year} {1979})}\BibitemShut {NoStop}%
\bibitem [{\citenamefont {Vanderbilt}(1985)}]{vanderbilt_optimally_1985}%
  \BibitemOpen
  \bibfield  {author} {\bibinfo {author} {\bibfnamefont {D.}~\bibnamefont
  {Vanderbilt}},\ }\href@noop {} {\bibfield  {journal} {\bibinfo  {journal}
  {Phys. Rev. B}\ }\textbf {\bibinfo {volume} {32}},\ \bibinfo {pages} {8412}
  (\bibinfo {year} {1985})}\BibitemShut {NoStop}%
\bibitem [{\citenamefont {Gygi}(2008)}]{gygi_architecture_2008}%
  \BibitemOpen
  \bibfield  {author} {\bibinfo {author} {\bibfnamefont {F.}~\bibnamefont
  {Gygi}},\ }\href@noop {} {\bibfield  {journal} {\bibinfo  {journal} {IBM J.
  RES. \& DEV.}\ }\textbf {\bibinfo {volume} {52}},\ \bibinfo {pages} {137}
  (\bibinfo {year} {2008})}\BibitemShut {NoStop}%
\bibitem [{\citenamefont {Bernasconi}\ \emph {et~al.}(1995)\citenamefont
  {Bernasconi}, \citenamefont {Chiarotti}, \citenamefont {Focher},
  \citenamefont {Scandolo}, \citenamefont {Tosatti},\ and\ \citenamefont
  {Parrinello}}]{bernasconi_first-principle-constant_1995}%
  \BibitemOpen
  \bibfield  {author} {\bibinfo {author} {\bibfnamefont {M.}~\bibnamefont
  {Bernasconi}}, \bibinfo {author} {\bibfnamefont {G.}~\bibnamefont
  {Chiarotti}}, \bibinfo {author} {\bibfnamefont {P.}~\bibnamefont {Focher}},
  \bibinfo {author} {\bibfnamefont {S.}~\bibnamefont {Scandolo}}, \bibinfo
  {author} {\bibfnamefont {E.}~\bibnamefont {Tosatti}}, \ and\ \bibinfo
  {author} {\bibfnamefont {M.}~\bibnamefont {Parrinello}},\ }\href@noop {}
  {\bibfield  {journal} {\bibinfo  {journal} {J. Phys. Chem. Solids}\ }\textbf
  {\bibinfo {volume} {56}},\ \bibinfo {pages} {501} (\bibinfo {year}
  {1995})}\BibitemShut {NoStop}%
\bibitem [{\citenamefont {Tassone}\ \emph {et~al.}(1994)\citenamefont
  {Tassone}, \citenamefont {Mauri},\ and\ \citenamefont
  {Car}}]{tassone_acceleration_1994}%
  \BibitemOpen
  \bibfield  {author} {\bibinfo {author} {\bibfnamefont {F.}~\bibnamefont
  {Tassone}}, \bibinfo {author} {\bibfnamefont {F.}~\bibnamefont {Mauri}}, \
  and\ \bibinfo {author} {\bibfnamefont {R.}~\bibnamefont {Car}},\ }\href@noop
  {} {\bibfield  {journal} {\bibinfo  {journal} {Phys. Rev. B}\ }\textbf
  {\bibinfo {volume} {50}},\ \bibinfo {pages} {10561} (\bibinfo {year}
  {1994})}\BibitemShut {NoStop}%
\bibitem [{\citenamefont {Kingma}\ and\ \citenamefont
  {Ba}(2015)}]{Kingma2015adam}%
  \BibitemOpen
  \bibfield  {author} {\bibinfo {author} {\bibfnamefont {D.}~\bibnamefont
  {Kingma}}\ and\ \bibinfo {author} {\bibfnamefont {J.}~\bibnamefont {Ba}},\
  }in\ \href@noop {} {\emph {\bibinfo {booktitle} {Proceedings of the
  International Conference on Learning Representations (ICLR)}}}\ (\bibinfo
  {year} {2015})\BibitemShut {NoStop}%
\bibitem [{\citenamefont {Wang}\ \emph {et~al.}(2018)\citenamefont {Wang},
  \citenamefont {Zhang}, \citenamefont {Han},\ and\ \citenamefont
  {E}}]{wang2018kit}%
  \BibitemOpen
  \bibfield  {author} {\bibinfo {author} {\bibfnamefont {H.}~\bibnamefont
  {Wang}}, \bibinfo {author} {\bibfnamefont {L.}~\bibnamefont {Zhang}},
  \bibinfo {author} {\bibfnamefont {J.}~\bibnamefont {Han}}, \ and\ \bibinfo
  {author} {\bibfnamefont {W.}~\bibnamefont {E}},\ }\href@noop {} {\bibfield
  {journal} {\bibinfo  {journal} {Comput. Phys. Commun.}\ }\textbf {\bibinfo
  {volume} {228}},\ \bibinfo {pages} {178} (\bibinfo {year}
  {2018})}\BibitemShut {NoStop}%
\bibitem [{\citenamefont {Abadi}\ \emph {et~al.}(2016)\citenamefont {Abadi},
  \citenamefont {Barham}, \citenamefont {Chen}, \citenamefont {Chen},
  \citenamefont {Davis}, \citenamefont {Dean}, \citenamefont {Devin},
  \citenamefont {Ghemawat}, \citenamefont {Irving}, \citenamefont {Isard},
  \citenamefont {Kudlur}, \citenamefont {Levenberg}, \citenamefont {Monga},
  \citenamefont {Moore}, \citenamefont {Murray}, \citenamefont {Steiner},
  \citenamefont {Tucker}, \citenamefont {Vasudevan}, \citenamefont {Warden},
  \citenamefont {Wicke}, \citenamefont {Yu},\ and\ \citenamefont
  {Zheng}}]{tensorflow2015-whitepaper}%
  \BibitemOpen
  \bibfield  {author} {\bibinfo {author} {\bibfnamefont {M.}~\bibnamefont
  {Abadi}}, \bibinfo {author} {\bibfnamefont {P.}~\bibnamefont {Barham}},
  \bibinfo {author} {\bibfnamefont {J.}~\bibnamefont {Chen}}, \bibinfo {author}
  {\bibfnamefont {Z.}~\bibnamefont {Chen}}, \bibinfo {author} {\bibfnamefont
  {A.}~\bibnamefont {Davis}}, \bibinfo {author} {\bibfnamefont
  {J.}~\bibnamefont {Dean}}, \bibinfo {author} {\bibfnamefont {M.}~\bibnamefont
  {Devin}}, \bibinfo {author} {\bibfnamefont {S.}~\bibnamefont {Ghemawat}},
  \bibinfo {author} {\bibfnamefont {G.}~\bibnamefont {Irving}}, \bibinfo
  {author} {\bibfnamefont {M.}~\bibnamefont {Isard}}, \bibinfo {author}
  {\bibfnamefont {M.}~\bibnamefont {Kudlur}}, \bibinfo {author} {\bibfnamefont
  {J.}~\bibnamefont {Levenberg}}, \bibinfo {author} {\bibfnamefont
  {R.}~\bibnamefont {Monga}}, \bibinfo {author} {\bibfnamefont
  {S.}~\bibnamefont {Moore}}, \bibinfo {author} {\bibfnamefont {D.~G.}\
  \bibnamefont {Murray}}, \bibinfo {author} {\bibfnamefont {B.}~\bibnamefont
  {Steiner}}, \bibinfo {author} {\bibfnamefont {P.}~\bibnamefont {Tucker}},
  \bibinfo {author} {\bibfnamefont {V.}~\bibnamefont {Vasudevan}}, \bibinfo
  {author} {\bibfnamefont {P.}~\bibnamefont {Warden}}, \bibinfo {author}
  {\bibfnamefont {M.}~\bibnamefont {Wicke}}, \bibinfo {author} {\bibfnamefont
  {Y.}~\bibnamefont {Yu}}, \ and\ \bibinfo {author} {\bibfnamefont
  {X.}~\bibnamefont {Zheng}},\ }in\ \href@noop {} {\emph {\bibinfo {booktitle}
  {12th USENIX Symposium on Operating Systems Design and Implementation (OSDI
  16)}}}\ (\bibinfo {year} {2016})\ pp.\ \bibinfo {pages}
  {265--283}\BibitemShut {NoStop}%
\bibitem [{\citenamefont {Plimpton}(1995)}]{plimpton_fast_1995}%
  \BibitemOpen
  \bibfield  {author} {\bibinfo {author} {\bibfnamefont {S.}~\bibnamefont
  {Plimpton}},\ }\href@noop {} {\bibfield  {journal} {\bibinfo  {journal} {J.
  Comput. Phys.}\ }\textbf {\bibinfo {volume} {117}},\ \bibinfo {pages} {1}
  (\bibinfo {year} {1995})}\BibitemShut {NoStop}%
\bibitem [{\citenamefont {Tuckerman}\ \emph {et~al.}(2006)\citenamefont
  {Tuckerman}, \citenamefont {Alejandre}, \citenamefont {L{\'o}pez-Rend{\'o}n},
  \citenamefont {Jochim},\ and\ \citenamefont
  {Martyna}}]{tuckerman_liouville-operator_2006}%
  \BibitemOpen
  \bibfield  {author} {\bibinfo {author} {\bibfnamefont {M.~E.}\ \bibnamefont
  {Tuckerman}}, \bibinfo {author} {\bibfnamefont {J.}~\bibnamefont
  {Alejandre}}, \bibinfo {author} {\bibfnamefont {R.}~\bibnamefont
  {L{\'o}pez-Rend{\'o}n}}, \bibinfo {author} {\bibfnamefont {A.~L.}\
  \bibnamefont {Jochim}}, \ and\ \bibinfo {author} {\bibfnamefont {G.~J.}\
  \bibnamefont {Martyna}},\ }\href@noop {} {\bibfield  {journal} {\bibinfo
  {journal} {J. Phys. A: Math. Gen.}\ }\textbf {\bibinfo {volume} {39}},\
  \bibinfo {pages} {5629} (\bibinfo {year} {2006})}\BibitemShut {NoStop}%
\bibitem [{\citenamefont {Skinner}\ \emph {et~al.}(2013)\citenamefont
  {Skinner}, \citenamefont {Huang}, \citenamefont {Schlesinger}, \citenamefont
  {Pettersson}, \citenamefont {Nilsson},\ and\ \citenamefont
  {Benmore}}]{skinner_benchmark_2013}%
  \BibitemOpen
  \bibfield  {author} {\bibinfo {author} {\bibfnamefont {L.~B.}\ \bibnamefont
  {Skinner}}, \bibinfo {author} {\bibfnamefont {C.}~\bibnamefont {Huang}},
  \bibinfo {author} {\bibfnamefont {D.}~\bibnamefont {Schlesinger}}, \bibinfo
  {author} {\bibfnamefont {L.~G.~M.}\ \bibnamefont {Pettersson}}, \bibinfo
  {author} {\bibfnamefont {A.}~\bibnamefont {Nilsson}}, \ and\ \bibinfo
  {author} {\bibfnamefont {C.~J.}\ \bibnamefont {Benmore}},\ }\href@noop {}
  {\bibfield  {journal} {\bibinfo  {journal} {J. Chem. Phys.}\ }\textbf
  {\bibinfo {volume} {138}},\ \bibinfo {pages} {074506} (\bibinfo {year}
  {2013})}\BibitemShut {NoStop}%
\bibitem [{\citenamefont {Zeidler}\ \emph {et~al.}(2011)\citenamefont
  {Zeidler}, \citenamefont {Salmon}, \citenamefont {Fischer}, \citenamefont
  {Neuefeind}, \citenamefont {Simonson}, \citenamefont {Lemmel}, \citenamefont
  {Rauch},\ and\ \citenamefont {Markland}}]{zeidler_oxygen_2011}%
  \BibitemOpen
  \bibfield  {author} {\bibinfo {author} {\bibfnamefont {A.}~\bibnamefont
  {Zeidler}}, \bibinfo {author} {\bibfnamefont {P.~S.}\ \bibnamefont {Salmon}},
  \bibinfo {author} {\bibfnamefont {H.~E.}\ \bibnamefont {Fischer}}, \bibinfo
  {author} {\bibfnamefont {J.~C.}\ \bibnamefont {Neuefeind}}, \bibinfo {author}
  {\bibfnamefont {J.~M.}\ \bibnamefont {Simonson}}, \bibinfo {author}
  {\bibfnamefont {H.}~\bibnamefont {Lemmel}}, \bibinfo {author} {\bibfnamefont
  {H.}~\bibnamefont {Rauch}}, \ and\ \bibinfo {author} {\bibfnamefont {T.~E.}\
  \bibnamefont {Markland}},\ }\href@noop {} {\bibfield  {journal} {\bibinfo
  {journal} {Phys. Rev. Lett.}\ }\textbf {\bibinfo {volume} {107}},\ \bibinfo
  {pages} {145501} (\bibinfo {year} {2011})}\BibitemShut {NoStop}%
\bibitem [{\citenamefont {Zeidler}\ \emph {et~al.}(2012)\citenamefont
  {Zeidler}, \citenamefont {Salmon}, \citenamefont {Fischer}, \citenamefont
  {Neuefeind}, \citenamefont {Simonson},\ and\ \citenamefont
  {Markland}}]{zeidler_isotope_2012}%
  \BibitemOpen
  \bibfield  {author} {\bibinfo {author} {\bibfnamefont {A.}~\bibnamefont
  {Zeidler}}, \bibinfo {author} {\bibfnamefont {P.~S.}\ \bibnamefont {Salmon}},
  \bibinfo {author} {\bibfnamefont {H.~E.}\ \bibnamefont {Fischer}}, \bibinfo
  {author} {\bibfnamefont {J.~C.}\ \bibnamefont {Neuefeind}}, \bibinfo {author}
  {\bibfnamefont {J.~M.}\ \bibnamefont {Simonson}}, \ and\ \bibinfo {author}
  {\bibfnamefont {T.~E.}\ \bibnamefont {Markland}},\ }\href@noop {} {\bibfield
  {journal} {\bibinfo  {journal} {J. Phys.: Condens. Matter}\ }\textbf
  {\bibinfo {volume} {24}},\ \bibinfo {pages} {284126} (\bibinfo {year}
  {2012})}\BibitemShut {NoStop}%
\bibitem [{\citenamefont {Cook}\ \emph {et~al.}(1974)\citenamefont {Cook},
  \citenamefont {De~Lucia},\ and\ \citenamefont
  {Helminger}}]{cook_molecular_1974}%
  \BibitemOpen
  \bibfield  {author} {\bibinfo {author} {\bibfnamefont {R.~L.}\ \bibnamefont
  {Cook}}, \bibinfo {author} {\bibfnamefont {F.~C.}\ \bibnamefont {De~Lucia}},
  \ and\ \bibinfo {author} {\bibfnamefont {P.}~\bibnamefont {Helminger}},\
  }\href@noop {} {\bibfield  {journal} {\bibinfo  {journal} {J. Mol.
  Spectrosc.}\ }\textbf {\bibinfo {volume} {53}},\ \bibinfo {pages} {62}
  (\bibinfo {year} {1974})}\BibitemShut {NoStop}%
\bibitem [{\citenamefont {Kuhs}\ and\ \citenamefont
  {Lehmann}(1986)}]{franks_structure_1986}%
  \BibitemOpen
  \bibfield  {author} {\bibinfo {author} {\bibfnamefont {W.~F.}\ \bibnamefont
  {Kuhs}}\ and\ \bibinfo {author} {\bibfnamefont {M.~S.}\ \bibnamefont
  {Lehmann}},\ }in\ \href@noop {} {\emph {\bibinfo {booktitle} {Water {Science}
  {Reviews} 2}}},\ \bibinfo {editor} {edited by\ \bibinfo {editor}
  {\bibfnamefont {F.}~\bibnamefont {Franks}}}\ (\bibinfo  {publisher}
  {Cambridge University Press},\ \bibinfo {address} {Cambridge},\ \bibinfo
  {year} {1986})\ pp.\ \bibinfo {pages} {1--66}\BibitemShut {NoStop}%
\bibitem [{\citenamefont {Zen}\ \emph {et~al.}(2015)\citenamefont {Zen},
  \citenamefont {Luo}, \citenamefont {Mazzola}, \citenamefont {Guidoni},\ and\
  \citenamefont {Sorella}}]{zen_ab_2015}%
  \BibitemOpen
  \bibfield  {author} {\bibinfo {author} {\bibfnamefont {A.}~\bibnamefont
  {Zen}}, \bibinfo {author} {\bibfnamefont {Y.}~\bibnamefont {Luo}}, \bibinfo
  {author} {\bibfnamefont {G.}~\bibnamefont {Mazzola}}, \bibinfo {author}
  {\bibfnamefont {L.}~\bibnamefont {Guidoni}}, \ and\ \bibinfo {author}
  {\bibfnamefont {S.}~\bibnamefont {Sorella}},\ }\href@noop {} {\bibfield
  {journal} {\bibinfo  {journal} {J. Chem. Phys.}\ }\textbf {\bibinfo {volume}
  {142}},\ \bibinfo {pages} {144111} (\bibinfo {year} {2015})}\BibitemShut
  {NoStop}%
\bibitem [{\citenamefont {Santra}\ \emph {et~al.}(2007)\citenamefont {Santra},
  \citenamefont {Michaelides},\ and\ \citenamefont
  {Scheffler}}]{santra_accuracy_2007}%
  \BibitemOpen
  \bibfield  {author} {\bibinfo {author} {\bibfnamefont {B.}~\bibnamefont
  {Santra}}, \bibinfo {author} {\bibfnamefont {A.}~\bibnamefont {Michaelides}},
  \ and\ \bibinfo {author} {\bibfnamefont {M.}~\bibnamefont {Scheffler}},\
  }\href@noop {} {\bibfield  {journal} {\bibinfo  {journal} {J. Chem. Phys.}\
  }\textbf {\bibinfo {volume} {127}},\ \bibinfo {pages} {184104} (\bibinfo
  {year} {2007})}\BibitemShut {NoStop}%
\bibitem [{\citenamefont {Santra}\ \emph {et~al.}(2008)\citenamefont {Santra},
  \citenamefont {Michaelides}, \citenamefont {Fuchs}, \citenamefont
  {Tkatchenko}, \citenamefont {Filippi},\ and\ \citenamefont
  {Scheffler}}]{santra_accuracy_2008}%
  \BibitemOpen
  \bibfield  {author} {\bibinfo {author} {\bibfnamefont {B.}~\bibnamefont
  {Santra}}, \bibinfo {author} {\bibfnamefont {A.}~\bibnamefont {Michaelides}},
  \bibinfo {author} {\bibfnamefont {M.}~\bibnamefont {Fuchs}}, \bibinfo
  {author} {\bibfnamefont {A.}~\bibnamefont {Tkatchenko}}, \bibinfo {author}
  {\bibfnamefont {C.}~\bibnamefont {Filippi}}, \ and\ \bibinfo {author}
  {\bibfnamefont {M.}~\bibnamefont {Scheffler}},\ }\href@noop {} {\bibfield
  {journal} {\bibinfo  {journal} {J. Chem. Phys.}\ }\textbf {\bibinfo {volume}
  {129}},\ \bibinfo {pages} {194111} (\bibinfo {year} {2008})}\BibitemShut
  {NoStop}%
\bibitem [{\citenamefont {Hart}\ \emph {et~al.}(2005)\citenamefont {Hart},
  \citenamefont {Benmore}, \citenamefont {Neuefeind}, \citenamefont {Kohara},
  \citenamefont {Tomberli},\ and\ \citenamefont
  {Egelstaff}}]{hart_temperature_2005}%
  \BibitemOpen
  \bibfield  {author} {\bibinfo {author} {\bibfnamefont {R.~T.}\ \bibnamefont
  {Hart}}, \bibinfo {author} {\bibfnamefont {C.~J.}\ \bibnamefont {Benmore}},
  \bibinfo {author} {\bibfnamefont {J.}~\bibnamefont {Neuefeind}}, \bibinfo
  {author} {\bibfnamefont {S.}~\bibnamefont {Kohara}}, \bibinfo {author}
  {\bibfnamefont {B.}~\bibnamefont {Tomberli}}, \ and\ \bibinfo {author}
  {\bibfnamefont {P.~A.}\ \bibnamefont {Egelstaff}},\ }\href@noop {} {\bibfield
   {journal} {\bibinfo  {journal} {Phys. Rev. Lett.}\ }\textbf {\bibinfo
  {volume} {94}},\ \bibinfo {pages} {047801} (\bibinfo {year}
  {2005})}\BibitemShut {NoStop}%
\bibitem [{\citenamefont {Wang}\ \emph {et~al.}(1994)\citenamefont {Wang},
  \citenamefont {Tripathi},\ and\ \citenamefont {Smith}}]{wang_chemical_1994}%
  \BibitemOpen
  \bibfield  {author} {\bibinfo {author} {\bibfnamefont {J.}~\bibnamefont
  {Wang}}, \bibinfo {author} {\bibfnamefont {A.~N.}\ \bibnamefont {Tripathi}},
  \ and\ \bibinfo {author} {\bibfnamefont {V.~H.}\ \bibnamefont {Smith}},\
  }\href@noop {} {\bibfield  {journal} {\bibinfo  {journal} {J. Chem. Phys.}\
  }\textbf {\bibinfo {volume} {101}},\ \bibinfo {pages} {4842} (\bibinfo {year}
  {1994})}\BibitemShut {NoStop}%
\bibitem [{\citenamefont {Schmidt}\ \emph {et~al.}(2009)\citenamefont
  {Schmidt}, \citenamefont {VandeVondele}, \citenamefont {Kuo}, \citenamefont
  {Sebastiani}, \citenamefont {Siepmann}, \citenamefont {Hutter},\ and\
  \citenamefont {Mundy}}]{schmidt_isobaricisothermal_2009}%
  \BibitemOpen
  \bibfield  {author} {\bibinfo {author} {\bibfnamefont {J.}~\bibnamefont
  {Schmidt}}, \bibinfo {author} {\bibfnamefont {J.}~\bibnamefont
  {VandeVondele}}, \bibinfo {author} {\bibfnamefont {I.-F.~W.}\ \bibnamefont
  {Kuo}}, \bibinfo {author} {\bibfnamefont {D.}~\bibnamefont {Sebastiani}},
  \bibinfo {author} {\bibfnamefont {J.~I.}\ \bibnamefont {Siepmann}}, \bibinfo
  {author} {\bibfnamefont {J.}~\bibnamefont {Hutter}}, \ and\ \bibinfo {author}
  {\bibfnamefont {C.~J.}\ \bibnamefont {Mundy}},\ }\href@noop {} {\bibfield
  {journal} {\bibinfo  {journal} {J. Phys. Chem. B}\ }\textbf {\bibinfo
  {volume} {113}},\ \bibinfo {pages} {11959} (\bibinfo {year}
  {2009})}\BibitemShut {NoStop}%
\bibitem [{\citenamefont {Wang}\ \emph {et~al.}(2011)\citenamefont {Wang},
  \citenamefont {Rom{\'a}n-P{\'e}rez}, \citenamefont {Soler}, \citenamefont
  {Artacho},\ and\ \citenamefont {Fern{\'a}ndez-Serra}}]{wang_density_2011}%
  \BibitemOpen
  \bibfield  {author} {\bibinfo {author} {\bibfnamefont {J.}~\bibnamefont
  {Wang}}, \bibinfo {author} {\bibfnamefont {G.}~\bibnamefont
  {Rom{\'a}n-P{\'e}rez}}, \bibinfo {author} {\bibfnamefont {J.~M.}\
  \bibnamefont {Soler}}, \bibinfo {author} {\bibfnamefont {E.}~\bibnamefont
  {Artacho}}, \ and\ \bibinfo {author} {\bibfnamefont {M.-V.}\ \bibnamefont
  {Fern{\'a}ndez-Serra}},\ }\href@noop {} {\bibfield  {journal} {\bibinfo
  {journal} {J. Chem. Phys.}\ }\textbf {\bibinfo {volume} {134}},\ \bibinfo
  {pages} {024516} (\bibinfo {year} {2011})}\BibitemShut {NoStop}%
\bibitem [{\citenamefont {Lin}\ \emph {et~al.}(2012)\citenamefont {Lin},
  \citenamefont {Seitsonen}, \citenamefont {Tavernelli},\ and\ \citenamefont
  {Rothlisberger}}]{lin_structure_2012}%
  \BibitemOpen
  \bibfield  {author} {\bibinfo {author} {\bibfnamefont {I.-C.}\ \bibnamefont
  {Lin}}, \bibinfo {author} {\bibfnamefont {A.~P.}\ \bibnamefont {Seitsonen}},
  \bibinfo {author} {\bibfnamefont {I.}~\bibnamefont {Tavernelli}}, \ and\
  \bibinfo {author} {\bibfnamefont {U.}~\bibnamefont {Rothlisberger}},\
  }\href@noop {} {\bibfield  {journal} {\bibinfo  {journal} {J. Chem. Theory
  Comput.}\ }\textbf {\bibinfo {volume} {8}},\ \bibinfo {pages} {3902}
  (\bibinfo {year} {2012})}\BibitemShut {NoStop}%
\bibitem [{\citenamefont {Gaiduk}\ \emph {et~al.}(2015)\citenamefont {Gaiduk},
  \citenamefont {Gygi},\ and\ \citenamefont {Galli}}]{gaiduk_density_2015}%
  \BibitemOpen
  \bibfield  {author} {\bibinfo {author} {\bibfnamefont {A.~P.}\ \bibnamefont
  {Gaiduk}}, \bibinfo {author} {\bibfnamefont {F.}~\bibnamefont {Gygi}}, \ and\
  \bibinfo {author} {\bibfnamefont {G.}~\bibnamefont {Galli}},\ }\href@noop {}
  {\bibfield  {journal} {\bibinfo  {journal} {J. Phys. Chem. Lett.}\ }\textbf
  {\bibinfo {volume} {6}},\ \bibinfo {pages} {2902} (\bibinfo {year}
  {2015})}\BibitemShut {NoStop}%
\bibitem [{\citenamefont {Miceli}\ \emph {et~al.}(2015)\citenamefont {Miceli},
  \citenamefont {de~Gironcoli},\ and\ \citenamefont
  {Pasquarello}}]{miceli_isobaric_2015}%
  \BibitemOpen
  \bibfield  {author} {\bibinfo {author} {\bibfnamefont {G.}~\bibnamefont
  {Miceli}}, \bibinfo {author} {\bibfnamefont {S.}~\bibnamefont
  {de~Gironcoli}}, \ and\ \bibinfo {author} {\bibfnamefont {A.}~\bibnamefont
  {Pasquarello}},\ }\href@noop {} {\bibfield  {journal} {\bibinfo  {journal}
  {J. Chem. Phys.}\ }\textbf {\bibinfo {volume} {142}},\ \bibinfo {pages}
  {034501} (\bibinfo {year} {2015})}\BibitemShut {NoStop}%
\bibitem [{\citenamefont {Pestana}\ \emph {et~al.}(2017)\citenamefont
  {Pestana}, \citenamefont {Mardirossian}, \citenamefont {Head-Gordon},\ and\
  \citenamefont {Head-Gordon}}]{pestana_ab_2017}%
  \BibitemOpen
  \bibfield  {author} {\bibinfo {author} {\bibfnamefont {L.~R.}\ \bibnamefont
  {Pestana}}, \bibinfo {author} {\bibfnamefont {N.}~\bibnamefont
  {Mardirossian}}, \bibinfo {author} {\bibfnamefont {M.}~\bibnamefont
  {Head-Gordon}}, \ and\ \bibinfo {author} {\bibfnamefont {T.}~\bibnamefont
  {Head-Gordon}},\ }\href@noop {} {\bibfield  {journal} {\bibinfo  {journal}
  {Chem. Sci.}\ }\textbf {\bibinfo {volume} {8}},\ \bibinfo {pages} {3554}
  (\bibinfo {year} {2017})}\BibitemShut {NoStop}%
\bibitem [{\citenamefont {Morrone}\ and\ \citenamefont
  {Car}(2008)}]{morrone_nuclear_2008}%
  \BibitemOpen
  \bibfield  {author} {\bibinfo {author} {\bibfnamefont {J.~A.}\ \bibnamefont
  {Morrone}}\ and\ \bibinfo {author} {\bibfnamefont {R.}~\bibnamefont {Car}},\
  }\href@noop {} {\bibfield  {journal} {\bibinfo  {journal} {Phys. Rev. Lett.}\
  }\textbf {\bibinfo {volume} {101}},\ \bibinfo {pages} {017801} (\bibinfo
  {year} {2008})}\BibitemShut {NoStop}%
\bibitem [{\citenamefont {Reilly}\ \emph {et~al.}(2016)\citenamefont {Reilly},
  \citenamefont {Cooper}, \citenamefont {Adjiman}, \citenamefont
  {Bhattacharya}, \citenamefont {Boese}, \citenamefont {Brandenburg},
  \citenamefont {Bygrave}, \citenamefont {Bylsma}, \citenamefont {Campbell},
  \citenamefont {Car}, \citenamefont {Case}, \citenamefont {Chadha},
  \citenamefont {Cole}, \citenamefont {Cosburn}, \citenamefont {Cuppen},
  \citenamefont {Curtis}, \citenamefont {Day}, \citenamefont {DiStasio~Jr},
  \citenamefont {Dzyabchenko}, \citenamefont {van Eijck}, \citenamefont
  {Elking}, \citenamefont {van~den Ende}, \citenamefont {Facelli},
  \citenamefont {Ferraro}, \citenamefont {Fusti-Molnar}, \citenamefont
  {Gatsiou}, \citenamefont {Gee}, \citenamefont {de~Gelder}, \citenamefont
  {Ghiringhelli}, \citenamefont {Goto}, \citenamefont {Grimme}, \citenamefont
  {Guo}, \citenamefont {Hofmann}, \citenamefont {Hoja}, \citenamefont {Hylton},
  \citenamefont {Iuzzolino}, \citenamefont {Jankiewicz}, \citenamefont
  {de~Jong}, \citenamefont {Kendrick}, \citenamefont {de~Klerk}, \citenamefont
  {Ko}, \citenamefont {Kuleshova}, \citenamefont {Li}, \citenamefont {Lohani},
  \citenamefont {Leusen}, \citenamefont {Lund}, \citenamefont {Lv},
  \citenamefont {Ma}, \citenamefont {Marom}, \citenamefont {Masunov},
  \citenamefont {McCabe}, \citenamefont {McMahon}, \citenamefont {Meekes},
  \citenamefont {Metz}, \citenamefont {Misquitta}, \citenamefont {Mohamed},
  \citenamefont {Monserrat}, \citenamefont {Needs}, \citenamefont {Neumann},
  \citenamefont {Nyman}, \citenamefont {Obata}, \citenamefont {Oberhofer},
  \citenamefont {Oganov}, \citenamefont {Orendt}, \citenamefont {Pagola},
  \citenamefont {Pantelides}, \citenamefont {Pickard}, \citenamefont
  {Podeszwa}, \citenamefont {Price}, \citenamefont {Price}, \citenamefont
  {Pulido}, \citenamefont {Read}, \citenamefont {Reuter}, \citenamefont
  {Schneider}, \citenamefont {Schober}, \citenamefont {Shields}, \citenamefont
  {Singh}, \citenamefont {Sugden}, \citenamefont {Szalewicz}, \citenamefont
  {Taylor}, \citenamefont {Tkatchenko}, \citenamefont {Tuckerman},
  \citenamefont {Vacarro}, \citenamefont {Vasileiadis}, \citenamefont
  {Vazquez-Mayagoitia}, \citenamefont {Vogt}, \citenamefont {Wang},
  \citenamefont {Watson}, \citenamefont {de~Wijs}, \citenamefont {Yang},
  \citenamefont {Zhu},\ and\ \citenamefont {Groom}}]{reilly_report_2016}%
  \BibitemOpen
  \bibfield  {author} {\bibinfo {author} {\bibfnamefont {A.~M.}\ \bibnamefont
  {Reilly}}, \bibinfo {author} {\bibfnamefont {R.~I.}\ \bibnamefont {Cooper}},
  \bibinfo {author} {\bibfnamefont {C.~S.}\ \bibnamefont {Adjiman}}, \bibinfo
  {author} {\bibfnamefont {S.}~\bibnamefont {Bhattacharya}}, \bibinfo {author}
  {\bibfnamefont {A.~D.}\ \bibnamefont {Boese}}, \bibinfo {author}
  {\bibfnamefont {J.~G.}\ \bibnamefont {Brandenburg}}, \bibinfo {author}
  {\bibfnamefont {P.~J.}\ \bibnamefont {Bygrave}}, \bibinfo {author}
  {\bibfnamefont {R.}~\bibnamefont {Bylsma}}, \bibinfo {author} {\bibfnamefont
  {J.~E.}\ \bibnamefont {Campbell}}, \bibinfo {author} {\bibfnamefont
  {R.}~\bibnamefont {Car}}, \bibinfo {author} {\bibfnamefont {D.~H.}\
  \bibnamefont {Case}}, \bibinfo {author} {\bibfnamefont {R.}~\bibnamefont
  {Chadha}}, \bibinfo {author} {\bibfnamefont {J.~C.}\ \bibnamefont {Cole}},
  \bibinfo {author} {\bibfnamefont {K.}~\bibnamefont {Cosburn}}, \bibinfo
  {author} {\bibfnamefont {H.~M.}\ \bibnamefont {Cuppen}}, \bibinfo {author}
  {\bibfnamefont {F.}~\bibnamefont {Curtis}}, \bibinfo {author} {\bibfnamefont
  {G.~M.}\ \bibnamefont {Day}}, \bibinfo {author} {\bibfnamefont {R.~A.}\
  \bibnamefont {DiStasio~Jr}}, \bibinfo {author} {\bibfnamefont
  {A.}~\bibnamefont {Dzyabchenko}}, \bibinfo {author} {\bibfnamefont {B.~P.}\
  \bibnamefont {van Eijck}}, \bibinfo {author} {\bibfnamefont {D.~M.}\
  \bibnamefont {Elking}}, \bibinfo {author} {\bibfnamefont {J.~A.}\
  \bibnamefont {van~den Ende}}, \bibinfo {author} {\bibfnamefont {J.~C.}\
  \bibnamefont {Facelli}}, \bibinfo {author} {\bibfnamefont {M.~B.}\
  \bibnamefont {Ferraro}}, \bibinfo {author} {\bibfnamefont {L.}~\bibnamefont
  {Fusti-Molnar}}, \bibinfo {author} {\bibfnamefont {C.-A.}\ \bibnamefont
  {Gatsiou}}, \bibinfo {author} {\bibfnamefont {T.~S.}\ \bibnamefont {Gee}},
  \bibinfo {author} {\bibfnamefont {R.}~\bibnamefont {de~Gelder}}, \bibinfo
  {author} {\bibfnamefont {L.~M.}\ \bibnamefont {Ghiringhelli}}, \bibinfo
  {author} {\bibfnamefont {H.}~\bibnamefont {Goto}}, \bibinfo {author}
  {\bibfnamefont {S.}~\bibnamefont {Grimme}}, \bibinfo {author} {\bibfnamefont
  {R.}~\bibnamefont {Guo}}, \bibinfo {author} {\bibfnamefont {D.~W.~M.}\
  \bibnamefont {Hofmann}}, \bibinfo {author} {\bibfnamefont {J.}~\bibnamefont
  {Hoja}}, \bibinfo {author} {\bibfnamefont {R.~K.}\ \bibnamefont {Hylton}},
  \bibinfo {author} {\bibfnamefont {L.}~\bibnamefont {Iuzzolino}}, \bibinfo
  {author} {\bibfnamefont {W.}~\bibnamefont {Jankiewicz}}, \bibinfo {author}
  {\bibfnamefont {D.~T.}\ \bibnamefont {de~Jong}}, \bibinfo {author}
  {\bibfnamefont {J.}~\bibnamefont {Kendrick}}, \bibinfo {author}
  {\bibfnamefont {N.~J.~J.}\ \bibnamefont {de~Klerk}}, \bibinfo {author}
  {\bibfnamefont {H.-Y.}\ \bibnamefont {Ko}}, \bibinfo {author} {\bibfnamefont
  {L.~N.}\ \bibnamefont {Kuleshova}}, \bibinfo {author} {\bibfnamefont
  {X.}~\bibnamefont {Li}}, \bibinfo {author} {\bibfnamefont {S.}~\bibnamefont
  {Lohani}}, \bibinfo {author} {\bibfnamefont {F.~J.~J.}\ \bibnamefont
  {Leusen}}, \bibinfo {author} {\bibfnamefont {A.~M.}\ \bibnamefont {Lund}},
  \bibinfo {author} {\bibfnamefont {J.}~\bibnamefont {Lv}}, \bibinfo {author}
  {\bibfnamefont {Y.}~\bibnamefont {Ma}}, \bibinfo {author} {\bibfnamefont
  {N.}~\bibnamefont {Marom}}, \bibinfo {author} {\bibfnamefont {A.~E.}\
  \bibnamefont {Masunov}}, \bibinfo {author} {\bibfnamefont {P.}~\bibnamefont
  {McCabe}}, \bibinfo {author} {\bibfnamefont {D.~P.}\ \bibnamefont {McMahon}},
  \bibinfo {author} {\bibfnamefont {H.}~\bibnamefont {Meekes}}, \bibinfo
  {author} {\bibfnamefont {M.~P.}\ \bibnamefont {Metz}}, \bibinfo {author}
  {\bibfnamefont {A.~J.}\ \bibnamefont {Misquitta}}, \bibinfo {author}
  {\bibfnamefont {S.}~\bibnamefont {Mohamed}}, \bibinfo {author} {\bibfnamefont
  {B.}~\bibnamefont {Monserrat}}, \bibinfo {author} {\bibfnamefont {R.~J.}\
  \bibnamefont {Needs}}, \bibinfo {author} {\bibfnamefont {M.~A.}\ \bibnamefont
  {Neumann}}, \bibinfo {author} {\bibfnamefont {J.}~\bibnamefont {Nyman}},
  \bibinfo {author} {\bibfnamefont {S.}~\bibnamefont {Obata}}, \bibinfo
  {author} {\bibfnamefont {H.}~\bibnamefont {Oberhofer}}, \bibinfo {author}
  {\bibfnamefont {A.~R.}\ \bibnamefont {Oganov}}, \bibinfo {author}
  {\bibfnamefont {A.~M.}\ \bibnamefont {Orendt}}, \bibinfo {author}
  {\bibfnamefont {G.~I.}\ \bibnamefont {Pagola}}, \bibinfo {author}
  {\bibfnamefont {C.~C.}\ \bibnamefont {Pantelides}}, \bibinfo {author}
  {\bibfnamefont {C.~J.}\ \bibnamefont {Pickard}}, \bibinfo {author}
  {\bibfnamefont {R.}~\bibnamefont {Podeszwa}}, \bibinfo {author}
  {\bibfnamefont {L.~S.}\ \bibnamefont {Price}}, \bibinfo {author}
  {\bibfnamefont {S.~L.}\ \bibnamefont {Price}}, \bibinfo {author}
  {\bibfnamefont {A.}~\bibnamefont {Pulido}}, \bibinfo {author} {\bibfnamefont
  {M.~G.}\ \bibnamefont {Read}}, \bibinfo {author} {\bibfnamefont
  {K.}~\bibnamefont {Reuter}}, \bibinfo {author} {\bibfnamefont
  {E.}~\bibnamefont {Schneider}}, \bibinfo {author} {\bibfnamefont
  {C.}~\bibnamefont {Schober}}, \bibinfo {author} {\bibfnamefont {G.~P.}\
  \bibnamefont {Shields}}, \bibinfo {author} {\bibfnamefont {P.}~\bibnamefont
  {Singh}}, \bibinfo {author} {\bibfnamefont {I.~J.}\ \bibnamefont {Sugden}},
  \bibinfo {author} {\bibfnamefont {K.}~\bibnamefont {Szalewicz}}, \bibinfo
  {author} {\bibfnamefont {C.~R.}\ \bibnamefont {Taylor}}, \bibinfo {author}
  {\bibfnamefont {A.}~\bibnamefont {Tkatchenko}}, \bibinfo {author}
  {\bibfnamefont {M.~E.}\ \bibnamefont {Tuckerman}}, \bibinfo {author}
  {\bibfnamefont {F.}~\bibnamefont {Vacarro}}, \bibinfo {author} {\bibfnamefont
  {M.}~\bibnamefont {Vasileiadis}}, \bibinfo {author} {\bibfnamefont
  {A.}~\bibnamefont {Vazquez-Mayagoitia}}, \bibinfo {author} {\bibfnamefont
  {L.}~\bibnamefont {Vogt}}, \bibinfo {author} {\bibfnamefont {Y.}~\bibnamefont
  {Wang}}, \bibinfo {author} {\bibfnamefont {R.~E.}\ \bibnamefont {Watson}},
  \bibinfo {author} {\bibfnamefont {G.~A.}\ \bibnamefont {de~Wijs}}, \bibinfo
  {author} {\bibfnamefont {J.}~\bibnamefont {Yang}}, \bibinfo {author}
  {\bibfnamefont {Q.}~\bibnamefont {Zhu}}, \ and\ \bibinfo {author}
  {\bibfnamefont {C.~R.}\ \bibnamefont {Groom}},\ }\href@noop {} {\bibfield
  {journal} {\bibinfo  {journal} {Acta Crystallogr., Sect. B: Struct. Sci.}\
  }\textbf {\bibinfo {volume} {72}},\ \bibinfo {pages} {439} (\bibinfo {year}
  {2016})}\BibitemShut {NoStop}%
\bibitem [{\citenamefont {Ko}\ \emph {et~al.}(2018)\citenamefont {Ko},
  \citenamefont {DiStasio~Jr.}, \citenamefont {Santra},\ and\ \citenamefont
  {Car}}]{ko_thermal_2018}%
  \BibitemOpen
  \bibfield  {author} {\bibinfo {author} {\bibfnamefont {H.-Y.}\ \bibnamefont
  {Ko}}, \bibinfo {author} {\bibfnamefont {R.~A.}\ \bibnamefont
  {DiStasio~Jr.}}, \bibinfo {author} {\bibfnamefont {B.}~\bibnamefont
  {Santra}}, \ and\ \bibinfo {author} {\bibfnamefont {R.}~\bibnamefont {Car}},\
  }\href@noop {} {\bibfield  {journal} {\bibinfo  {journal} {Phys. Rev.
  Materials}\ }\textbf {\bibinfo {volume} {2}},\ \bibinfo {pages} {055603}
  (\bibinfo {year} {2018})}\BibitemShut {NoStop}%
\bibitem [{\citenamefont {Hoja}\ \emph {et~al.}(2019)\citenamefont {Hoja},
  \citenamefont {Ko}, \citenamefont {Neumann}, \citenamefont {Car},
  \citenamefont {DiStasio},\ and\ \citenamefont
  {Tkatchenko}}]{hoja_reliable_2019}%
  \BibitemOpen
  \bibfield  {author} {\bibinfo {author} {\bibfnamefont {J.}~\bibnamefont
  {Hoja}}, \bibinfo {author} {\bibfnamefont {H.-Y.}\ \bibnamefont {Ko}},
  \bibinfo {author} {\bibfnamefont {M.~A.}\ \bibnamefont {Neumann}}, \bibinfo
  {author} {\bibfnamefont {R.}~\bibnamefont {Car}}, \bibinfo {author}
  {\bibfnamefont {R.~A.}\ \bibnamefont {DiStasio}}, \ and\ \bibinfo {author}
  {\bibfnamefont {A.}~\bibnamefont {Tkatchenko}},\ }\href@noop {} {\bibfield
  {journal} {\bibinfo  {journal} {Sci. Adv.}\ }\textbf {\bibinfo {volume}
  {5}},\ \bibinfo {pages} {eaau3338} (\bibinfo {year} {2019})}\BibitemShut
  {NoStop}%
\bibitem [{\citenamefont {Tkatchenko}\ \emph {et~al.}(2012)\citenamefont
  {Tkatchenko}, \citenamefont {DiStasio~Jr.}, \citenamefont {Car},\ and\
  \citenamefont {Scheffler}}]{tkatchenko_accurate_2012}%
  \BibitemOpen
  \bibfield  {author} {\bibinfo {author} {\bibfnamefont {A.}~\bibnamefont
  {Tkatchenko}}, \bibinfo {author} {\bibfnamefont {R.~A.}\ \bibnamefont
  {DiStasio~Jr.}}, \bibinfo {author} {\bibfnamefont {R.}~\bibnamefont {Car}}, \
  and\ \bibinfo {author} {\bibfnamefont {M.}~\bibnamefont {Scheffler}},\
  }\href@noop {} {\bibfield  {journal} {\bibinfo  {journal} {Phys. Rev. Lett.}\
  }\textbf {\bibinfo {volume} {108}},\ \bibinfo {pages} {236402} (\bibinfo
  {year} {2012})}\BibitemShut {NoStop}%
\bibitem [{\citenamefont {Ambrosetti}\ \emph {et~al.}(2014)\citenamefont
  {Ambrosetti}, \citenamefont {Reilly}, \citenamefont {DiStasio~Jr.},\ and\
  \citenamefont {Tkatchenko}}]{ambrosetti_long-range_2014}%
  \BibitemOpen
  \bibfield  {author} {\bibinfo {author} {\bibfnamefont {A.}~\bibnamefont
  {Ambrosetti}}, \bibinfo {author} {\bibfnamefont {A.~M.}\ \bibnamefont
  {Reilly}}, \bibinfo {author} {\bibfnamefont {R.~A.}\ \bibnamefont
  {DiStasio~Jr.}}, \ and\ \bibinfo {author} {\bibfnamefont {A.}~\bibnamefont
  {Tkatchenko}},\ }\href@noop {} {\bibfield  {journal} {\bibinfo  {journal} {J.
  Chem. Phys.}\ }\textbf {\bibinfo {volume} {140}},\ \bibinfo {pages} {18A508}
  (\bibinfo {year} {2014})}\BibitemShut {NoStop}%
\bibitem [{\citenamefont {Blood-Forsythe}\ \emph {et~al.}(2016)\citenamefont
  {Blood-Forsythe}, \citenamefont {Markovich}, \citenamefont {DiStasio~Jr.},
  \citenamefont {Car},\ and\ \citenamefont
  {Aspuru-Guzik}}]{blood-forsythe_analytical_2016}%
  \BibitemOpen
  \bibfield  {author} {\bibinfo {author} {\bibfnamefont {M.~A.}\ \bibnamefont
  {Blood-Forsythe}}, \bibinfo {author} {\bibfnamefont {T.}~\bibnamefont
  {Markovich}}, \bibinfo {author} {\bibfnamefont {R.~A.}\ \bibnamefont
  {DiStasio~Jr.}}, \bibinfo {author} {\bibfnamefont {R.}~\bibnamefont {Car}}, \
  and\ \bibinfo {author} {\bibfnamefont {A.}~\bibnamefont {Aspuru-Guzik}},\
  }\href@noop {} {\bibfield  {journal} {\bibinfo  {journal} {Chem. Sci.}\
  }\textbf {\bibinfo {volume} {7}},\ \bibinfo {pages} {1712} (\bibinfo {year}
  {2016})}\BibitemShut {NoStop}%
\end{thebibliography}

%

\end{document}